
\input phyzzx
\nopagenumbers

%
\def\refout{\par\penalty-400\vskip\chapterskip
   \spacecheck\referenceminspace
   \ifreferenceopen \Closeout\referencewrite \referenceopenfalse \fi
   \noindent{\bf References\hfil}\vskip\headskip
   \input \jobname.refs }
\def\chapter#1{\par \penalty-300
   \chapterreset \noindent{\bf \chapterlabel.~~#1}
   \nobreak \penalty 30000 }

\def\IR{\relax{\rm I\kern-.18em R}}

\def\npb#1#2#3{{\it Nucl. Phys.} {\bf B#1} (#2) #3 }
\def\plb#1#2#3{{\it Phys. Lett.} {\bf B#1} (#2) #3 }
\def\prd#1#2#3{{\it Phys. Rev. } {\bf D#1} (#2) #3 }
\def\prl#1#2#3{{\it Phys. Rev. Lett.} {\bf #1} (#2) #3 }
\def\mpla#1#2#3{{\it Mod. Phys. Lett.} {\bf A#1} (#2) #3 }
\def\ijmpa#1#2#3{{\it Int. J. Mod. Phys.} {\bf A#1} (#2) #3 }

\def\cmp#1#2#3{{\it Commun. Math. Phys.} {\bf #1} (#2) #3 }

\def\ptp#1#2#3{{\it Prog. Theor. Phys.} {\bf #1} (#2) #3 }
\def\bb#1{{\tt hep-th/#1}}
\REF\Klebreview{I. R. Klebanov, {\it ``String theory in two
dimensions'',} in ``String Theory and Quantum Gravity'',
Proceedings of the Trieste Spring School 1991, eds. J. Harvey et al.,
(World Scientific, Singapore, 1992).}
\REF\Kutreview {D. Kutasov, {\it ``Some properties of (non) critical
strings'',} in ``String Theory and Quantum Gravity'',  Proceedings of
the Trieste Spring School 1991, eds. J. Harvey et al., (World Scientific,
Singapore, 1992).}
\REF\Polone {J. Polchinski, \npb {346}{1990}{253.}}
\REF\WBH {E. Witten, \prd {44} {1991} {314.}}
\REF\BaNe {I. Bars and B. Nemeschansky, \npb {348} {1991} {89.}}
\REF\RSS {M. Ro\v cek, K. Schoutens and A. Sevrin, \plb {265} {1991}
{303.}}
\REF\MaWa {G. Mandal, A. Sengupta and S. Wadia, \mpla {6}{1991}{1685.}}
\REF\El {S. Elizur, A. Forge and E. Rabinovici, \npb {359}{1991}{581.}}
\REF\Po{A.~M. Polyakov, \mpla {6}{1991}{635.}}
\REF\GoLi {M. Goulian and M. Li, \prl {66}{1990}{2051.}}
\REF\DiFrKu{P. DiFrancesco and D. Kutasov, \plb {261}{1991}{385;}
\npb {375}{1991}{119;} Y. Kitazawa, \plb {265}{1991}{262;}
Y. Tanii, \ptp {86}{1991}{547;}
V.~S. Dotsenko, \mpla {6}{1991}{3601.}}
\REF\Bab{O. Babelon, \plb {215}{1988}{523.}}
\REF\Ge {J.-L. Gervais, \cmp {130}{1990}{257;} \npb {391}{1993}{287.}}
\REF\JeSa {A. Jevicki and B. Sakita, \npb {165} {1980} {511.}}
\REF\DaJe {S.~R. Das and A. Jevicki, \mpla {5} {1990} {1639.}}
\REF\DJR {K. Demeterfi, A. Jevicki and J.~P. Rodrigues, \npb {362}{1991}
{173;}  \npb {365} {1991} {499.}}
\REF\Poltwo { J. Polchinski, \npb {362} {1991} {125.}}

\REF\AvJe {J. Avan and A. Jevicki, \plb {266}{1991}{35;}
\plb {272}{1991}{17.}}
\REF\MoSe {G. Moore and N. Seiberg, \ijmpa{7}{1992}{2601.}}
\REF\GKN {D.~J. Gross, I.~R. Klebanov and M. Newman,
\npb {350} {1991} {671.}}
\REF\LM {J. Lee and P.~F. Mende, \plb {312} {1993} {433.}}
\REF\winf {D. Minic, J. Polchinski and Z. Yang, \npb {369}{1992}{324;}}
\REF\WiGR {E. Witten, \npb {373}{1992}{187.}}
\REF\KlPo {I.~R. Klebanov and A.~M. Polyakov, \mpla {6}{1991}{3273.}}
\REF\JRvT {A. Jevicki, J.~P. Rodrigues and A. van Tonder,
\npb {404} {1993} {91.}}
\REF\Kl {I.~R. Klebanov, \mpla {7}{1992}{723.}}
\REF\ZW {E.Witten and B.Zwiebach, \npb {377}{1992}{55.}}
\REF\Ve {E. Verlinde, \npb {381}{1992}{141.}}
\REF\KlPa{I.~R. Klebanov and A. Pasquinucci, \npb {393}{1993}{261.}}
\REF\Ba {J.~L.~F. Barbon, \ijmpa {7}{1992}{7579;}
Y. Kazama and H. Nicolai, {\it ``On the exact operator formalism of
two-dimensional Liouville quantum gravity in Minkowski space-time,''}
DESY-93-043, \bb{9305023};
V.~S. Dotsenko, {\it ``Remarks on the physical states and the chiral
algebra of 2D gravity coupled to $c\le 1$ matter,''}
PAR-LPTHE-92-4, \bb{9201077}; \mpla {7}{1992}{2505.}}
\REF\GrKl {D.~J. Gross and I.~R. Klebanov, \npb {359} {1991} {3.}}
\REF\MoPl {G. Moore and R. Plesser, \prd {46} {1992} {1730.}}
\REF\MaSh {E. Martinec and S. Shatashvili, \npb {368}{1992}{338.}}
\REF\JeYo {A. Jevicki and T. Yoneya, {\it ``A deformed matrix model and
the black hole background in two-dimensional string theory'',}
NSF-ITP-93-67, BROWN-HEP-904, UT-KOMABA/93-10, \bb{9305109}.}
\REF\DDMW {S.~R. Das, \mpla {8} {1993} {69;}
A. Dhar, G. Mandal and S. Wadia, \mpla {7} {1992} {370.}}
\REF\MuVa {S. Mukhi and C. Vafa, {\it ``Two-dimensional black hole
as a topological coset model of $c=1$ string theory,''}
HUTP-93/A002, TIRF/TH/93-01, \bb{9301083}.}
\REF\DVV {R. Dijkgraaf, H. Verlinde and E. Verlinde,
\npb {371} {1992} {269.}}
\REF\new{M. Bershadsky and D. Kutasov, \plb {266}{1991}{345;}
T. Eguchi, H. Kanno and S.-K. Yang, \plb {298}{1993}{73;}
H. Ishikawa and M. Kato, \plb {302}{1993}{209.}}
\REF\AJcmp {J. Avan and A. Jevicki, \cmp {150}{1992}{149.}}
\REF\DeRo {K. Demeterfi and J.~P. Rodrigues, {\it ``States and
quantum effects in the collective field theory of a deformed matrix
model'',} PUPT-1407, CNLS-93-06, \bb {9306141};
K. Demeterfi, I.~R. Klebanov and J.~P. Rodrigues,
{\it The exact $S$-matrix of the deformed $c=1$ matrix model,''}
PUPT-1416, CNLS-93-09, \bb {9308036}.}
\REF\Da{U. Danielsson, {\it ``A matrix-model black hole,''}
CERN-TH.6916/93, \bb{9306063}.}
\REF\VV {E. Verlinde and H. Verlinde, {\it ``A unitary $S$-matrix
and 2D black hole formation and evaporation'',} IASSNS-HEP-93/18,
PUPT-1380, \bb {9302022}.}
\REF\Eguchi{T. Eguchi, {\it ``$c=1$ Liouville theory perturbed by the
black-hole mass operator,''} UT 650, \bb{9307185}.}

\singlespace
\hsize=6.0in
\vsize=8.5in
\voffset=0.0in
\hoffset=0.0in
\overfullrule=0pt

\line{}
\line{\hfill HET-918}
\line{\hfill TA-502}
\vskip .75in
\centerline{{\bf DEVELOPMENTS IN 2D STRING THEORY}}
\vskip .40in
\centerline{ANTAL JEVICKI}
\smallskip
\centerline{{\it Physics Department, Brown University}}
\centerline{{\it Providence, Rhode Island 02912, USA}}
\vskip .25in
\centerline{(Lectures presented at the Spring School }
\centerline{in String Theory, Trieste, Italy, April, 1993)}

\vskip 0.70in

{\chapter{Introduction}}
\medskip

Recent years have witnessed a remarkable progress in 2d  string theory
and quantum gravity. Beginning with matrix models one found a new and
computationally powerful description of the theory, free of mathematical
complexities. The relevance of these models to
string theory comes through a $1/N$ expansion where $1/N$ plays the role of a
bare string coupling constant $g_{\rm st}^0 = 1/N$.  This classifies
Feynman diagrams according to their topology; for a fixed topology the
sum of all graphs in a dual picture becomes a sum of triangulated
surfaces.  The continuum theory is then approached by sending
the value of lattice spacing to zero.

This heuristic picture was completely carried out in one dimension giving
an exactly solvable theory of two-dimensional strings
(for an earlier review see [\Klebreview]).
It lead first to a series of explicit results including
the computation of free energy and correlation functions at any order in
the loop expansion. The new formulation also offered a framework for
non-perturbative investigations. It provided  a new fundamental insight into
the origin of metric fluctuations and the physical nature of the Liouville
mode. Through a critical scaling limit a two-dimensional theory is
generated where the logarithmic scaling violation is seen to be the origin of
the extra dimension.

Most of the interesting features of 2d strings were
clearly exhibited in the field-theoretic description achieved in
terms of  collective field theory. Starting from matrix models one
builds a field theory describing the dynamics of observable (Wilson) loop
variables. The collective Hamiltonian describes the processes of
joining and splitting of loops, giving
A cubic interaction and a linear (tadpole) term were shown to
successfully produce all tree and loop diagrams. The theory is naturally
integrable and exactly solvable. Its
integrable nature  leads to understanding of a
$w_\infty$ algebra as a space-time symmetry of the theory. This algebra acts
in a nonlinear way on the basic collective field representing the tachyon.
It is interpreted as a spectrum-generating algebra allowing to build
an infinite sequence of discrete imaginary energy states which turn
out to be remnants of higher string modes in two dimensions.
The presence and interplay of discrete modes with the scalar tachyon are
particularly interesting. The $w_\infty $ symmetry is seen to serve as
an organizational principle specifying the dynamics.

Two-dimensional physics is made even richer by the existence of other
nontrivial backgrounds. Most interesting is the black hole type classical
solution described by an exact $SL(2,\IR)/U(1)$ sigma model. Its quantum
mechanical interpretation is of major interest and was the object of
various recent studies.

Even though there is a wealth of results coming from detailed studies of
matrix models and conformal field theories a full understanding of the theory
and its dynamics is still not available. In particular, a clear correspondence
between the two fundamentally different methods is lacking. One has an
(excellent) comparison of results and a pattern of similarities and analogies
hinting at a more unified  framework. Prospects for such a framework are
particularly exciting since this would eventually represent a
new formulation of string field theory.

A need for such a general framework is most clear already when
addressing the question of the black hole. In general one would like
to command sufficient insight to be able to go from one solution to
another. This, at present, is also one of the fundamental
challenges of string theory.

In this series of lectures we describe the progress already achieved.
The emphasis is on a unified understanding of the subject.
We will try to bridge the two major approaches: the matrix model
and conformal field theory, as much as possible describing analogies
and similarities that one has between them.  In this process a
dictionary  emerges; it is most visible in the discussion
of the infinite $w_\infty $ symmetry and the associated Ward identities.
The question of incorporating the black hole background is then addressed
and some preliminary results in this direction are described.

The selection of topics covered is as follows: In sect.~2 we give a
summary of basic two-dimensional string theory (for a more detailed
review see [\Kutreview]). In sect.~3 we describe the matrix model
and a transition to field theory. We discuss the integrability of
the theory and the construction of exact states
and their string interpretation. In sect.~4 the corresponding $w_\infty$
symmetry  is described.  A detailed comparison of Ward identities and a
description of the agreement between matrix
model and conformal field constructions is given. Sect.~5 contains
the discussion of the $S$-matrix of the theory. The latter is described by
an exact generating function, connection of which to matrix model
harmonic oscillator states we emphasize. In sect.~6 we discuss the
black hole background.

\bigskip
\chapter {String Theory in Two Dimensions}
\medskip

The conceptually simplest way to discuss the dynamics of strings
is through a $\beta$-function approach which provides
effective equations for low-lying fields . In the case of a closed
string in two dimensions these are the $m^2 = 0$ scalar $T(X^{\mu}$) (the
would-be tachyon), the graviton $G_{\mu\nu}(X)$ and the dilaton $D(X)$.
The leading $\beta$-function Lagrangian reads:
$$S_{\rm eff} = {1\over 2\pi} \int d^2 X \sqrt{G} \, e^{-2D(X)}
\Bigl\{ {1\over 2} \left[ \nabla_{\mu} T \nabla^{\mu} T + 2T^2 - V\right] +
R + 4 \nabla D\cdot\nabla D + \ldots\Bigr\}\,\,. \eqno\eq$$
The tachyon potential $V(T)$ is not so well known and neither are the
couplings to possibly higher--spin fields.  But this effective Lagrangian
exhibits several simple solutions which can serve as classical
configurations of two-dimensional string theory.

Denoting $X^{\mu} \equiv (X^0 = t , X^1= \varphi)$ one
has the {\it linear dilaton vacuum} solution
$$\eqalign{&T(X)  = 0\,\,,\cr
\noalign{\vskip 0.1cm}
&G_{\mu\nu}(X)  = \eta_{\mu\nu}\,\,,\cr
\noalign{\vskip 0.1cm}
&D(X) = - \sqrt{2}\, \varphi\,\,. }\eqno\eq$$
The scalar (tachyon) effective Lagrangian in this linear dilaton background
reads
$$S_{\rm eff} (T)={1\over 2}\int d^2 X\,e^{2\sqrt{2}\varphi}\,\,\Bigl\{
{1\over 2} \,T \left( - \partial_t^2 + \partial_{\varphi}^2 + 2\sqrt{2}
\partial_{\varphi} + 2 \right) T-V\Bigr\}\,\,.$$
Rescaling the scalar fields
$$e^{\sqrt{2}\varphi}\,\,T (t,\varphi )=\tilde{T} (t,\varphi)\eqno\eq$$
yields a massless theory
$$S = {1\over 2} \int dt d\varphi \Bigl\{ {1\over 2}\,\tilde{T}
\left( - \partial_t^2 + \partial_{\varphi}^2 \right) \tilde{T} -
{e^{-\sqrt{2}\varphi}\over 3!}\,\tilde{T}^3 + \ldots
\Bigr\}\,\,,\eqno\eq$$
with a spatially dependent string coupling constant
$$g_{\rm st} (\varphi) = e^{-\sqrt{2} \varphi}\eqno\eq$$
(we have taken for simplicity a cubic interaction).

This coupling grows and becomes infinite at $\varphi \rightarrow -
\infty$. This is usually taken as a signal that the linear dilaton
vacuum should be modified (at least in the region $\varphi \rightarrow
- \infty$).  Indeed the linearized static tachyon equation
$$\left( \partial_{\varphi}^2 + 2 \sqrt{2} \partial_{\varphi}
+ 2\right) T_0 (\varphi ) = 0\eqno\eq$$
already has two linearly independent solutions $T_0 (\varphi)=
e^{- \sqrt{2} \varphi}, \varphi e^{-\sqrt{2}\varphi}$.
This would imply that the correct vacuum is given by a tachyon
condensate [\Polone].  An (incomplete) analysis indicates that this
vacuum is then described by a $c=1$ conformal field theory coupled
to a Liouville field:
$${\cal L}={1\over 8\pi}\int d^2z\,\Bigl(\partial X \bar{\partial} X
+ \partial \varphi \bar{\partial} \varphi - 2 \sqrt{2} \varphi
(z,\bar{z}) R^{(2)}+\mu \, e^{-\sqrt{2} \varphi (z,\bar z)}\Bigl)\,\,.
\eqno\eq$$
Here the central charge $c_{X} = 1$ refers to the (matter) coordinate
$X(z,\bar{z})$ while the Liouville field with $Q = 2\sqrt{2}$ carries a
central charge $c_{\varphi} = 1 + 3Q^2 = 25$ leading to the required total of
$c=c_{X}+c_{\varphi}=26$. It is very interesting that in two dimensions
one  has another conformally invariant background, the WZW
$SL(2,\IR)/U(1)$ sigma model representing a black hole
(BH) [\WBH--\El].
Its physical properties are of major
interest as is the general question of describing different string theory
backgrounds in a single field-theoretic framework.

The presence of the cosmological term in the Liouville  theory (and of
the mass term in the black hole conformal field theory) leads to
computational difficulties when evaluating the correlation
functions (these actually become quite untractable
for the BH case).  It is a remarkable fact that the matrix model
formulation succeeds in handling  the first problem with ease and has some
promise for addressing the second as well.

The spectrum of states is usually obtained by neglecting the nonlinear
terms $\mu = 0$ (or $M=0$ for the black hole) in which case one has a
free field representation for the Virasoro generators.  In the above limit
the spectra of two theories are the same.  They consist
of a massless tachyon and an infinite sequence of discrete states.

We begin with the zero mode or tachyon states:
$$\eqalign{ & (L_0 - 1) \,\, V_{k,\beta} = 0 \,\,,\cr
\noalign {\vskip 0.1cm}
& L_0 = {1\over 2} \Bigl(  {\partial^2\over\partial X^2} +
{\partial^2\over\partial\varphi^2} + Q
{\partial\over\partial\varphi}\Bigr)\,\,,}\eqno\eq$$
with two branches of solutions
$$V_{\pm}= e^{ikX + \beta_{\pm} \varphi}\,\,,\qquad\quad
\beta_{\pm} = - \sqrt{2} \pm \vert k \vert\,\,,\eqno\eq$$
following from the on-shell condition
$$k^2 - \beta (Q+\beta ) = 0\,\,.\eqno\eq$$
Here we have taken an Euclidean (space) signature for $X$ and
$\varphi$ which is a convention in conformal field theory discussions.
One can take $X$ to be the space variable and $\varphi$ to be the
(Euclidean) time variable.
It will be more physical, and from the matrix model viewpoint more
natural, to treat $\varphi$ as a space coordinate and continue $X$ to
Minkowski time:
$$X \rightarrow - it\,\,,\qquad\quad
k \rightarrow i p\,\,.\eqno\eq$$

In the context of full Liouville
theory, the second branch with $\beta_- = - \sqrt{2} -\vert k\vert$ has a
questionable meaning since the wave functions grow at $\varphi
\rightarrow - \infty$ which is the location of the infinitely high
Liouville wall $\mu e^{-\sqrt{2} \varphi}$.  These vertex operators
are termed ``wrongly" dressed.  Operators with positive Liouville
dressing have a clear meaning.  Depending on the sign of the momentum,
$\pm =$ sign $k$, these are either right- or left-moving waves.  It is
sensible to use them to compute scattering processes and denote them
as
$$T_k^{\pm} = e^{ikX+(-\sqrt{2} \pm k)\varphi}\,,\qquad
\pm = {\rm sign} \, \, k\,\,. \eqno\eq$$
The Minkowskian continuation is $k = \pm ip$ and
$$\eqalign { T_p^+ &=e^{i p(t + \varphi )} e^{-\sqrt{2} \varphi}\,\,,\cr
\noalign {\vskip 0.1cm}
T_p^- &= e^{-ip(t-\varphi )} e^{-\sqrt{2}\varphi}\,\,, }\eqno\eq$$
for $p>0$ describe left-  and right-moving waves, respectively.

In addition one has an infinite sequence of nontrivial discrete
states specified by discrete (imaginary) values of energy and Liouville
momenta [\Po]:
$$ip_{\varphi}= -\sqrt{2} (1-j)\,,\qquad
ip = \sqrt{2} m \,\,,\eqno\eq$$
with $j = 0, {1\over 2} , 1, \ldots$ and $m = -j, \ldots , j$.
Clearly the states with $m = j$ and $m=-j$ are just special tachyon
states.  The simplest way then to reach the other states is to use the
$SU(2)$ generators as raising and lowering operators on the $m=\pm j$
tachyon states.  The $SU(2)$ generators are given by
$$\eqalign{ t_+ & = e^{i\sqrt{2} X (z)}\,\,,\cr
t_- & = e^{-i\sqrt{2} X(z)} \,\,,\cr
t_3 & = i\sqrt{2} \, \partial X (z)\,\,.}\eqno\eq$$
Denoting now the highest weight state as
$$W_{jj}^{(+)}= e^{i\sqrt{2}\,jX (z)}\,e^{-\sqrt{2}\,(1-j) \varphi
(z)}\eqno\eq$$
one gets the vertex operator for  general discrete states
$$W_{jm}^{(+)} = \left( \oint d\omega e^{-i\sqrt{2} X (\omega ) }
\right)^{j-m} W_{jj}^{(+)}\,\,,\quad -j\leq m\leq + j\,\,.
\eqno\eq$$
These can also be found in the Fock  space where they solve the Virasoro
conditions of the $c=1$ theory
$$\eqalign{ &(L_0 - 1) \vert jm \rangle =0\,\,,\cr
\noalign {\vskip 0.1cm}
&L_n \vert jm \rangle =0\,\,,\cr
&\vert jm \rangle = \int dz \, W_{jm}(z)\vert 0\rangle\,\,.}\eqno\eq$$
One also has operators with the opposite (negative) Liouville dressing
$$W_{jm}=V_{jm}(X) \,e^{-\sqrt{2} (1+j) \varphi (z)}\,\,,\eqno\eq$$
whose physical meaning is again questionable in the full Liouville
theory.  These states, however, turn out to play an important role as black
hole mass perturbations.

Evaluation of correlation functions in the continuum approach is
rather nontrivial and often relies on a number of educated guesses
involving various analytic continuations.  The problem lies
in the nontrivial Liouville potential term.  By separating and
integrating out the zero mode $\varphi (z,\bar{z} ) = \varphi_0 +
\tilde{\varphi}$ one finds, through a functional integral formulation,
the representation
$$ \langle\,\prod_{i=1}^{N} \, T_i \,\rangle =
\left( {\mu\over\pi}\right)^s \Gamma (-s)\,\bigl\langle
\Bigl( \prod_i T_i \Bigr)
\Bigl( \int d^2 z \,e^{\sqrt{2}\tilde{\varphi}}\Bigr)^s\,
\bigr\rangle_{\mu=0}\,\,.\eqno\eq$$
Here the $\Gamma$-function is a result of $\varphi_0$ integration
$$\int d\varphi_0 \, e^{Q\varphi_{0}} \Bigl( \prod_i
e^{\beta_{i}\varphi_{0}} \Bigr)\,\, {\rm exp}\,
\Bigl(-{\mu\over\hbar}\,e^{- \sqrt{2} \varphi_0} \int e^{-\sqrt{2}
\tilde{\varphi}}\Bigr)\,\,,\eqno\eq$$
and
$$-\sqrt{2} s \equiv \sum_i \, \beta_i + Q\,\,.\eqno\eq$$
The remaining correlation function is at $\mu=0$ but has a nontrivial
power of the Liouville term given by $s$. It can be evaluated only for
$s$ = integer with the full result  to be obtained by some analytic
continuation [\GoLi]. The $s=0$ amplitude is termed a ``bulk" amplitude
since the condition $s=0$ coincides with a Liouville momentum conservation.
Nontrivial computation involving major cancellation between matter and
Liouville contributions gives the simple result
$$T(k_1, k_2 , \ldots ,k_N ) = (N-3)!\,\prod_{i=1}^{N} \,\,
{\Gamma(-\sqrt{2}\,\vert k_i\vert )\over\Gamma(\sqrt{2}\,\vert k_i\vert )}
\,\,. \eqno\eq$$
At $s=0$ one has both energy and momentum conservation laws:
$$\sum_{i=1}^{N} k_i = 0\,,\qquad\quad
\sum_{i=1}^N \vert k_i \vert = -2 \sqrt{2}\,\,.\eqno\eq$$
Choosing $k_1 , k_2, \ldots ,k_{N-1} > 0$, one finds that the
$N$-th particle momentum is totally determined
$$k_N = - {N-2\over \sqrt{2}}\,\,, \eqno\eq$$
implying that the $N$-th leg factor diverges
$${\Gamma (-N+2)\over \Gamma (N-2)} \sim {1\over 0}\,\,.\eqno\eq$$
This is  in agreement with the previous $\Gamma (0) \sim {1\over 0}$
divergence.  This divergence is related to the length of the Liouville
line $\int d\varphi_0$ and is only fully understood in the matrix
description.

The final result for these $s=0$ bulk amplitudes is that they consist
of purely external leg factors
$\Delta = \Gamma (-\sqrt{2}\,\vert k\vert )/
\Gamma (\sqrt{2}\,\vert k \vert )$ and that only
$T_{++\ldots +-}$ and $T_{-\ldots - +}$ amplitudes contain a diverging
factor playing the role  of the Liouville volume.  These bulk
amplitudes can then lead to the full $s\not= 0$ amplitudes by an
appropriate continuation  [\DiFrKu]. This had to await  developments
given by the matrix model formalism. Concerning the full treatment
of Liouville theory one has the interesting algebraic approach
of [\Bab,\Ge].

\bigskip
\chapter{Matrix Model and Field Theory}
\medskip
The manner in which a simple matrix dynamics gives rise to nonlinear
two-dimensional string theory is rather interesting and is related
to collective phenomena.  The major tool employed is a field-theoretic
representation given by collective field theory [\JeSa]. We shall now
give the main features of the field-theoretic approach and describe
its significance to string theory [\DaJe]. The field theory
turns out to correctly describe interactions of strings, it therefore
represents a
very simple string field theory. It provides some major insight into
the physics of noncritical strings allowing the computation of
scattering processes [\DJR] and giving the exact $S$-matrix [\Poltwo].
New higher space-time symmetries are seen to emerge [\AvJe] with
further implications on  general string field theory being likely.

The simple model that one considers is a Hermitian matrix
$M^\dagger (t) = M(t)$ in one time dimension ($X^0 = t$) with a Lagrangian
$$L = {1\over 2} \Tr \left(\dot{M}^2 - u(M)\right)\,\,.\eqno\eq$$
It has an associated $U(N)$ conserved (matrix) charge $J = i [M,\dot{M}]$.
Restricting oneself to the singlet subspace $\hat{J}\vert\,\,\,\rangle=0$
turns this model into a gauge theory. The matrix can be diagonalized:
$M(t) \to {\rm diag}\,\, (\lambda_i (t) )$ with the eigenvalues
describing a system of nonrelativistic fermions.

The collective variables of the model are the gauge invariant (Wilson)
loop operators
$$\phi_k (t) = \Tr \left( e^{ikM(t)}\right) = \sum_{i=1}^N \,
e^{ik\lambda_{i}(t)}\,\,,\eqno\eq$$
which, after a Fourier transform
$$\phi (x,t) = \int {dk\over 2\pi} \, e^{-ikx} \, \phi_k (t) =
\sum_{i=1}^N \, \delta \left( x-\lambda_i (t)\right)\,\,,\eqno\eq$$
have a physical interpretation of a density field (of
fermions).  Introduction of a conjugate field $\Pi (x,t)$ with
Poisson brackets
$$\left\{ \phi (x), \Pi (y)\right\} = \delta (x-y)\eqno\eq$$
gives a canonical phase space.

The dynamics of this field theory is directly induced from the simple
dynamics of the matrix model variables
$M(t)$ and  $P(t) = \dot{M} (t)$. It is found to be given by the
Hamiltonian
$$H_{\rm coll}=\int dx\,\,\Bigl\{ {1\over 2}\,\Pi_{,x}\,\phi\,\Pi_{,x}+
{\pi^2\over6} \phi^3 + u (x) \phi \Bigr\}\,\,,\eqno\eq$$
where the first two terms come from the kinetic term of the matrix model
$\Tr (P^2/2 )$ while the last term represents the potential
(the latter can be easily seen through the density representation):
$${1\over 2}\,\Tr P^2 \rightarrow {1\over 2}\,\Pi_{,x}\,\phi\,\Pi_{,x}
+ {\pi^2\over 6} \phi^3\,\,, \qquad
\Tr u(M) \rightarrow \int dx \, u (x) \phi(x,t)\,\,.\eqno\eq$$
The Hamiltonian constructed in this way consists of a cubic (interaction)
term and a linear (tadpole) term.  In terms of basic loops (and strings)
the cubic interaction has the effect of splitting and joining  strings.
The linear tadpole term represents a process of string annihilation
into the vacuum. It contains the classical background potential. This
potential is tuned to get a particular string theory background;
the  noncritical $c=1$ string theory is obtained for example with
an inverted oscillator potential.  Two relevant facts are immediate
in this transition to collective
field theory:

(1)  The field $\phi (x,t)$ is two-dimensional with
the extra spatial dimension $x$ being related to the eigenvalue space
$\lambda_i$.  The appearance of an extra dimension is the first
sign that this theory will be describing $D=2$ strings.

(2)  The equations of motion for the induced fields are nonlinear while
the matrix equations (in particular for the physically relevant
oscillator potential $u(M)= -M^2/2$) are  linear
$$\ddot{M} (t) - M(t) = 0\,\,.\eqno\eq$$
Through a nonlinear transformation, $\phi (x,t) = \Tr \delta
(x-M(t))$ the matrix model provides an exact solution to the
nonlinear field theory.
 The feature of integrability and the collective transformation
itself is very similar to the well known inverse
scattering transformation in integrable field theories.  Actually
introduction of left- and right-moving chiral components
$\alpha_{\pm} (x,t) = \Pi_{,x} \pm \pi\phi (x,t)$ with Poisson
brackets
$$\left\{ \alpha_{\pm} (x) , \alpha_{\pm} (y) \right\} = \pm 2\pi
\partial_x \delta (x-y)\eqno\eq$$
brings the Hamiltonian to the form
$$H_{{\rm coll}} = {1\over 2} \int \, {dx\over 2\pi}\,\Bigl\{
{1\over 3} \left( \alpha_+^3 - \alpha_-^3\right) - \left( x^2 - \mu
\right) \left( \alpha_+ - \alpha_- \right) \Bigr\}\,\,.\eqno\eq$$
The equations of motion
$$\partial_t \alpha_{\pm} + \alpha_{\pm} \partial_x \alpha_{\pm} - x =
0\eqno\eq$$
are then seen to be two copies of a large-wavelength KdV type equation
with an external $(-x^2 )$ potential.
Collective field theory shares with some other field theories in
two dimensions the feature of exact solvability.
One can indeed write down an infinite sequence of conserved commuting
quantities (Hamiltonians).  They are simply given by [\AvJe]:
$$H_n = {1\over 2\pi} \int dx \int_{\alpha_- (x,t)}^{\alpha_+ (x,t)}
d\alpha \,\, \left( \alpha^2 - x^2 \right)^n\,\,,\eqno\eq$$
and are related to the matrix model quantities $\Tr (P^2 - M^2 )^n$.
In fact one has a simple set of transition rules between the two
descriptions. These are useful when constructing
exact eigenstates and symmetry generators of the theory.

One  easily checks that the Poisson brackets vanish
$$\left\{ H_n , H_m \right\} = 0\,\,, \eqno\eq$$
and that these charges are formally conserved
$${d\over dt}\, H_n = \int dx\,\partial_x \left( \alpha^2 - x^2 \right)
\left( \alpha^2 - x^2 \right)^n = 0\,\,. \eqno\eq$$
This naturally is correct only up to surface terms which are present and
will allow particle production.

Before continuing with the integrability features of
the theory one can study perturbation theory and small fluctuations to
clarify at this simple level the connection to string theory.
The static (ground state) equation reads
$${1\over 2} \left( \pi \phi_0 (x) \right) ^2 + u (x) =
\mu_{F}\,\,,\eqno\eq$$
where $\mu_{F}$ is the Fermi energy introduced as a linear term in the
Hamiltonian
$$\Delta H = - \int dx\,\,\mu_{F}\, \phi (x,t)\,\,.\eqno\eq$$
Denoting $\pi\phi_0 = p_0$ we see this as being simply the equation
specifying the Fermi surface:  ${1\over 2} p_0^2 + u(x)=\mu_{F}$ with
the solution
$$\pi \phi_0 = p_0 (x) = \sqrt{2(\mu_{F} - u(x) )}\,\,.\eqno\eq$$
Introducing small fluctuations with a shift $\phi (x,t) = \phi_0 (x) +
{1\over\sqrt{{\pi}}}
\partial_x \eta (x,t)$ the Hamiltonian becomes
$$H = \int dx\,\, \Bigl\{(\pi\phi_0) \Bigl( {1\over 2}\,
\Pi^2 + {1\over 2}\, \eta_{,x}^2\Bigr) + {\pi^2\over 6}\,
(\eta_{,x})^3 {\pi\over 2} \Pi^2 \eta_{,x}\Bigr\}\,\,, \eqno\eq$$
with the quadratic term (in the Lagrangian form):
$$L_2 = \int dt \int dx\,\, {1\over 2}\,\Bigl( {\dot\eta^{2}\over
\pi\phi_{0} (x) } \, - \left(\pi \phi_0 \right) \eta_{,x}^2\Bigr)\,\,.
\eqno\eq$$
This is a free massless particle in an external gravitational
background
$$g_{\mu \nu}^0=\Bigl( 1/\pi \phi_0 (x)\,,\,\,\pi\phi_0(x)\,\Bigr)
\eqno\eq$$
specified by our potential $u(x)$.  However, this
metric is removable by a coordinate transformation.
In terms of the time-of-flight coordinate
$$\tau = \int^x \, {dx\over\pi\phi_0} \qquad {\rm or} \qquad
{dx(\tau)\over d\tau} = p_0 \eqno\eq$$
one has
$$H = \int d\tau\,\,\Bigl\{ {1\over 2}\,\Bigl(\Pi^2 + (\partial_{\tau}
\eta )^2 \Bigr) + {1\over 6 p_0^2}\,\Bigl(
(\partial_{\tau} \eta ) ^3 +
3\Pi^2(\partial_\tau \eta)\Bigr) \Bigr\} \,\,,\eqno\eq$$
which describes a massless theory with a spatially dependent coupling
constant
$$g_{\rm st} (\tau) = {1\over p^2_0(\tau )}\,\,.\eqn\stringcc$$
The continuum $c=1$ string theory is approached for a special choice
of the potential $v(x)= -x^2/2$.  In this case one has a
critical theory near $\mu_F = - \mu \rightarrow 0$.
For the oscillator we have
$$ \eqalign { x(\tau) & = \sqrt{2\mu}\, \cosh\,\tau\,\,,\cr
\noalign {\vskip 0.1cm}
p_0 (\tau) & = \sqrt{2\mu} \, \sinh \, \tau\,\,.}\eqno\eq$$
The length of the (physical) $\tau$-space diverges at the turning
point $x_0 = \sqrt{2\mu}$.  The string coupling constant \stringcc\
is now
$$g_{\rm st}(\tau)={1\over 2\mu \, \sinh^2 \tau}\,\,.\eqno\eq$$
It depends on the Fermi level as $g_{\rm st} \sim 1/\mu$.
This is in parallel with the dependence of the string coupling on the
cosmological constant of the $c=1$ string theory. We also see that
asymptotically $ g_{\rm st} \sim {1\over \mu}\,e^{-2\tau}$
as  $\tau \rightarrow + \infty$.
Comparing it to the expected behavior in $c=1$ string
theory  $g_{\rm st} \sim {1\over\mu}\,e^{-\sqrt{2} \varphi}$ one has
the (asymptotic) identification
$$\tau \leftrightarrow {1\over \sqrt{2}}\,\varphi\,,\qquad\quad
t_M \leftrightarrow {1\over\sqrt{2}}\,t_{c=1} \,\,.\eqno\eq$$
One can  now identify $\eta (\tau,t)$ with the tachyon
field $T(\varphi, t)$.  Remembering that $e^{\sqrt{2}\varphi}\,
T(\varphi, t)$ was the field satisfying the massless Klein-Gordon
equation, one  also has the identification of the
energy-momenta:
$$ip_{\tau}\leftrightarrow 2 + i\sqrt{2} \, p_\varphi\,,\qquad\quad
i\epsilon \leftrightarrow i \sqrt{2}\,p \,\,,\eqno\eq$$
where $\epsilon$ is the energy in the matrix model picture.

The above identification of the collective field
$\eta(\tau,t)$ was only done asymptotically when the
$\mu e^{- \sqrt{2}\varphi}$ term in the Liouville equation is ignored.
A much more precise identification can be performed with the cosmological
term also present.

In the matrix model the time-of-flight coordinate is
introduced to bring the quadratic mass operator of the
collective field into a Klein-Gordon form:
$$\Bigl[\,\partial_t^2 -\sqrt{x^2-2\mu}\,\,\partial_x\,
\sqrt{x^2-2\mu}\,\,\partial_x\Bigr]\,\eta =
(\partial_t^2 -\partial_\tau^2)\,\eta(t,\tau)\,\,, \eqno\eq$$
with $x=\sqrt{2\mu}\,\cosh\tau$. If we use a basis
conjugate to $x$: $p=-i\,(\partial/\partial x)$ the spatial
operator reads
$$\omega^2 =p^2x^2-2\mu p^2\,\,,\eqno\eq$$
and after a change of variables
$p=\sqrt{2}\,e^{-\varphi/\sqrt{2}}$ this gives the Liouville
operator
$$\hat\omega^2 =-{1\over 2}\,(\partial_\varphi)^2
-4\mu\,e^{-\varphi/\sqrt{2}}\equiv {\cal H}_L\,\,.\eqno\eq$$
We see that the Liouville coordinate is to be identified more
precisely [\MoSe] with the variable $p$ conjugate to the matrix
eigenvalue $\lambda$. The conjugate basis is not unnatural
in collective theory, it is associated with the
(Wilson) loop operator itself
$$W(\ell,t)= \Tr\,(e^{-\ell M}) =
\int dx\,e^{-\ell x}\,\phi(x,t) \,\,.\eqno\eq$$
which at the linearized level
$$W(\ell,t)=
\int_0^\infty  d\tau\,e^{-\sqrt{2\mu}\,\ell \cosh\tau}\,
\partial_\tau \eta\,\,, \eqno\eq$$
is  seen to obey
$$(\partial_t^2 -\hat\omega^2)\,\hat W(\ell,t)=0\,\,,\eqno\eq$$
with
$$\hat\omega^2=\partial_\tau^2 \quad
\Rightarrow\quad  -(\ell\,\partial_\ell)^2 +
2\mu\,\ell^2\,\,.\eqno\eq$$
After a change $\ell=2e^{-\varphi/\sqrt{2}}$
one has the Liouville operator ${\cal H}_L$.
In this conjugate momentum basis the connection to Liouville
theory is therefore manifest. One could obviously write all
equations in this representation  but formulae are
much simpler in terms of time-of-flight coordinate $\tau$.
The (Wilson) loop field and its natural connection to the
Liouville picture will be relevant for defining the string
theory $S$-matrix.

To further clarify the identification of the Liouville mode
let us write the transformation between the matrix
eigenvalue $x=\lambda$ and the time-of-flight coordinate
$\tau$ as a point canonical transformation:
$$\eqalign{x&=\sqrt{2\mu}\,\cosh\tau\,\,,\cr
\noalign{\vskip 0.2cm}
p&={1\over \sqrt{2\mu}\,\sinh\tau}\,\,p_\tau\,\,,}\eqno\eq$$
where $p$ and $p_\tau$ are the conjugates:
$\{x,p\}=\{\tau,p_\tau\} =1$.
Introducing $p=\sqrt{2}\,e^{-\varphi/\sqrt{2}}$
we have
$$\eqalign{&p_\varphi=\sqrt{2\mu}\, e^{-\varphi/\sqrt{2}}\,
\cosh\tau\,\,,\cr
\noalign{\vskip 0.2cm}
&p_\tau=\sqrt{2}\,\sqrt{2\mu}\, e^{-\varphi/\sqrt{2}}\,
\sinh\tau \,\,,}\eqno\eq$$
as a canonical transformation between the Liouville and
time-of-flight coordinates. The property of this
transformation is that
$${1\over 2}\,\omega^2 ={1\over 2}\,p_\tau^2 =
p_\varphi^2 -2\mu\, e^{-\varphi/\sqrt{2}}\,\,.$$
Now in Liouville theory one also usually deals
with two alternate descriptions and two different fields:
the original Liouville field $\varphi(z,\bar z)$
and a free field $\psi(z,\bar z)$.
They are related by a canonical (B\"acklund) transformation
$$\eqalign{\dot\varphi &= \psi'+\sqrt{2\mu}\,
e^{-\varphi/\sqrt{2}}\, \cosh (\psi/\sqrt{2})\,\,,\cr
\noalign{\vskip 0.2cm}
\dot\psi &=\varphi'+\sqrt{2\mu}\,e^{-\varphi/\sqrt{2}}\,
\sinh (\psi/\sqrt{2})\,\,,}\eqno\eq$$
where the two derivatives correspond to the two-dimensional space
$z=\sigma+i\xi$. The above transformation relates the
Liouville action to the action of a free field
$\psi(z,\bar z)$. Clearly for the center of mass mode
$(\varphi' = \psi' =0)$ one sees
$$\eqalign{\Pi_\varphi &= \sqrt{2\mu}\,
e^{-\varphi/\sqrt{2}}\, \cosh (\psi/\sqrt{2})\,\,,\cr
\noalign{\vskip 0.2cm}
\Pi_\psi &=\sqrt{2\mu}\,e^{-\varphi/\sqrt{2}}\,
\sinh (\psi/\sqrt{2})\,\,. }\eqno\eq$$

The transformation between $\varphi$ and $\psi$ is
identical to the one in the matrix model. We have then the
fact that the time-of-flight coordinate $\tau$ is to be
identified with the free field zero mode $\psi_0:\,\psi_0=
\sqrt{2}\,\tau$. In most of the vertex operator construction
it is the free field which is used.

We shall now continue and discuss the exact classical solution
of the theory and exhibit its integrability.
Consider first  the physical meaning of the
component fields $\alpha_{\pm}$ and the nature of boundary conditions
at the turning point or wall $\tau = 0$.  Shifting by the classical
solution, $\alpha_{\pm} = \pm p_0 + \epsilon_{\pm}\,$, the equations
of motion linearize to
$$\partial_{\pm} \epsilon_{\pm} \pm \left( p_0 \partial_x + \partial_x
p_0 \right) \epsilon_{\pm} = 0\,\,. \eqno\eq$$
Denoting $\epsilon_{\pm} \equiv \mp {1\over p_0} \psi_{\mp}$ we have
$$\left( \partial_t \pm \partial_{\tau} \right) \psi_{\mp}=0\,\,.
\eqno\eq$$
So indeed, $\psi_{\pm} = \psi_{\pm} ( t\pm \tau)$ are left- and
right-moving waves, respectively.  There is however a nontrivial
boundary condition in the theory which comes in as follows:  The
eigenvalue density $\phi = {1\over 2\pi} \left( \alpha_+ - \alpha_-
\right)$ gives the conserved fermion number
$$\dot{N} = \int dx \,\dot{\phi} = {1\over 2\pi} \int dx \left(
\alpha_+^2 - \alpha_-^2 \right)=0\,\,.\eqno\eq$$
At the boundary point for $x$ (or $\tau = 0)$, this implies
$$\left( \alpha_+^2 - \alpha_-^2 \right)\Bigl\vert_{{\rm boundary}} =
0\,\,,\eqno\eq$$
so that there is no leakage into the region under the barrier (this
may have to be given up in nonperturbative discussion [\LM]).  For the
small fluctuations we then have
$$\psi_+ (x) = \psi_- (x)\,\,,\eqno\eq$$
implying Dirichlet boundary conditions. In terms of Fourier modes
$$\psi_{\pm} = \int_{-\infty}^{\infty} dk\,\,\alpha_k^{\pm}\,\,
e^{ik (t\pm \tau)}\eqno\eq$$
with $\alpha_{-k} = \alpha_k^+$ our boundary condition implies that one
has only one set of oscillators with positive momenta
$$\alpha_k^+ = \alpha_k^- = \alpha_k\,,\qquad k>0\,\,.\eqno\eq$$
This is appropriate for a theory defined on a half-line
$\tau \in [0,\infty)$.

A very simple form for the exact solution of the collective equations
was given by Polchinski [\Poltwo].
At  the classical level one has a  phase space picture of
the eigenvalues $\lambda (\sigma,t)$ and their momenta $p(\sigma,t) =
\dot{\lambda}$.  They obey the classical equations of motion
$$\dot{p}(\sigma,t) = - u'\left(\lambda(\sigma,t)\right)\,\,.
\eqno\eq$$
The information that the particles are fermions is contained in the
statement that the equation $x = \lambda (\sigma,t)$ is invertible: $\sigma
= \sigma (x,t)$ so that for each $\sigma$ there is only one particle
(actually there is a degeneracy corresponding to the upper and lower
Fermi surface).  Consider in particular the inverted oscillator: the
solution is immediately written as
$$\eqalign{ x & = a(\sigma) \cosh (t - \sigma )\,\,,\cr
p & = a(\sigma ) \sinh(t - \sigma )\,\,. }\eqn\parsol$$
Here $a(\sigma )$ is an arbitrary function giving an arbitrary initial
condition.  The simplest configuration is obtained for
$a (\sigma ) = \sqrt{\mu} = {\rm const.}$ and we have
$$p_{\pm}=p(\sigma_{\pm},t)=\pm\sqrt{x^2 -\mu}\,\,.\eqno\eq$$
This is recognized as the static ground state collective field
configuration $\pi \phi_0 (x)$. It is easy to see that the general
configuration leads to the solution of collective
equations. The collective field is identified with the Fermi momentum
densities
$$\alpha_{\pm}(x,t)\equiv p_{\pm}=p (\sigma_{\pm}(x,t),t)\,\,.\eqno\eq$$
Conversely, $p(\sigma,t)=\alpha(x(\sigma,t),t)$. Using the chain rule
$${\partial p\over \partial t} = {\partial\alpha\over \partial t} +
{\partial\alpha\over \partial x}\,\, {\partial x\over\partial t} = -
u'(x)\,\,,\eqno\eq$$
and the equation of motion obeyed by $p(\sigma,t)$, there follows the
equation
$${\partial\alpha\over\partial t}= -u'(x)-\alpha\,\partial_x \alpha
\,\,.\eqno\eq$$
These are the decoupled quadratic equations for the collective fields
$\alpha_{\pm} (x,t)$ associated with the cubic Hamiltonian.

Knowledge of the exact solution can be directly  used to determine
scattering amplitudes. One considers and follows the time evolution of
an incoming left-moving lump.
A point parametrized by $\sigma$ which passes through $x$
at some (early) time $t$ will reflect on the boundary and pass through
the same point $x$ at some later time $t'$ as a right-moving lump.
The time evolution of the particle coordinates is known
explicitly~\parsol\
so one can determine the relationship between the two times $t$ and
$t'$. Consider the exact solution given by Eq.~\parsol,
at a distance $\tau$ large enough one has
$$x=e^{\tau} = \cases{a(\sigma)\, e^{-(t-\sigma )}\,, \quad &
$\quad t\rightarrow - \infty$\cr
a(\sigma)\, e^{+(t' + \sigma )}\,, \quad & $\quad t' \rightarrow +
\infty$\cr }$$
from where
$$ t' - \tau = t + \tau + \ln a^2 (\sigma)\,\,.\eqno\eq$$
On the other hand $a^2 (\sigma ) $ is related to $\alpha_{\pm}$:
$$a^2 = x^2 - p^2 = x^2 - \alpha_{\pm}^2 \approx 1 +
\psi_{\mp}\,\,.\eqno\eq$$
The outgoing particle momentum $p_+ (t', \sigma)$ is equal in magnitude
(but opposite in sign) to the incoming momentum of the particle
$p_- (t, \sigma)$:
$$p_+ (t', \sigma ) = - p_- (t,\sigma )\,\,.\eqno\eq$$
This elementary relationship provides a relationship between the
incoming and outgoing wave and therefore yields the $S$-matrix.
Collecting the above formulas we have
$$\psi_-(z)=\psi_+ \Bigl(z-\ln(1+\psi_-(z))\Bigr)\,\,.\eqno\eq$$
This functional equation determines the relation between the left- and
right-moving (incoming and outgoing fields). It represents a nonlinear
version of our Dirichlet boundary conditions and is characteristic of
scattering problems involving a wall. An expansion in power series can
be performed determining explicitly the outgoing modes in terms of the
incoming ones.  This is then sufficient to give the
scattering amplitudes. We shall return to this subject in sect.~5.

In general, all features of the exactly solvable matrix model
translate into string theory.  More precisely there is a direct
translation of matrix model quantities into the collective field
theory which itself is then completely integrable as we have
emphasized. We  end this section by summarizing the  set of
translation rules between the matrix model and collective field
theory representations.

At the  classical level one thinks of matrix variables
as coordinates in a fermionic phase space $M\rightarrow \lambda ,
P\rightarrow p$. Collective field theory
represents a second quantization according to $p\rightarrow
\alpha (x,t)$.  So we have the correspondences:
$$\eqalign{ M & \leftrightarrow \lambda \leftrightarrow x\,\,,\cr
P &\leftrightarrow p \leftrightarrow \alpha (x,t)\,\,.}\eqno\eq$$
The $U(N)$ trace becomes a phase space integration in the fermionic
picture and in  the collective representation:
$$\Tr \left\{ \,\,\right\}\quad \rightarrow \quad\int{dx\over 2\pi}
\int_{\alpha_{-}(x,t)}^{\alpha_{+}(x,t)} d\alpha \,\, \left\{ \,\,
\right\}\,\,, \eqno\eq$$
where $\alpha_{\pm} (x,t)$ are the chiral components of the scalar
field density. For example the collective Hamiltonian comes out
as follows:
$${1\over 2}\,\Tr \left( P^2 - M^2 \right)\rightarrow{1\over 2}
(p^2 - x^2) \rightarrow \int {dx\over 2\pi} \int d\alpha \,
{1\over 2}(\alpha^2- x^2) ={1\over 2} \int {dx\over 2\pi}
\Bigl[ {\alpha\over 3}^3 - x^2 \alpha \Bigr]_-^+ \,\,.\eqno\eq$$
The above transition rules summarize the statement that
the Poisson brackets of single particle quantities in the Fermi (or
matrix) phase space
$$\left\{ f_1 (x,p), f_2 (x,p)\right\}_{\rm P.B.}\eqno\eq$$
remain preserved in the field theory.  For example, the field-theoretic
operator inferred from the  oscillator states is
$$B_n^{\pm}=\int{dx\over 2\pi}\int d\alpha \,\,(\alpha \pm x)^n\,\,.
\eqno\eq$$
We can now use the $\alpha$-field Poisson brackets
$\left\{ \alpha_{\pm} (x), \alpha_{\pm} (y) \right\} = \mp 2\pi i
\delta '(x-y)$  to verify that indeed
$$\left\{ H_{{\rm coll}} , B_n^{\pm} \right\} = \pm n \,
B_n^{\pm}\,\,.\eqno\eq$$
This represents an eigenstate of the collective field theory Hamiltonian.
At the quantum level a normal ordering prescription is used to
completely define the operators.

The outlined string field theory gives a systematic perturbation theory
in the string coupling constant. The Feynman rules that are constructed
are characterized by a nontrivial cubic vertex exhibiting discrete poles
in the momenta. Most importantly a fully quantized Hamiltonian is achieved
through normal ordering with the counter-terms being supplied by the original
collective formalism. So what one has is a totally finite string field
theory capable of reproducing string theory diagrams to all orders. It
works at loop level without further counter-terms giving a single covering of
modular space. This, as is well known, has always been quite nontrivial in a
string-theoretic framework. For more details of the quantum theory and
explicit calculations at the loop level the reader should
consult [\DJR].

\bigskip
\chapter{$w_{\infty}$ Symmetry}
\medskip

The matrix model description has the virtue of great simplicity:  it
is linear and trivially exactly solvable.  For the matrix Hamiltonian
$$H = {1\over 2}\, \Tr\left(P^2-M^2\right)\eqno\eq$$
one can write down exact creation--annihilation operators
$$B_n^{\pm} = \Tr \left( P \pm M\right)^n\,,
\quad\quad n = 0,1,2,\ldots\eqno\eq$$
creating imaginary energy eigenstates
$$\left[\,H, B_n^{\pm}\,\right]= \mp in B_n^{\pm}\,,\qquad
\epsilon_n = \pm in \,\,.\eqno\eq$$
The whole point here is to be able to translate this exact information
into physical results which, as we have emphasized, is achieved through
collective field theory. The direct connection of the space-time string
field theory with the matrix model leads then further insight. The simple
oscillator structure with its creation--annihilation basis implies the
presence of a similar structure in the field theory and therefore string
theory.

To understand the physical meaning of the (oscillator) states
recall that in the collective field theoretic description we have another
spatial quantum number in addition to the energy.
This feature arose as a consequence of scaling invariance.
The coordinate and the fields transform as
$$x\rightarrow ax\,,\qquad\quad
\alpha (x,t)\rightarrow {1\over a}\,\alpha(ax,t)\,\,,\eqno\eq$$
and the Hamiltonian, without the chemical potential term,
$-\mu \alpha$, scales as
$$H \rightarrow {1\over a^4}\,H\,\,.\eqno\eq$$
The classical equations of motion are consequently scale invariant.
One then defines the scaling momentum as
$$i p_s = - 4 + s\,\,,\eqno\eq$$
where $s$ is the naive scaling dimension $s[x] = s[\alpha] = 1$.  The
creation--annihilation operators
$$\tilde T_n^{\pm} ={1\over n}\,\int {dx \over 2\pi}\,\,
{(\alpha\pm x)^{n+1}\over n+1}\,\,\Big\vert_{\alpha_{-}}^{\alpha_{+}}
\eqno\eq$$
consequently have the following energy-momentum:
$$i \epsilon = n \,,\qquad\quad
i p_s = -2 + n \,\,.\eqno\eq$$
We find these to be in precise agreement with the discrete tachyon
vertex operator states since there
$$i \sqrt{2}\, p  = \pm 2 j\,,\qquad\quad
i \sqrt{2}\, p_{\varphi} = -2 + 2j \,\,,\eqno\eq$$
and we have already noted the relations $\sqrt{2}\, p = \epsilon,\,\,
\sqrt{2}\,p_{\varphi}=p_s$. We then have a one-to-one correspondence
between oscillator states of the matrix model and discrete tachyon
vertex operators of the conformal description of $c=1$ string theory
$$B_n^{\pm}=\Tr\left(P\pm M\right)^n \quad \leftrightarrow \quad
T_{p}^{(\pm)}=e^{\pm i\sqrt{2} jX}\,e^{-\sqrt{2}(1-j)\varphi}\eqno\eq$$
with $n=2j$ or $j=n/2$.

An analytic continuation of discrete imaginary momenta to real
values $(n = i\kappa,\, p_s = 2i - \kappa )$
gives the scattering operators
$$\eqalign{ B_{-i\kappa}^- & = \Tr \left( P - M\right)^{-i\kappa}
\sim e^{-i\kappa (t+\tau )}\,\,,\cr
B_{-i\kappa}^+ & = \Tr \left( P + M\right)^{-i\kappa}
\sim e^{-i\kappa (t-\tau )}\,\,, }\eqno\eq$$
describing left- and right-moving waves, respectively. These operators
can be used to construct the in- and out-states of scattering theory
$$\eqalign{\Tr \left( P-M \right)^{-i\kappa} \vert 0 \rangle & = \vert
\kappa;\,{\rm in} \rangle \,\,,\cr
\Tr\left( P + M \right)^{+i\kappa}\vert 0\rangle & = \vert\kappa;\,
{\rm out} \rangle \,\,. }\eqno\eq$$
Namely, for an in-state, one needs a {\it left}-moving wave while the
out-state is necessarily given by a {\it right}-moving one. Here we
have used the picture where the wall is at $\tau = -\infty$
corresponding to the physical space being defined on the right semi-axis
$x = e^{\tau} \geq 0$. Had we chosen to define the theory on the
other side of the barrier, the states
$$\eqalign{\Tr\left(P+M\right)^{i\kappa}&=e^{i\kappa(t+\tau)}\,\,,\cr
\Tr\left(P-M\right)^{i\kappa}&=e^{i\kappa(t-\tau)}\,\,, }\eqno\eq$$
would be physical since they have the meaning of a right-moving
{\it in}-wave and a left-moving {\it out}-wave.
Hence there is a one-to-one correspondence between the scattering
operators in the matrix model and the string theory vertex operators
$$\Tr\,(P\pm M)^{-i \kappa}\quad\leftrightarrow \quad
T^{\pm}=e^{\pm {\kappa\over\sqrt{2}} X}\,e^{-\sqrt{2}+i{\kappa\over\sqrt{2}}
\varphi}\,\,.\eqno\eq$$
A typical transition amplitude reads
$$S = \langle {\rm out} \vert {\rm in}\rangle=\langle 0\vert
\Tr (P+M)^{{\kappa\over i}}\Tr(P-M)^{{\kappa\over i}}\vert 0\rangle\,\,.
\eqno\eq$$
It only contains operators with the same (Liouville)-exponential
dressing. This is in total agreement with the continuum string theory
situation.

In addition to the tachyon states, the matrix oscillator description
immediately allows a construction of an infinite sequence of discrete
states [\GKN,\AvJe].
They are created by the operators
$$B_{n,\bar n}=\Tr\Bigl((P+M)^n (P-M)^{\bar{n}}\Bigr)\,\,,\eqno\eq$$
with energies and momenta given by
$$i\epsilon=n-\bar{n}\,,\qquad\quad
i p_s =-2+(n+ \bar{n})\,\,.\eqno\eq$$
Comparing this with the discrete spectrum of the string theory given by
$$i \sqrt{2} p = 2m\,,\qquad\quad
i \sqrt{2} p_{\varphi} =-2+2j\,\,,\eqno\eq$$
we find the correspondence
$$m ={n-\bar{n}\over 2}\,,\qquad\quad
j ={n + \bar{n}\over 2} \,\,.\eqno\eq$$
These are indeed half-integers once $n,\bar{n}$ are integers.  The
field theory operators
$$B_{jm}=\int {dx\over 2\pi}\int_{\alpha_-}^{\alpha_+} d\alpha\,
(\alpha + x )^{j+m}\,(\alpha - x )^{j-m}\eqno\eq$$
can be shown (again by using the Poisson brackets or the commutators)
to generate discrete imaginary energy eigenstates of the Hamiltonian
$$\left[ H, B_{jm} \right] = -2i m B_{jm}\,\,.\eqno\eq$$
This commutator shows that the operators $B_{jm}$ are
spectrum-generating operators for the Hamiltonian $H$;
but it also signals the existence of a large symmetry algebra which
operates in this theory [\AvJe,\MoSe,\winf,\WiGR,\KlPo,].

First we had the  sequence of conserved quantities
$$H_l =\Tr\left( P^2 - M^2 \right)^{l+1}\eqno\eq$$
commuting among themselves
$$\left[ H_l , H_{l'} \right] = 0\,\,.\eqno\eq$$
These are  particular cases of the spectrum-generating operators
$B_{jm}$. One is then lead to consider the complete algebra of all the
operators. Introducing the more standard notation
$$O_{JM} = (p+x)^{J+M+1} (p-x)^{J-M+1}\,\,,\eqno\eq$$
with the associated collective field realization
$$O_{JM} = \int {dx\over 2\pi} \int_{\alpha_{-}}^{\alpha_{+}} d\alpha
\,\, (\alpha + x )^{J+M+1} (\alpha - x )^{J-M+1}\,\,,\eqno\eq$$
one checks that they obey the $w_{\infty}$ commutation relations
$$\left[ O_{J_{1}M_{1}} , O_{J_{2} M_{2}} \right] = 4i
\Bigl((J_2+1)M_1-(J_1+1)M_2\Bigr)\,O_{J_{1}+J_{2},M_{1}+M_{2}}\,\,.
\eqn\winfalg$$
We note that this commutator results if no special ordering is taken
for the noncommuting factors.  At the full operator
quantization level, field theory requires special normal ordering.
It is likely that this modifies the simple $w_{\infty}$ algebra to a
$W_{1+\infty}$ algebra.

Recalling the form of tachyon operators $T_n^{\pm}=\Tr(P\pm M )^n$ one sees
a special relationship between the tachyon operators and the
$w_{\infty}$ generators.  A simple computation shows that
$$O_{JM} = {1\over 2i} \, {1\over (J+M+2) (J-M+2)} \, \bigl[
T_{J+M+2}^+ , T_{J-M+2}^- \bigr] \,\,.\eqno\eq$$
This first sheds some light on the nature of higher discrete modes in
the collective formalism:  they are composite states of the tachyon.
More importantly, one then expects a simple realization of the
$w_{\infty}$ algebra on the tachyon sector.

To understand the role played by the scalar collective field with
respect to the $w_\infty$ algebra one can first look at the
following Virasoro subalgebra:
$$O_{l} \equiv O_{{l\over 2},{l\over 2}} =
\int {dx \over 2\pi}\,\int d\alpha\,\, (\alpha+x)^{l+1}\,
(\alpha -x)\,\,,\eqno\eq$$
with
$$[ O_{l},  O_{l'} ]= 2i(l-l')\,O_{l+l'} \,\,.  \eqno\eq$$
We can then determine the transformation property of the tachyon
field under this subalgebra. Actually, the exact tachyon
creation operator $T_n$ can itself be written as an
extension of the whole algebra
$$\tilde T_n^{+} ={1\over n}\,\int {dx \over 2\pi}\,
{(\alpha+x)^{n+1}\over n+1} =
{1\over n}\, O_{{n\over 2}-1, {n\over 2}}\,\,,\eqno\eq$$
and this determines the commutator (with the indices extended  outside
the standard range $|m| \leq j$). Alternatively one  can also directly
use the basic commutation relations to find
$$[O_l, \tilde T_n^{+}] = 2i (n+l)\,\tilde T_{n+l}^{+}\,\,,\eqno\eq$$
which shows that the tachyon transforms as a field of conformal weight 1.
This is understood to be a space-time and not a world sheet feature.
The fact that an infinite space-time  symmetry appears
in the collective field theory explains many similarities that it
has with conformal field theory.
 The $w_\infty$ generators act in a
nonlinear way  on the tachyon field.
This implies that this symmetry can be used to write down Ward
identities for correlation functions and the $S$-matrix.

Let us study in more detail the nonlinearity involved in
the collective representation(here we summarize the results
achieved in [\JRvT]).
One is in general interested in comparison with similar
nonlinearities (and Ward identities) obtainable in the
world sheet conformal field theory analysis.
The latter is only performed in the approximation neglecting
the cosmological constant term $(\mu \to 0)$ which  represents the
strong coupling regime of the field theory
$(g_{\rm st} = 1/\mu \to \infty)$.
In this limit one simply expands
$$\alpha_{\pm}(x,t) =\pm x+{1\over 2x}\,\hat\alpha_{\pm}\,\,,\eqno\eq$$
which is an approximate form when
$$\pi\phi_0(x) =\sqrt{x^2-\mu} \to x = {e^{\tau}\over 2}\,\,.
\eqno\eq$$
The exact tachyon operators reduce
in the leading (linear) approximation to
$$\tilde T_n^{\pm} ={1\over n}\,\int {d\tau \over 2\pi}\,\,
e^{n\tau}\,\hat\alpha_\pm \,\,,\eqno\eq$$
This is as it should be since they are to describe left- and right-moving
waves,
respectively. Consider now the $w_\infty$ generators
in the same approximation. With the
above background shift one easily finds that they reduce to
$$O_{JM}={1\over J-M+2}\,\int {d\tau\over 2\pi}\,\,
e^{2M\tau}\,\hat\alpha_{+}^{J-M+2}
+{(-1)^{J-M}\over J+M+2}\,\int {d\tau\over 2\pi}\,\,
e^{-2M\tau}\,\hat\alpha_{-}^{J+M+2}\,\,.\eqno\eq$$
Here we see that the operator $O_{JM}$ behaves as the
$(J-M+2)$th power of the right-moving tachyon $\alpha_+$ and
also the $(J+M+2)$th power of the left-moving tachyon $\alpha_-$.
These are the leading polynomial powers in the left- and
right-moving components of the tachyon; even in this
strong coupling limit the theory is nonlinear and one has
further higher order terms. Concerning these one can go to the
in (out) fields (where it is likely that only leading terms
remain). The in (out) fields are simply limits of the component fields
$\alpha_\pm$:
$$\alpha_{\rm out}(t-\tau) = \lim_{t\to +\infty} \alpha_+\,,
\qquad\quad \alpha_{\rm in}(t+\tau) = \lim_{t\to -\infty} \alpha_-\,\,,
\eqn\inoutfields$$
Since the operators $O_{JM}$ are conserved (up to a phase),
looking at the $t\to \pm\infty$ limit of $e^{2Mt}\,O_{JM}$ we
obtain an identity (between the in- and out-representation):
$$\eqalign{O_{JM}&={1\over J-M+2}\,\int {dz\over 2\pi}\,\,
\alpha_{\rm out}^{J-M+2} (z)\cr
\noalign{\vskip 0.2cm}
&={(-1)^{J-M}\over J+M+2}\,\int {dz\over 2\pi}\,\,
\alpha_{\rm in}^{J+M+2} (z) \,\,.}\eqno\eq$$
Introducing creation--annihilation operators
$$\alpha_{\rm in}(z)=\int dz\,e^{-ikz}\,\alpha(k)\,,
\qquad\quad \alpha_{\rm out}(z)= \int dz\,e^{-ikz}\,\beta(k)\,\,,
\eqno\eq$$
with $\alpha(k)=a(k)$ and $\alpha(-k)=ka(k)^\dagger$;$\beta(k)=b(k)$ and
$\beta(-k)=kb(k)^\dagger$ we have the expressions
after a continuation $k \to ik$:
$$\eqalign{O_{J,-M}&={1\over J-M+2}\,\int dk_1\ldots dk_{J-M+2}\,\,
\alpha(k_1)\ldots\alpha(k_{J-M+2})
\delta(\sum k_i +2M)\cr
\noalign{\vskip 0.2cm}
&={(-1)^{J-M}\over J+M+2}\,\int dp_1\ldots dp_{J+M+2}\,\,
\beta(p_1)\ldots\beta(p_{J+M+2})
\delta(\sum p_i +2M)\,\,.}\eqno\eq$$
These representations can be compared with analogue expressions
found in conformal field theory [\KlPo].

The Ward identities essentially follow from the in--out
representations of the generators in terms of the tachyon
field. A typical $S$-matrix element is given by
$$S(\{k_i\};\{p_j\})=\langle 0| \prod_j \,\beta(p_j)\,\prod_i \, \alpha(-k_i),
|0\rangle \,\,.\eqno\eq$$
Consider a general matrix element of the $w_\infty$
generator $O_{JM}$,
$$\langle \,\,0|\beta\, O_{JM}\, \alpha |0\,\,\rangle \,\,.$$
It can be evaluated by commuting to the left or to the right
using alternatively the above in--out representations. The
two different evaluations give the identity
$$\langle 0| [\beta, O_{JM}]\, \alpha|0\rangle =
\langle 0| \beta \,[O_{JM}, \alpha]|0\rangle\,\,,\eqno\eq$$
which summarizes the general Ward identities. These, when
written out explicitly using the representations for
$O_{JM}$, have the form of recursion relations reducing the
$N$-point amplitude to lower point ones. Specifically, the
one creation operator term of the $\alpha$-representation
for $O_{JM}$ gives:
$$O_{M+1,-M}\,a^\dagger(k_1)\,a^\dagger(k_2) |0\rangle =
4\pi (k_1+k_2+2M)\,a^\dagger(k_1+k_2+2M) |0\rangle\,\,,\eqno\eq$$
which turns a two-particle state into one-particle state.
Generally,
$$O_{M+N,-M}\,a^\dagger(k_1)\ldots a^\dagger(k_{N+1}) |0\rangle =
2\pi^N\,(N+1)! (\sum k_i+2M)\,a^\dagger(\sum k_i+2M)
|0\rangle\,\,,\eqno\eq$$
showing a reduction of the $(N+1)$-particle state into a
single-particle state.

As an example let us calculate the $3\to 1$ amplitude
$$S_{3,1}=\langle 0|b(p)\,a^\dagger(k_1)\,
a^\dagger(k_2)\,a^\dagger(k_3) |0\rangle\,\,.\eqno\eq$$
The energy-momentum conservation laws imply
(recall that for $B^\pm_n$, $\epsilon=\pm n,
p_s =-2+n$):
$$\eqalign{&k_1+k_2+k_3-p=0\,\,,\cr
&(-2+k_1)+(-2+k_2)+(-2+k_3)+(-2+p)=-4\,\,.}$$
The latter is a specific case of the general
Liouville conservation (or bulk condition):
$$\sum_{i=1}^N = p_s^i = -4\,\,.\eqno\eq$$
The energy-momentum relations specify the momentum of
the 4th particle
$$p=2 \qquad {\rm or} \qquad k_1+k_2+k_3=2\,\,.$$
Use now the operator
$$\eqalign{O_{{1\over 2},-{1\over 2}} &= {1\over \sqrt{\mu}}\,
\int dk_1 dk_2 dk_3\,\, k_1 \,a^\dagger(k_1)\,
a(k_2)\,a(k_3)\,\delta(k_1-k_2-k_3+1) \cr
\noalign{\vskip 0.2cm}
&= {\sqrt{\mu}}\, \int dp_1 dp_2 \,\, p_1\,
b^\dagger(p_1)\,b(p_2)\,\delta(p_1-p_2+1) }$$
to deduce
$$\eqalign{S_{3,1}&=\langle \,b(2)\,a^\dagger(k_1)\,
a^\dagger(k_2)\,a^\dagger(k_3)\,\rangle\cr
\noalign{\vskip 0.2cm}
&={\pi\over \mu}\,(k_1+k_2-1)\,
\langle \,b(1)\,a^\dagger(k_1+k_2-1)\,
a^\dagger(k_3)\,\rangle\cr
\noalign{\vskip 0.2cm}
&+{\pi\over \mu}\,(k_2+k_3-1)\,
\langle \,b(1)\,a^\dagger(k_2+k_3-1)\,
a^\dagger(k_1)\,\rangle\cr
\noalign{\vskip 0.2cm}
&+{\pi\over \mu}\,(k_1+k_3-1)\,
\langle \,b(1)\,a^\dagger(k_1+k_3-1)\,
a^\dagger(k_2)\,\rangle \,\,.}$$
Taking the normalized three-point function to be
$S_3 = 1/\mu$, the result
$$S_{3,1} ={\pi\over \mu^2}\,\bigl(2(k_1+k_2+k_3)-3\bigr)=
{\pi\over \mu^2} $$
follows. One can iteratively repeat the same reduction for higher
point amplitudes and find
$$S_{N,1}= {\pi^{N-1}\over (N-2)!}\,\,\mu^{-N+1}\,\,,\eqno\eq$$
which is the $(N+1)$-point  ``bulk'' scattering amplitude.

In describing the infinite symmetry  we have followed the
matrix model approach were the appearance of the symmetry structure
is most natural. The features described arise also in the continuum
conformal field theory language where the Ward identities take a
particularly elegant form.

Of crucial importance in establishing continuum
quantities that are analogous with those of the matrix model is Witten's
identification of the ground ring [\WiGR]. This consist of ghost
number zero, conformal spin zero operators ${\cal O}_{JM}$
which are closed under operator products
${\cal O'}\cdot {\cal O''} \sim{\cal O''}$ (up to BRST
commutators). The basic generators are
$${\cal O}_{0,0}=1\,,\qquad
{\cal O}_{{1\over 2},\pm{1\over 2}} =
\Bigl[ c\,b \pm {i\over\sqrt{2}}\,\partial X -
{1\over\sqrt{2}}\,\partial\varphi\Bigr]\,
e^{(\pm iX+\varphi)/\sqrt{2}}\,\,.\eqno\eq$$
The suggestion (of Witten) was that
${\cal O}_{{1\over 2},\pm{1\over 2}}$ are the
variables which correspond to the phase space coordinates of the
matrix model
$$\eqalign{
{\cal O}_{{1\over 2},+{1\over 2}} &=a_+ \equiv p+x\,\,,\cr
\noalign{\vskip 0.2cm}
{\cal O}_{{1\over 2},-{1\over 2}} &=a_- \equiv p-x\,\,,}\eqno\eq$$
with ${\cal O}_{0,0}=1$ being the cosmological constant
operator. Once the (fermionic) matrix eigenvalue coordinates have
been identified one could study the action of discrete states
vertex operators upon them. They turn out to act as vector
fields on the scalar ring
$$\Psi_{JM}={\partial h\over \partial a_+}\,\,
{\partial\over\partial a_-} -
{\partial h\over \partial a_-}\,\,
{\partial\over\partial a_+} \,\,,\eqno\eq$$
with the familiar matrix model form
$h_{JM}=a_+^{J+M}\,a_-^{J-M}$. In the continuum
approach the $w_\infty$ generators are integrals of
conserved currents which are for closed string theory
constructed as
$$\eqalign{
&Q_{JM}=\oint {dz\over 2\pi i}\,\,W_{JM}\,(z,\bar z)\,\,,\cr
\noalign{\vskip 0.2cm}
&W_{JM}\,(z,\bar z)=\Psi^+_{J+1,M}(z)\,{\cal O}_{JM}(\bar z)\,\,.}
\eqno\eq$$
One can study the action of these operators on the
the tachyon vertex operators.
A formula derived by Klebanov [\Kl] reads
$$Q_{M+N-1}\,T^+_{k_1}(0)\int T^+_{k_2}\ldots\int T^+_{k_N}=
F_{N,M}(k_1,\ldots,k_N)\, T^+_{-\sum k_i +M}\,\,,\eqno\eq$$
where
$$F_{N,M}(k_1,\ldots,k_N)=2\pi^{N-1}\,N!\,k\,
{\Gamma(2k)\over \Gamma(1-2k)}\,
\prod_{i=1}^N\,{\Gamma(1-2k_i)\over \Gamma(2k_i)}\,\,,\eqno\eq$$
with a similar formula for the action of
$Q_{-M+N-1}$ on $N$ oppositely moving $T^-$ tachyons.
These representations of the $w_\infty$ generators on
tachyon vertex operators are clearly comparable to the
direct representation obtained in the matrix model
(or more precisely collective field formalism). The
comparison and agreement of these representations
 is the closest one comes in being
able to identify the two approaches.

The conformal (vertex operator) formalism gives a very
elegant summary of Ward identities in the form of general
(master) equation. We end this section with a short description
of this equation [\ZW,\Ve,\KlPa]. It  follows from BRST
invariance of the discrete state vertex operators
$$\{ Q_{\rm BRST}, c(z) W_{JM}(z)\}=0\,\,,\eqno\eq$$
which implies that for general tachyon correlation function
$$\big\langle \{ Q_{\rm BRST}, c\, W_{JM}\}\,
V^\pm_{k_1} \ldots V^\pm_{k_n}\,\big\rangle =0\,\,.\eqno\eq$$
Changing to operator formalism
$$\sum_{{\rm perm}} \big\langle V^\pm_{k_1}\,
\{ Q_{\rm BRST}, c\, W_{JM}\}\Delta V^\pm_{1} \ldots
\Delta V^\pm_{1}\,V^\pm_{k_n} \big\rangle =0\,\,,\eqno\eq$$
allows one to eliminate $ Q_{\rm BRST}$. The vertex operators are all
BRST invariant while the propagator is essentially the
inverse of $Q$:
$$[Q,\Delta]=\Pi_{L_0 -\bar L_0}\,b^-_0\,\,,\eqno\eq$$
where the $\Pi$ projects on the subspace
$(L_0 -\bar L_0)|\Phi_i \rangle =0$. The final form of the
Ward identity then follows
$$\sum_{{\rm partitions}\atop {m+m'=n}}\,
\langle V_{i_1}\ldots  V_{i_m}\,\Phi \rangle
\langle \Phi  V_{j_1}\ldots  V_{j_{m'}}\,c W_{JM} \rangle =0\,\,.
\eqno\eq$$

Most of the considerations of this section and most of the studies
of the $w_\infty$ symmetry are performed in the extreme limit where the
cosmological term is ignored. This is particularly the case for the
continuum, conformal field theory approach. Some attempts to
extensions and inclusion of the nontrivial cosmological constant
effect were made however. In the matrix model the cosmological
term is introduced in a simple and elegant way corresponding
 to nonzero Fermi energy
$$h_0={1\over 2}\,(p^2 -x^2)\quad \to \quad
h_\mu={1\over 2}\,(p^2 -x^2)+\mu\,\,.\eqno\eq$$
Since the ground ring generators $a_\pm$ were identified
to be analogues to $p\pm x$ it is then expected that an
equivalent deformation from $a_+ a_- =0$ can be established in
the continuum conformal field theory approach. This is seen
by considering the action of a ground ring on tachyons. For
$\mu =0$ it reads
$$\eqalign{&a_+\,c\,\bar c \,\tilde T^+_k = c\,\bar c \,
\tilde T^+_{k+1}\,\,,\cr
\noalign{\vskip 0.2cm}
&a_-\,c\,\bar c \,\tilde T^+_k = 0\,\,.}\eqno\eq$$
The effect of cosmological perturbation
$\mu\,T^+_{k=0}$ is found by evaluating the first order
perturbation theory contribution
$$a_-\,c\,\bar c \,\tilde T^+_k =
-a_-\,c\,\bar c \,\tilde T^+_k(0)\,
\Bigl( \mu\int d^2 z\,,\tilde T^+_{k=0}(z) \Bigr)\,\,.\eqno\eq$$
On the right-hand side $a_-$ essentially fuses the two
tachyon operators into one giving (to first order)
$$a_- \,\tilde T^+_k = -\mu \,\tilde T^+_{k-1}\,\,.\eqno\eq$$
This replaces the second relation above and now the nonzero
Fermi level condition $a_+ a_- =-\mu$ results. To first
order [\WiGR,\Ba] one then has an agreement with the matrix model.
This is encouraging and one would clearly like to establish the complete
agreement at an exact level.

\bigskip
\chapter {$S$-matrix}
\medskip

Let us now describe the complete tree-level $S$-matrix of the
$c=1$ theory. In the previous sections we have seen the ``bulk''
scattering amplitudes which follow from the Ward identities or
are computed in conformal field theory. The complete
$N$-point scattering amplitude
$S_N=\langle T_{p_1}  T_{p_2}\ldots  T_{p_N} \rangle$
takes the (factorized) form
$$S_N=\prod_{i=1}^N (-)\mu^{i p_i}\,
{\Gamma(-ip_i)\over \Gamma(+ip_i)}\,\,A_{\rm coll}(p_1,\ldots,p_N)\,\,.
\eqno\eq$$
The external leg factors are associated with a field
redefinition [\GrKl] of vertex operators
$$T_k^\pm = {\Gamma(\mp k)\over \Gamma(\pm k)}\,\,
\tilde T_k^\pm\,\,.\eqno\eq$$
It is the redefined tachyon vertex operator $\tilde T$
that found its natural role in collective field theory.

The external leg factors of the full $S$-matrix have a
very relevant physical meaning which we now discuss. In
Minkowski space-time, $(k=ip)$, one has
$$\Delta =\mu^{\mp ip}\,\,
{\Gamma(\pm ip)\over \Gamma(\mp ip)}\,\,.\eqno\eq$$
So, the factors $\Delta =e^{i\theta_p}$ are pure phases.
As such they give no contributions to the actual
transition amplitudes and could be ignored.
The fact is however that they carry physical information
on the nature of tachyon background. The factors exhibit
poles at discrete imaginary energy
$$p\sqrt{2}=in\,,\quad n=1,2,3,\ldots \eqno\eq$$
If we consider a process with an incoming tachyon and
$N$ outgoing ones, the discrete imaginary value of the
incoming momenta signifies the resonant on-shell
process in which a certain number  $r$ of Liouville
exponentials participate
$$\langle T_-\,(\mu\,e^{-\sqrt{2}\varphi})^r\, T_+ \ldots
t_+ \rangle\,\,.\eqno\eq$$
The on-shell condition in this case indeed gives
$$i\sqrt{2}\,p =-(r+N-1) \,\,,\eqno\eq$$
in agreement with the discrete imaginary energy poles noted above.

In collective field theory the
external leg factors are associated with
a field redefinition given by an integral transformation. The transformation
comes from the change of coordinates between the Liouville  and the
time-of-flight variables. It is the later that appears naturally in the
collective field formalism  and as we have seen provides a simple description
of the theory.
Let us recall the basic  (Wilson) loop operator of the matrix model
with its Laplace transform
$$\hat W(\ell,t)\equiv \Tr\,(e^{-\ell M}) =
W_0 +\int dx\,e^{-\ell x}\,\partial_x \eta \,\,.\eqno\eq$$
After the change to the time of flight coordinate
$x=\sqrt{2\mu}\,\cosh\tau$ and the explicit identification of the Liouville
$\ell =2e^{-\varphi/\sqrt{2}}$ the integral
transformation results
$$\hat W(\ell,t)=\int_0^\infty d\tau\,\,
{\rm exp}\,\Bigl[-2\sqrt{2\mu}\,e^{-\varphi/\sqrt{2}}\,
\cosh \tau\Bigr]\,\partial_\tau\,\eta(\tau,t)\,\,,\eqno\eq$$
We have seen in our earlier study of the linearized theory that
this integral transformation takes the
Liouville operator into a Klein-Gordon operator
$$(\partial_t^2 -\partial_\tau^2)\,\eta
\quad \Longleftrightarrow \quad
\bigl(\partial_t^2 -{1\over 2}\,\partial_\varphi^2
+4\mu e^{-\sqrt{2}\,\varphi}\,\bigr) \hat W\,\,.\eqno\eq$$

The integral transformation therefore expresses the tachyon field in
terms of a simple Klein-Gordon field $\eta(\tau,t)$:
$$T(\varphi,X)\equiv e^{-\sqrt{2}\,\varphi}\,
\hat W(\ell,t) = \int_0^\infty d\tau\,\,
{\rm exp}\,\,\Bigl[-2\sqrt{2\mu}\,e^{-\varphi/\sqrt{2}}\,
\cosh \tau\Bigr]\,\partial_\tau\,\eta \,\,,\eqno\eq$$
with the expected relation between the matrix model and
string theory times $X=\sqrt{2}\,t$.
The correlation functions of the
tachyon field $T$  are then expressible in terms of correlation
functions of the collective field
$\eta$. The transformation described takes plane wave solutions of the
Klein-Gordon equation
$$\eta(\tau,t)=\int_{-\infty}^\infty\,{dp\over p}\,\,
\tilde\eta (p)\,e^{-ipt}\,\sin (p\tau) \eqno\eq$$
into Liouville solutions
$$T(\varphi,t)=\int dp\,e^{-ipt}\,\gamma(p)\,
K_{ip}(2\sqrt{\mu}\,e^{-\varphi/\sqrt{2}}\,)\,
\tilde\eta(p)\,\,.\eqno\eq$$
The above redefinition of the
in--out fields does have an effect in supplying external leg
factors. The asymptotic behavior of $T(\varphi,t)$ reads
$$T\sim \int dp\,e^{-ipt}\,\Bigl(\Gamma(ip)\mu^{-ip/2}\,
e^{ip\varphi/\sqrt{2}} +
\Gamma(-ip)\mu^{ip/2}\, e^{-ip\varphi/\sqrt{2}} \Bigr)\,\,,\eqno\eq$$
giving the reflection coefficient
$$R(p)=-\mu^{ip}\,\,{\Gamma(-ip)\over \Gamma(ip)} \eqno\eq$$
for each external leg of the $S$-matrix.

After this redefinition the problem is reduced to
calculating amplitudes in collective field theory:
$ A_{\rm coll}\,(p_1,\ldots,p_N)$. There, as we have already seen,
one can derive an exact relationship between the in- and
out-field which contains the complete information about the $S$-matrix.
The solution to the scattering problem
can also be directly deduced from the exact oscillator
states. It is this procedure that turns out to be the most
straightforward and we now describe it in detail.

Consider the exact (tachyon) creation--annihilation operators
in collective field theory
$$B_{\pm ip}^\pm = \int {dx\over 2\pi}\,
\left\{ {(\alpha_+ \pm x)^{1\pm ip}\over 1\pm ip} -
{ (\alpha_- \pm x)^{1\pm ip}\over 1\pm ip} \right\}\,\,.
\eqno\eq$$
Previously we have seen that at fixed time these operators serve
as exact creation--annihilation operators of the
nonlinear collective Hamiltonian. Let us now follow the
time-dependent formalism (we describe here the derivation given
in [\JRvT]). The exact creation--annihilation operators have a simple
time evolution
$$B_{\pm ip}^\pm (t)= e^{-ipt}\, B_{\pm ip}^\pm (0)\,\,.
\eqno\eq$$
Consequently the quantity
$$\hat B_{\pm ip}^\pm \equiv e^{ipt}\, B_{\pm ip}^\pm (t)
\eqno\eq$$
is time-independent. We can simply look at the operator
$\hat B_{\pm ip}^\pm$ at asymptotic times $t= \pm\infty$
and obtain a relationship between the in and out fields~\inoutfields.
The  operator
$$B_{\pm ip}^\pm = \int {dx\over 2\pi}\,
\left\{ {(\alpha_+ \pm x)^{1\pm ip}\over 1\pm ip} -
{ (\alpha_- \pm x)^{1\pm ip}\over 1\pm ip} \right\}
\eqno\eq$$
contains contributions from both $\alpha_+$ and
$ \alpha_-$. At $t=\pm \infty$ only one of the terms
survives and we have an identity
(recall that $\hat B$ is time-independent),
$\hat B(+\infty) =\hat B(-\infty)$, which reads
$$\int {dx\over 2\pi}\,(\alpha_\pm \pm x)^{1\pm ip} =
\int {dx\over 2\pi}\,(\alpha_\mp \pm x)^{1\pm ip} \,\,.
\eqn\scatteq$$
It relates in and out fields ($\alpha_\pm$ now represent the
asymptotic fields). This is the scattering equation. It contains the
full specification of the $S$-matrix.

One can evaluate
and expand the left- and right-hand side of the Eq.~\scatteq.
Shifting by the static background
$$\alpha_\pm(t,x) \approx \pm \bigl(x-{1\over 2x}\bigr)
+{1\over 2x}\,\hat\alpha_\pm(t\mp \tau)\,\,,\eqno\eq$$
the left-hand side becomes
$$L=\int dx\,(-2x)^{1\pm ip}\,\Bigl\{
1-{(1\pm ip)\over 4x^2}\,(1\pm \hat\alpha_\pm) +
{\cal O}({1\over x^4})\Bigr\}\,\,.\eqno\eq$$
After a change of integration variable
$x=\cosh \tau \approx e^\tau/2$ the
${\cal O}(1/x^4)$ terms are seen to decay away
exponentially and what remains is only the term linear in
$\hat\alpha_\pm$. For the right-hand side we
simply find
$$R=\int dx\,\Bigl(-{1\over 2x}\Bigr)^{1\pm ip}\,
(1\pm \hat\alpha_\mp)^{1\pm ip}\,\,.\eqno\eq$$
The scattering equation then becomes
$$\int dz\,e^{-ipz}\,{1\over \mu}\,\alpha_\pm (z)=
{1\over 1\pm ip}\,\int dz\, e^{-ipz}\,
\left\{ \Bigl(1\pm {1\over\mu}\,\alpha_\mp \Bigr)^{1\mp ip}
-1 \right\}\,\,,\eqno\eq$$
giving the solution for the in-field as a function of the out-field
and vice versa. We have also explicitly restated the string
coupling constant $g_{\rm st}= 1/\mu$. This solution was
originally obtained [\MoPl] by explicitly solving the functional
relationships between the left and right collective
field components $\alpha_+$ and $\alpha_-$ given
earlier. We see here that it directly follows from the
exact oscillator states.

Before proceeding with the consideration of the $S$-matrix
let us note that the solution found has a reasonable
strong coupling limit. Indeed, for $\mu \to 0$ one is
lead to choose $ip$ to be an integer, $ip=N$, and the strong coupling
relation
$$\int dz\,e^{-Nz}\,\alpha_+ (z)=
{1\over 1+N}\,\int dz\, e^{-Nz}\,\Bigl({1\over\mu}\,
\alpha_- \Bigr)^{N+1}\eqno\eq$$
results. It is recognized as a statement specifying the bulk
amplitude where the $S_{1,N}$ and $S_{N,1}$ amplitudes
were nonzero with the momenta of the first (last) particle
being equal to $-1+N$. We have already used
relations of the above type (and their $w_\infty$ generalizations)
in our discussion of the Ward identities at (strong coupling)
$\mu=0$.

One can explicitly perform the series expansion in
$g_{\rm st}=1/\mu$, it reads
$$\hat\alpha_\pm (z) =\sum_{l=1}^{\infty}\,
{(-g_{\rm st})^{l-1}\over l!}\,\,
{\Gamma(\mp\partial +1)\over \Gamma(\mp\partial +2-l)}\,\,
\hat\alpha_\mp^l (z)\,\,.\eqno\eq$$
The $S$-matrix is defined in terms of momentum space
creation--annihilation operators
$$\pm\alpha_\pm (z) =\int {dp\over 2\pi}\,e^{-ipz}\,
\tilde\alpha_\pm (p)\,,\qquad\quad
[\tilde\alpha_\pm (p),\tilde\alpha_\pm (p')]=p\,\delta(p+p')\,\,,
\eqno\eq$$
with $\tilde\alpha_- (-p)$ and  $\tilde\alpha_+ (-p)$
being the in--out creation operators, respectively
($\alpha(-p)$ and $\beta(-p)$ in our earlier notation).
In momentum space
$$\tilde\alpha_\mp (p)= \sum_{l=1}^{\infty}\,
{(-g_{\rm st})^{l-1}\over l!}\,\,
{\Gamma(1 \pm ip)\over \Gamma(2\pm ip -l)}\,\,
\int dp_i\,\,\delta(p-\sum p_i)\,
\tilde\alpha_\pm(p_1)\ldots \tilde\alpha_\pm(p_l)\,\,,\eqn\momsp$$
and the $n\to m$ $S$-matrix element is defined by
$$A_{\rm coll} (\{p_i\}\to\{p'_j\}) =
\langle 0|\,\prod_{j=1}^m\,\tilde\alpha_+ (p'_j)\,\,
\prod_{i=1}^n\,\tilde\alpha_- (p_i)\,|0\rangle\,\,.\eqno\eq$$
Consider for example $n=1, m=3$ which is the four-point amplitude
$$A_{1,3}=\langle 0| \alpha_+(p'_1)\, \alpha_+(p'_2)\,
\alpha_+(p'_3)\,\alpha_-(-p_1)\,|0\rangle\,\,.\eqno\eq$$
It is given by the cubic term ${\cal O}(g_{\rm st}^2)$
in the expansion of \momsp\ which equals
$$\alpha_- (-p_1)= {1\over 3!}\,\,g_{\rm st}^2\,\,
{\Gamma(1-ip_1)\over \Gamma(-1-ip_1)}\, \int dp'_i\,
\delta(p_1-\sum p'_i)\,
\alpha_+(p'_1)\, \alpha_+(p'_2)\, \alpha_+(p'_3)\eqno\eq$$
and
$$A_{1,3}(p_1;p'_1,p'_2,p'_3)=
i\, g_{\rm st}^2 \,p_1 p'_1 p'_2 p'_3\,(1+ip_1)\,\,.
\eqno\eq$$
In general, an arbitrary amplitude is given in [\MoPl] to read
$$A_{n,m}=i\,(- g_{\rm st})^{n+m-2}\,
\Bigl( \prod_{i=1}^n p_i \Bigr) \Bigl( \prod_{j=1}^m p'_j \Bigr)\,
{\Gamma(-ip_n)\over \Gamma(1-m-ip_n)}\,\,
{\Gamma(1-m-i\Omega)\over \Gamma(-3-n-m-i\Omega)}\eqno\eq$$
where $\Omega=\sum_{i=1}^n\, p_i$ and the result is valid
in the kinematic region $p_n> p'_k > \sum_{j=1}^{n-1}\,p_j$.
This completes the derivation of the collective tree-level
amplitudes.

\bigskip
\chapter {Black Hole}
\medskip

Two-dimensional string theory possesses another interesting curved
space solution taking the form of a black hole. It is
described exactly by the $SL(2,\IR)/U(1)$
nonlinear $\sigma$-model
$$S_{\rm WZW} = {k\over 8\pi}\int d^2 z\,\Tr\left( g^{-1}
\partial g g^{- 1} \bar{\partial} g\right) -
ik \Gamma_{\rm WZW} + {\rm Gauge}\,\,,\eqno\eq$$
with $k = {9\over 4}$. This then gives the required central charge
$c = {3k\over k-2} -1 = 26$.
As such the model should be thought of as a different classical solution of the
same theory. We have in the earlier lectures seen that the flat
space-time string theory is very nicely and very completely described by a
matrix model. The black hole solution is however markedly different from the
$c=1$ theory. It is characterized by the absence of tachyon
condensation and a nontrivial metric and dilaton field:
$$\eqalign{&T(X) = 0\,\,,\cr
\noalign{\vskip 0.1cm}
&(ds)^2= - {k\over 2} \, {dudv\over M-uv}\,\,\cr
\noalign{\vskip 0.1cm}
&D = \log (M-uv)\,\,. }\eqno\eq$$
Here a particular $SL(2,\IR)$ parametrization is chosen:
$g=\left( {\alpha\atop -v} \,\,{u\atop \beta}\right) $,
$\alpha\beta +uv=1$ and $M$
is the black hole mass.  There is actually a parametrization (related
to the $c=1$ theory) in which the $\sigma$-model Lagrangian reads
$$\eqalign{S_{\rm eff}={1\over 8\pi} \int d^2 z\,\, \Bigl\{
&(\partial X')^2 + (\partial\varphi')^2 - 2\sqrt{2} \phi' R^{(2)} \cr
&+ M \vert{1\over 2\sqrt{2}} \partial\varphi' + i \sqrt{{k\over2}}
\partial X'\vert^2\, e^{-2\sqrt{2} \varphi'} \Bigr\}\,\,.}\eqno\eq$$
This parametrization corresponds to a linear dilaton but
in contrast to the $c=1$ theory, one has a black hole mass term
perturbation represented by a gravitational vertex operator instead of
the cosmological constant term given by a
tachyon operator $e^{-\sqrt{2} \varphi}$.
One of the surprising facts is however that there exists a classical
duality transformation that can be used to relate the two sigma models to
each other [\MaSh]. From this there arises a hope that one could
possibly be able to describe the black hole by a matrix model also.
More generally from a string field theory viewpoint one would hope to
be able to describe different classical solutions in the same setting.
In what follows we will present some joint work done with T. Yoneya
on this subject [\JeYo]. For other different attempts see
[\DDMW, \MuVa].

First insight into the black hole problem is gained by
considering the linearized tachyon [\DVV] field in the external background.
In the conformal field theory  this is given by the zero mode Virasoro
condition. The Virasoro operator $L_0(u,v)$ consists of
two parts,
$$L_0= -\Delta_0 + {1\over 4}\,(u\partial_u-v\partial_v)^2\,\,,
\eqno\eq$$
where $\Delta_0$ is the Casimir operator of $SL(2,\IR)$.
The Virasoro condition for the linear tachyon field (vertex operator)
reads:
$$L_0(u,v)T\equiv {1 \over k-2}\Bigl[(1-uv)\partial_u \partial_v -
{1 \over 2}(u\partial_u + v\partial_v)-{1 \over 2k}
(u\partial_u-v\partial_v)^2 \Bigr]\,T=T\,\,.\eqn\Vircond$$

The on-shell  tachyon corresponds to the continuous
representation of $SL(2,\IR)$ which has eigenvalues
$\Delta_0= -\lambda^2 - {1 \over 4} \, \quad (\lambda = {\rm  real})$
and  $-i\partial_t= 2i\omega$ with the on-shell condition
$\lambda^2 = 9\omega^2$ at $k=9/4$.
The above equation can be interpreted as corresponding to a covariant
Laplacian $L_0=-{1 \over 2e^{D}\sqrt{G}}\,\partial^{\mu}\,e^{D}\,
\sqrt{G}\,G_{\mu\nu}\,\partial_{\nu}$ in the  background space-time metric
$G_{\mu\nu}$ and dilaton $D$,
which can be read off  from Eq.~\Vircond
$$\eqalign{
&ds^2={k-2 \over 2}\,[\,dr^2 - \beta^2(r)\,d\bar t^2\,]\,\,,\cr
\noalign{\vskip 0.15cm}
&D = \log \,\bigl(\sinh {r\over\beta(r)}\,\bigr) + a\,\,,\cr
\noalign{\vskip 0.15cm}
&\beta(r)= 2\,(\coth^2{r \over 2} -{2 \over k})^{-1/2}\,\,.}\eqno\eq$$
These are  then candidates for the
``exact'' background.
Here the new coordinate $r$ and time $\bar t$ are defined by
$$u= \sinh{r \over 2} \,e^{\bar t}\,,\qquad\quad
v= -\sinh{r \over 2} \,e^{-\bar t}\,\,.\eqno\eq$$
These variables describe the static exterior region outside the
event horizon located  at $r=0$. The constant $a$
determines the  mass of the black hole
$$M_{{\rm bh}}= \sqrt{{2 \over k-2}}\,\,e^a\,\,.\eqno\eq$$

The  exact metric can be shown to be free of curvature singularity.
However, one still has a ``dilaton singularity'' at $uv=1$
where the string coupling $g_{\rm st}\sim e^{-D/2}$
diverges. In terms of the variables $u$ and $v$, the dilaton reads
$$D=\log\,\Bigl[4\Bigl(-uv(1-uv)(-{(1-uv) \over uv}-{2 \over k})
\Bigr)^{1/2}\Bigr]+a\,\,,\eqno\eq$$
and the region $uv >1$ corresponds to a disjoint
region with a naked singularity.

The free parameter $a$ can be eliminated by a scale transformation
$$u \rightarrow M^{-1/2}\,u\,,\qquad v \rightarrow M^{-1/2}\,v\,,
\qquad M\equiv e^a\,\,.\eqno\eq$$
This introduces the black hole mass parameter in more explicit way,
where one replaces $(1-uv)$ by $(M-uv)$ in the expressions for the
dilaton and the metric. An important relation is the connection of the
string coupling constant with the parameter
$a$, or rather the black hole mass. In general, the dilaton field
determines the string coupling constant and in the present case one
obtains
$$g_{\rm st}\,(r=0)  \propto e^{-a/2}= M^{-1/2}\,\,.\eqno\eq$$
This is to be compared  with the dependence of
$g_{\rm st} \propto \mu^{-1}$
on the cosmological constant in flat space-time. One notes the
different power which comes from the different scaling dimensions
of the two parameters.
The two backgrounds become identical in the asymptotic region.
Consider the asymptotic  behavior of the Virasoro operator
and the dilaton  when $r\rightarrow \infty$. Using
$u\sim e^{{r\over 2}+\bar t}\,, \,\,
v \sim e^{{r\over 2}-\bar t}$ one finds
$$\eqalign{&L_0\sim {1 \over 4(k-2)}\,(\partial_r^2+\partial_r)+
{1 \over 4k} \,\partial_{\bar t}^2\,\,,\cr
\noalign{\vskip 0.1 cm}
&D\sim r + a - \log 4\,\,.}\eqno\eq$$
This is the form of Virasoro operator in the linear dilaton case,
the parameters $r, \bar t$ are identified asymptotically with the
$\varphi$ and $t$ for the linear dilaton background as
$$\eqalign{\bar t &\leftrightarrow\sqrt{{1 \over 2k}}\,\,t=
{\sqrt{2} \over 3}\,\,t\,\,,\cr
\noalign{\vskip 0.1 cm}
r &\leftrightarrow\sqrt{{2 \over k-2}}\,\,\varphi=2\sqrt{2}\,\varphi\,\,.}
\eqno\eq$$
For the conjugate  momentum and energy, the correspondence is then
$$\eqalign{&ip_{\varphi}= -\sqrt{2}+i2\sqrt{2}\,\lambda = -\sqrt{2} +
{i \over \sqrt{2}}\,\,p_{\tau}\,\,,\cr
\noalign{\vskip 0.1 cm}
&ip = i{2\sqrt{2} \over 3}\,\omega={i \over\sqrt{2}}\,p_{t}\,\,.}
\eqn\enmom$$
This implies a  one-to-one correspondence of tachyon states in the
black hole and linear dilaton backgrounds. There is also a correspondence
between the discrete states spectra in the two theories.

In the Minkowski metric, the spectrum of the
discrete states  for the black hole is isomorphic to that in the linear
dilaton background. In particular,
the first nontrivial discrete state with zero energy
($j=1, m=0$ or $ip_{\varphi}= -2\sqrt{2}, \, p =0$) is
identified with the operator associated with the mass
of black hole, as can be seen from  the first correction to the
asymptotic behavior of the exact space-time metric
$$ds^2 \sim {k-2 \over 2}\,\Bigl[dr^2 - {4k \over k-2}\,\Bigl(1-
{4k \over k-2}\,e^{-r} + {\cal O}(e^{-2r}) \Bigr)dt^2\Bigr]\,\,.
\eqno\eq$$
It is important to note that the $\varphi$ momentum
is twice that of the operator corresponding to
tachyon condensation.

The solutions of the tachyon Virasoro conditions describe the scattering of
a single tachyon on the black hole. It represents
one of the few quantities that has been rigorously computed
in black hole string theory [\DVV]. The amplitude provides some nontrivial
physical insight and is obtained as follows. One writes an  integral
representation for the solution with definite energy $\omega$ and momentum
$\lambda$ as
$$\int_C {dx\over x}\, x^{-2i\omega}\,\bigl(\sqrt{M-uv}
+{u \over x}\bigr)^{-\nu_-}\, \bigl(\sqrt{M-uv} -vx\bigr)^{-\nu_+}\,\,,
\eqn\intrep$$
with $\nu_{\pm}={1 \over 2}-i(\lambda \pm \omega)$. In general, one has
four different contours of integration with two linearly independent
solutions corresponding, for example, to the coutours
$C_2\equiv[-u\sqrt{M-uv}, 0\,],\,
C_4\equiv(-\infty,\nu^{-1}\sqrt{M-uv}\,]$ as
($y\equiv uv =-\sinh^2 {r \over 2}$):
$$\eqalign{&T_{C_2}= U_{\omega}^{\lambda}=e^{-2i\omega \bar t}\,
F_{\omega}^{\lambda} (y)\,\,,\cr
\noalign{\vskip 0.2cm}
&T_{C_4}= V_{\omega}^{\lambda}=e^{-2i\omega \bar t}\,
F_{-\omega}^{\lambda} (y)\,\,,}\eqno\eq$$
where
$$F_{\omega}^{\lambda}(y)= (-y)^{-i\omega}B(\nu_+, \nu_-)
F(\nu_+, \nu_-, 1-2i\omega, y)\,\,.\eqno\eq$$
The asymptotic behaviors of the solutions are,
for $r \rightarrow 0 \, ({\rm horizon})$:
$$\eqalign{&U_{\omega}^{\lambda} \sim \beta(\lambda, \omega)
({u \over \sqrt{M}})^{-2i \omega}\,\,,\cr
\noalign{\vskip 0.1cm}
&V_{\omega}^{\lambda} \sim \beta(\lambda, -\omega)
(-{v \over \sqrt{M}})^{-2i \omega}\,\,}\eqno\eq$$
while for null-infinity $r \rightarrow \infty$:
$$F_{\omega}^{\lambda}\sim \alpha (\lambda, \omega)(-y)
^{-{1 \over 2}+ i\lambda} + \alpha (-\lambda,\omega)(-y)
^{-{1 \over 2}- i\lambda}\,\,,\eqn\asymp$$
where
$$\eqalign{\alpha(\lambda, \omega) &={\Gamma(\nu_+)\,
\Gamma(\bar\nu_- - \nu_+)\over \Gamma(\bar\nu_-)}\,\,,\cr
\noalign{\vskip 0.1cm}
\beta(\lambda, \omega) &= B(\nu_+, \bar \nu_-)\,\,.}\eqno\eq$$
We  see  that $U_{\omega}^{\lambda}$ describes
a wave coming from past null-infinity scattering on
the black hole, while $V_{\omega}^{\lambda}$ describes a wave emitted
by the white hole crossing the past event horizon.
The solution $U_{\omega}^{\lambda}$ gives the $S$-matrix elements of
tachyons, incoming from the asymptotic flat region at past null-infinity and
scattered out to future null-infinity. On-shell
$\omega = 3\lambda \,(>0)$, and this solution gives the reflection
and transmission coefficients as ratios of the coefficients appearing
in the above asymptotic forms:
$$\eqalign{R_B(\lambda)&= {\alpha(\lambda, \omega)\over
\alpha (-\lambda, \omega)}\,\,,\cr
\noalign{\vskip 0.1cm}
T_B(\lambda)&={\beta(\lambda, \omega)\over\alpha (-\lambda,\omega)}\,\,.}
\eqn\rtcoeff$$
The reflection and abosorbtion coefficients satisfy the unitarity relation
$$|R_B|^2 + {\omega \over \lambda} |T_B|^2 = 1\,\,.\eqno\eq$$
This describes the two-point correlation function and it
is of major interest to formulate a full quantum field theory in
the presence of a black hole which would be capable of giving
general $N$-point scattering amplitudes and correlation functions.
One is also very interested in being able to evaluate loop effects
and even to discuss formation and evaporation of black holes in
the general field theoretic framework.
In the absence of a general theory one can try to follow the analogy
with the $c=1$ theory and attempt to guess the structure required for the
black hole. This is what was done in [\JeYo]. Persuing the above
analogy we can postulate again that string theory in the black hole
background is described by
a factorized $S$-matrix. It is then reasonable to expect that the
external leg factors of the full $S$-matrix
are again determined through a non-local field redefinition whose role
is to connect
the Virasoro equations in the black hole background with the free
massless Klein-Gordon equation. The main part of the $S$-matrix is then
to be determined. The suggestion based on the analogy with the $c=1$
theory is that one again has a description  in terms of a
matrix model and the associated  collective field theory.
To simulate the black hole background the matrix model is expected to
include a deformation from the standard inverted oscillator potential.
 There as yet exist no general principles for constructing the theory but
one can make certain concrete suggestions on the eventual form of the
matrix model.
Let us describe first  the expected form for the external leg factors.
These are supplied by a field redefinition whose
purpose is to reduce the black hole background Virasoro condition
to the scalar free field equation. We have summarized the  black hole
Virasoro equation and its solutions in detail, so let us
consider the integral representation \intrep\
with the contour $C_2$ which is appropriate for
the scattering problem in the exterior region $(u>0, v<0)$:
$$U_{\omega}^{\lambda}(u,v)=
\int_{C_2}{dx \over x} x^{-2i\omega}(\sqrt{M-uv}+ {u \over x})^{-\nu_-}
(\sqrt{M-uv}-vx)^{-\nu_+}\,\,.\eqno\eq$$
Since the spectrum of the on-shell solution has a one-to-one
correspondence through \enmom\ with that of the free Klein-Gordon
equation, it is natural to make the following change of integration
variable:
$$\eqalign{&(\sqrt{M-uv} + {u \over x})^{-1}\,(\sqrt{M-uv} - vx)
= e^{-4t /3} \,\,,\cr
\noalign{\vskip 0.1cm}
&(\sqrt{M-uv} + {u \over x})\,(\sqrt{M-uv} - vx)
=e^{-4\tau} \,\,.}\eqno\eq$$
The integral formula for the solution takes  now the form
$$U_{\omega}^{\lambda}
=\int_{-\infty}^{\infty}dt \int_0^{\infty} d\tau\,
\delta({u\,e^{-2t/3}+v\,e^{2t/3} \over 2}-\sqrt{M}\cosh 2\tau)\,
e^{-4i\omega t/3}\cos 4\lambda \tau\,\,.\eqno\eq$$
This is seen to be an integral transform of a  Klein-Gordon plane wave with
momentum and energy
$$p_{\tau}=4\lambda\,,\qquad\quad p_{t}={4\over 3}\,\omega\,\,.
\eqno\eq$$
Since the plane waves are recognized as natural eigenstates of the
linearied collective field  terms of time-of-flight variable
we have the candidate for the non-local field redefinition
$$T(u,v) =
\int_{-\infty}^{\infty}dt \int_0^{\infty} d\tau\,
\delta({u\,e^{-2t/3}+v\,e^{2t/3} \over 2} - \sqrt{M} \cosh 2\tau)
\gamma (i\partial_{t})\partial_{\tau}\eta (t, \tau)\,\,,
\eqn\bhfieldred$$
where $\gamma(i\partial_{t})^* = \gamma (-i\partial_{t})$
is an arbitrary weight function to be fixed by normalization
condition.

In terms of the Fourier decomposition
$$\eta(t,\tau)=\int_{-\infty}^\infty {dp\over p}\,\,
\tilde\eta (p)\,e^{-ipt}\,\sin\,p\tau\,\,,\eqno\eq$$
it reads
$$T(u,v) = \int_{-\infty}^{\infty}dp\,\,
\tilde\eta(p)\,\gamma (p)\,U_{\omega (p)}^{\lambda(p)}(u,v)\,\,,
\eqno\eq$$
with $\omega (p) = 3p/2, \lambda (p) = p/2$. In particular, the
asymptotic behavior for $y\rightarrow \infty$ is
$$\eqalign{T(u,v) = \int_{-\infty}^{\infty}dp\,\,
\tilde\eta(p)\,\gamma(p)
\Bigl[&(-y)^{-{1 \over 2}+i\lambda(p)}\,\alpha\bigl(\lambda (p),\omega(p)
\bigr) +\cr
&(-y)^{-{1 \over 2} - i \lambda (p)}\,\alpha\bigl(-\lambda (p),\omega(p)
\bigr)\Bigr]\,\, e^{-2i\omega(p) t}\,\,.}\eqno\eq$$
This shows that an asymptotic wave packet of $\eta$ field is
transformed into a deformed wave packet of the tachyon field.
The integral transformation that we have given suplies the leg factors of the
conjectured black hole $S$-matrix.
Even if the factorization becomes only an
approximate feature of the full theory one could expect that the
factorization holds near the poles of the $S$-matrix.

Let us now study the possible resonance poles produced by the external leg
factors. It turns out that studying the location of these poles
gives useful and nontrivial constraints on the full $S$-matrix.
 From the asymptotic behavior of \asymp\ and the associated
reflection coefficient \rtcoeff, we see that the positions
of the resonance poles are
$$i4\lambda = i{4 \over 3}\omega=i\sqrt{2}\,p_t =-2,-4,-6,\ldots\,\,.
\eqno\eq$$
This contrasts with the case of the usual $c=1$ model where we
have poles at all negative integers of the corresponding energy.
On the other hand, if we consider an amplitude for
an incoming tachyon with producing $N-1$ outgoing
tachyons, the energy and momentum conservation laws
are satisfied when the energy of incoming tachyon obeys
$$i\sqrt{2}p_t = -(2r + N-2)\,\,,\eqno\eq$$
where $r$ now counts the number of insertions of the
black hole mass operator. The factor $2$ multiplying  $r$
comes about because the momentum carried by the
black hole mass is twice that of the tachyon condensation.
Comparing next the two expressions for the location of the poles
we seee that these are consistent only if $N$ is even. More precisely,
only the even $N=2k$ point amplitudes are to be nonzero while the odd
$N=2k+1$ point amplitudes should vanish. This represents a strong requirement
on the form of the complete theory.

We are than lead to the main problem of specifying the full dynamics
in the form of a generalized matrix model.
In the limit of vanishing black hole mass, the black hole
background reduces to the linear dilaton vacuum.
This is a singular limit in the sense that
the string coupling diverges, corresponding to the $c=1$ matrix
model with vanishing 2d cosmological constant $\mu = 0$,
or zero Fermi energy. Since, according to our
hypothesis, the deformation corresponding to non-vanishing black hole
mass cannot be described by the usual matrix model,
we have to seek for other possible
deformations than the one given by  the
Fermi energy.  We  assume that the Fermi energy is
kept exactly at zero, while the Hamiltonian itself is modified.

{}From the earlier analysis we have several hints or constraints which the
modified Hamiltonian should obey. The first is that there is a double scaling
limit and that the resulting string coupling constant squared should be given
by the black hole mass $M$. The second constraint is the required vanishing
of all odd $N$-point amplitudes. Finally, in agreement with the world sheet
description of the black hole string theory one should have a natural
$SL(2,\IR)$ symmetry.

Consider a general modification of the inverted oscillator Hamiltonian
$$h(p,x) \quad\rightarrow \quad h_M (p, x)={1\over 2}\,(p^2 -  x^2)
+ M \delta h (p, x)\,\,.\eqno\eq$$
We have assumed that the deformation is
described by a term linear in $M$.
The first requirement for $\delta h$ is a scaling property
to ensure that the string coupling is proportional to $M^{-1/2}$.
In collective field theory after a shift by the classical ground state one has
that the string coupling  generally
proportional to $({dx \over d\tau})^{-2}$. Thus the above
requirement is satisfied if the deformation operator
$\delta h$ scales as
$\delta h(p, x) \rightarrow \rho^{-2}\delta h(p,x)$
under scale transformations $(p, x) \rightarrow
(\rho p, \rho x)$. This leads to
$$\delta h(p,x)={1\over 2 x^2}\,\,f({p \over x})\,\,.\eqno\eq$$

To further specify the  general function $f(p/x)$,
one invokes the requirement of $SL(2,\IR)$ symmetry. We have seen
in sect.~4, that the usual $c=1$
Hamiltonian $h= (p^2 - x^2)/2$ allows a set of
eigenoperators $O_{j,m}$ satisfying  the
the $w_{\infty}$ algebra~\winfalg.
The origin of this  algebraic structure, which is supposed to encode
the extended nature of strings, can be traced to  existence
of an $SL(2,\IR)$ algebra consisting of
$$\eqalign{&L_1={1 \over 4}\,(p^2-x^2)= h(p,x)\,\,,\cr
&L_2= -{1 \over 4}\,(px + xp)\,\,,\cr
&L_3= {1 \over 4}\,(p^2+x^2)\,\,.}\eqno\eq$$
The  eigenoperators satisfying the $w_{\infty}$
algebra are constructed in terms of the  $SL(2,\IR)$ operators according to
$$O_{j,m}=L_+^{{j+m \over 2}}\,L_-^{{j-m \over 2}}\,,
\qquad L_{\pm}=L_3 \pm L_2\,\,,\eqno\eq$$
which close under the Poisson bracket since the Casimir invariant has
a fixed value
$$L_1^2 + L_2^2 - L_3^2= {3\hbar \over 16}\,\,.\eqno\eq$$
(the Planck constant indicates the effect of operator ordering).

Since the spectrum of  discrete states in the
black hole background is expected to be the same
as that of the usual $c=1$ model in the Minkowski metric,
it is natural to require that the deformed model
should also share a similar algebraic structure.

There is actually a very simple model with the above structure. It is given
for $f=1$ in which case one has the extra term represented by a well known
singular potential. One has  the $SL(2,\IR)$ generators of the form:
$$\eqalign{&L_1(M)= {1\over 2}\,h_M (p,x)
={1 \over 4}\,(p^2-x^2+{M \over x^2})\,\,,\cr
&L_2(M)= -{1 \over 4}\,(px + xp)\,\,,\cr
&L_3(M)= {1 \over 4}\,(p^2 + x^2 + {M \over x^2})\,\,,}\eqno\eq$$
which satisfy
$$L_1(M)+L^2_2(M)-L^2_3(M)= -{M\over 2}+{3\hbar\over 16}\,\,.\eqno\eq$$
We  note that because of  different constraint
for the Casimir invariant  the  algebra of eigenoperators is now
modified in an $M$-dependent way. The algebraic properties of the
model with the singular potential have been investigated in detail
in [\AJcmp].
The model is exactly solvable and possesses some features characteristic of
black hole background.

Let us proceed to describe the properties of the  deformed model:
$$h_M(p,x) = {1 \over 2}(p^2 - x^2) + {M \over 2 x^2}\,\,.
\eqn\defham$$
We assume here that $M>0$. Then the genus zero free energy in the limit of
vanishing scaling parameter, $\bar M \rightarrow 0$, behaves like
$F\sim {N^2\over 8\pi\sqrt{2}}\,\bar M \log {\bar M \over \sqrt{2}}$.
The double scaling  limit is thus the limit $\bar M \rightarrow 0,
\,N\rightarrow \infty$
with $M \equiv N^2 \bar M$ being kept fixed. After  the
usual rescaling, $\,x\equiv \sqrt{N}\times$  matrix eigenvalue,
the system is reduced to the free fermion system with the one-body
potential $-{1 \over 2}x^2 + {M \over 2x^2}$.
Note that in the limit $M \rightarrow 0$ the potential
approaches the usual inverted harmonic oscillator potential with
a repulsive $\delta$-function-like singularity.

The solution of the classical equations
with energy $\epsilon$ reads
$$x^2(t) =-\epsilon +  \sqrt{M + \epsilon^2} \cosh 2t\,\,.\eqno\eq$$
The ground state corresponding to zero Fermi energy is obtained
by setting $\epsilon =0$ and replacing the time variable $t$
by the time-of-flight coordinate $\tau$, $x^2=\sqrt{M}\cosh 2\tau$.
This is recognized as  precisely the quantity appearing in
the integral transformation \bhfieldred. The
$\delta$-function present in the transformation gives a relation
between the black hole and the matrix model variables.
It serves to identify the matrix eigenvalue as
$$x^2=(u e^{-2t/3}+v e^{2t/3})/2\,\,.$$
The string coupling is now space dependent
$$g(\tau) \equiv {\sqrt{\pi} \over 12}\,\Bigl({dx \over
d\tau}\Bigr)^{-2} = {1 \over 48}\sqrt{{\pi \over M}}\,
\Bigl({1 \over \sinh^2\tau } +{1 \over \cosh^2 \tau}\Bigr)\,\,,
\eqno\eq$$
with the required relation with the black hole mass and
the asymptotic behavior at large $\tau$.

The tree  level scattering amplitudes are generally obtained
from the exact solution of the classical equations. The exact
solution to the collective  equations  has the following
parametrized  form
$$\eqalign{&x(t,\sigma)=\Bigl[-a(\sigma)+\sqrt{M+a^2(\sigma)}\,
\cosh 2(\sigma-t)\Bigr]^{1/2}\,\,,\cr
&\alpha(t,\sigma)={1 \over x(t,\sigma)}\,\sqrt{M+a^2(\sigma)}\,
\sinh 2(\sigma-t)\,\,.}\eqn\bhparasol$$
It contains an arbitrary function
$a(\sigma)$ describing the deviation of the
Fermi surface from its ground state form.
The asymptotic behavior for large $x$, of the profile
function reads
$$\alpha_{\pm}(t,\tau)=\pm x(\tau)\Bigl(1- {\psi_{\pm}
(t\pm \tau) \over x^2(\tau)}\Bigr) + {\cal O}({1 \over x^2})
\,\,.\eqno\eq$$
The functions $\psi_{\pm}(t\pm \tau)$
represent incoming and outgoing waves, respectively.
In terms of the $\eta$ field, we have
$$(\partial_{t}\pm \partial_{\tau})\eta =
\pm {1 \over \sqrt{\pi}}\,\,\psi_{\pm}(t\pm \tau)\eqno\eq$$
for $t \rightarrow \mp \infty$.

A nonlinear relation between incoming and outgoing fields
$\psi_+$ and $\psi_-$ can be established
by studying the time delay. Take the times at which a parametrized
point $\sigma$ is passed by the incoming and outgoing waves at a fixed
value of large $\tau$ be $t_1 (\rightarrow -\infty)$ and
$t_2 (\rightarrow \infty)$, respectively.
{}From \bhparasol\ we have then
$$\eqalign{\bigl(&M+a^2(\sigma)\bigr)^{1/4}\,e^{\sigma-t_1}=
M^{1/4}\, e^{\tau}\,\,,\cr
\big(&M+a^2(\sigma)\bigr)^{1/4}\,e^{t_2-\sigma}=M^{1/4}\,e^{\tau}\,\,.}
\eqno\eq$$
This implies
$$t_1+\tau=t_2 -\tau +{1\over 2}\log\Bigl(1+{a^2(\sigma)\over M}
\Bigr)\,\,,\eqno\eq$$
and hence
$$a(\sigma) = \psi_+(t_1 +\tau) = \psi_-(t_2 -\tau)\,\,.\eqno\eq$$
This then gives  functional scattering equations connecting
the incoming and outgoing waves
$$\psi_{\pm}(z)= \psi_{\mp}\Bigl(z \mp {1 \over 2}\log\,
(1+ {1 \over M}\,\psi^2_{\pm}(z))\Bigr)\,\,. \eqn\scatt$$
The result is  similar  in form to that of the usual $c=1$ model.
However, one notes a crucial difference that
Eq.~\scatt\ is even, i.e. it is invariant under
the change of sign of $\psi_{\pm}\rightarrow -\psi_{\pm}$.
This  ensures that the number of particles participating in
the scattering is even. All the odd point amplitudes do vanish in
the deformed model.

The explicit power series solution of \scatt\ is
$$\psi_{\pm}(z) = \sum_{p=0}^{\infty}{M^{-p} \over p!\,(2p+1)}\,\,
{\Gamma(1\pm {1 \over 2}\partial_z) \over \Gamma(1-p \pm
{1 \over 2}\partial_z)}\,\,\psi^{2p+1}_{\mp}(z)\,\,,\eqno\eq$$
which shows that the amplitudes are essentially polynomial with respect
to the momenta without any singularity.

The scattering equation \scatt\ can also be derived using directly the
exact states [\DeRo,\Da], as was done in sect.~5 for the $c=1$ model.
First, one recalls the symmetry structure of the collective
theory with Hamiltonian~\defham\ given in [\AJcmp]:
$$\eqalign{
\bigl[ O_{j_1,m_1}^{a_1},  O_{j_2,m_2}^{a_2} \bigr] =
&-4i(j_1 m_2 - m_1 j_2)\,O_{j_1+j_2-2, m_1+m_2}^{a_1+a_2+1}\cr
\noalign{\vskip 0.2cm}
&-4i(a_1 m_2 - m_1 a_2)\,O_{j_1+j_2, m_1+m_2}^{a_1+a_2-1}\,\,,}\eqno\eq$$
where
$$\eqalign{
O_{j,m}^{a} \equiv \int {dx\over 2\pi} \int_{\alpha_-}^{\alpha_+}
d\alpha &\Bigl( \alpha^2 - x^2 +{M\over x^2} \Bigr)^a
\Bigl( (\alpha + x)^2 +{M\over x^2} \Bigr)^{{j+m\over 2}}\cr
\noalign{\vskip 0.1cm}
&\Bigl( (\alpha - x)^2 +{M\over x^2} \Bigr)^{{j-m\over 2}}\,\,.}\eqn\op$$
The operators $T^{(-)}_{-ip}\, (T^{(+)}_{ip})$
which create exact tachyon in (out) states
are obtained by analytic continuation
$j\to \pm ip/2$ of some special operators \op:
$$\eqalign{
&O_{j,j}^{a=0} = \int  {dx\over 2\pi} \int_{\alpha_-}^{\alpha_+}
d\alpha
\Bigl[ (\alpha +x)^2 +{M\over x^2}\Bigr]^j \equiv T_{2j}^{(+)}\,\,, \cr
\noalign{\vskip 0.2truecm}
&O_{j,-j}^{a=0} = \int  {dx\over 2\pi} \int_{\alpha_-}^{\alpha_+}
d\alpha \Bigl[ (\alpha - x)^2 + {M\over x^2}\Bigr]^j
\equiv T_{2j}^{(-)}\,\,.}\eqno\eq$$
Eq.~\scatt\ then easily follows from an asymptotic expansion of
$$T^{(+)}_{ip,+} = -T^{(+)}_{ip,-}\,\,,$$
for large $\tau$, where $T^{(+)}_{ip,+}$ and $T^{(+)}_{ip,-}$
are defined by
$$T^{(+)}_{ip}=T^{(+)}_{ip,+}-T^{(+)}_{ip,-}\,\,.\eqno\eq$$

The scattering equation
can also be rewritten in terms of  energy-momentum tensor
$$T_{\pm\pm}(z) = {1 \over 2\pi}\, \psi^2_{\pm}(z)\,\,,\eqno\eq$$
as
$$\int dz\,\,e^{i\omega z} \,T_{\pm\pm}(z)
={M \over 2\pi}\int dz \,\,e^{i\omega z}
{1 \over 1 \pm {i\omega \over 2}}\,
\Bigl[\Bigl(1+{2\pi\over M}\,T_{\mp\mp}(z)\Bigr)^{1\pm{i\omega \over 2}}
-1\Bigr]\,\,.\eqno\eq$$
One can easily check that this defines a canonical
transformation by confirming that the Virasoro algebra
(at the level of Poisson bracket) is preserved by this transformation.
This relation for  the energy momentum tensor is very
similar to the one obtained recently by Verlinde and Verlinde
[\VV] for the $S$-matrix of the $N=24$ dilaton gravity.A slight difference is
that in the case of dilaton gravity one has the derivative of the energy
momentum tensor participating in the equation.

In conclusion, the framework presented above gives some initial
picture of a black hole in the matrix model. It contains some basic
requirements for a consistent formalism. In particular, the scaling
properties of the black hole mass deformation are in agreement with
the corresponding vertex operators (see also [\Eguchi]). The
particular singular matrix model studied has an interesting double
scaling limit with an $SL(2,\IR)$ algebraic structure. This clearly
is not enough to completely describe black hole and further
generalizations and studies are likely to lead to further interesting
results.

\noindent
{\bf Acknowledgement}

These notes were written while the author was visiting LPTHE, Paris 6,
Paris, France.
He is grateful to the members of the high energy group for
their hospitality.

\refout

\end

======================================================================== 2200
Return-Path: <@BROWNVM.BROWN.EDU:kresimir@PUHEP1.PRINCETON.EDU>
Received: from BROWNVM (NJE origin SMTP@BROWNVM) by BROWNVM.BROWN.EDU (LMail
          V1.1d/1.7f) with BSMTP id 1733 for <maryannr@BROWNVM>; Mon,
          20 Sep 1993 12:40:58 -0400
Received: from Princeton.EDU by BROWNVM.brown.edu (IBM VM SMTP V2R2) with TCP;
   Mon, 20 Sep 93 12:40:16 EDT
Received: from puhep1.Princeton.EDU by Princeton.EDU (5.65b/2.98/princeton)
        id AA05296; Mon, 20 Sep 93 12:42:02 -0400
Received: by puhep1.Princeton.EDU (5.52/1.113)
        id AA03431; Mon, 20 Sep 93 12:42:01 EDT
Date: Mon, 20 Sep 93 12:42:01 EDT
{}From: "Kresimir Demeterfi" <kresimir@puhep1.Princeton.EDU>
Message-Id: <9309201642.AA03431@puhep1.Princeton.EDU>
To: maryannr@brownvm.brown.edu

\input phyzzx
\nopagenumbers

%
\def\refout{\par\penalty-400\vskip\chapterskip
   \spacecheck\referenceminspace
   \ifreferenceopen \Closeout\referencewrite \referenceopenfalse \fi
   \noindent{\bf References\hfil}\vskip\headskip
   \input \jobname.refs }
\def\chapter#1{\par \penalty-300
   \chapterreset \noindent{\bf \chapterlabel.~~#1}
   \nobreak \penalty 30000 }

\def\IR{\relax{\rm I\kern-.18em R}}

\def\npb#1#2#3{{\it Nucl. Phys.} {\bf B#1} (#2) #3 }
\def\plb#1#2#3{{\it Phys. Lett.} {\bf B#1} (#2) #3 }
\def\prd#1#2#3{{\it Phys. Rev. } {\bf D#1} (#2) #3 }
\def\prl#1#2#3{{\it Phys. Rev. Lett.} {\bf #1} (#2) #3 }
\def\mpla#1#2#3{{\it Mod. Phys. Lett.} {\bf A#1} (#2) #3 }
\def\ijmpa#1#2#3{{\it Int. J. Mod. Phys.} {\bf A#1} (#2) #3 }

\def\cmp#1#2#3{{\it Commun. Math. Phys.} {\bf #1} (#2) #3 }

\def\ptp#1#2#3{{\it Prog. Theor. Phys.} {\bf #1} (#2) #3 }
\def\bb#1{{\tt hep-th/#1}}
\REF\Klebreview{I. R. Klebanov, {\it ``String theory in two
dimensions'',} in ``String Theory and Quantum Gravity'',
Proceedings of the Trieste Spring School 1991, eds. J. Harvey et al.,
(World Scientific, Singapore, 1992).}
\REF\Kutreview {D. Kutasov, {\it ``Some properties of (non) critical
strings'',} in ``String Theory and Quantum Gravity'',  Proceedings of
the Trieste Spring School 1991, eds. J. Harvey et al., (World Scientific,
Singapore, 1992).}
\REF\Polone {J. Polchinski, \npb {346}{1990}{253.}}
\REF\WBH {E. Witten, \prd {44} {1991} {314.}}
\REF\BaNe {I. Bars and B. Nemeschansky, \npb {348} {1991} {89.}}
\REF\RSS {M. Ro\v cek, K. Schoutens and A. Sevrin, \plb {265} {1991}
{303.}}
\REF\MaWa {G. Mandal, A. Sengupta and S. Wadia, \mpla {6}{1991}{1685.}}
\REF\El {S. Elizur, A. Forge and E. Rabinovici, \npb {359}{1991}{581.}}
\REF\Po{A.~M. Polyakov, \mpla {6}{1991}{635.}}
\REF\GoLi {M. Goulian and M. Li, \prl {66}{1990}{2051.}}
\REF\DiFrKu{P. DiFrancesco and D. Kutasov, \plb {261}{1991}{385;}
\npb {375}{1991}{119;} Y. Kitazawa, \plb {265}{1991}{262;}
Y. Tanii, \ptp {86}{1991}{547;}
V.~S. Dotsenko, \mpla {6}{1991}{3601.}}
\REF\Bab{O. Babelon, \plb {215}{1988}{523.}}
\REF\Ge {J.-L. Gervais, \cmp {130}{1990}{257;} \npb {391}{1993}{287.}}
\REF\JeSa {A. Jevicki and B. Sakita, \npb {165} {1980} {511.}}
\REF\DaJe {S.~R. Das and A. Jevicki, \mpla {5} {1990} {1639.}}
\REF\DJR {K. Demeterfi, A. Jevicki and J.~P. Rodrigues, \npb {362}{1991}
{173;}  \npb {365} {1991} {499.}}
\REF\Poltwo { J. Polchinski, \npb {362} {1991} {125.}}

\REF\AvJe {J. Avan and A. Jevicki, \plb {266}{1991}{35;}
\plb {272}{1991}{17.}}
\REF\MoSe {G. Moore and N. Seiberg, \ijmpa{7}{1992}{2601.}}
\REF\GKN {D.~J. Gross, I.~R. Klebanov and M. Newman,
\npb {350} {1991} {671.}}
\REF\LM {J. Lee and P.~F. Mende, \plb {312} {1993} {433.}}
\REF\winf {D. Minic, J. Polchinski and Z. Yang, \npb {369}{1992}{324;}}
\REF\WiGR {E. Witten, \npb {373}{1992}{187.}}
\REF\KlPo {I.~R. Klebanov and A.~M. Polyakov, \mpla {6}{1991}{3273.}}
\REF\JRvT {A. Jevicki, J.~P. Rodrigues and A. van Tonder,
\npb {404} {1993} {91.}}
\REF\Kl {I.~R. Klebanov, \mpla {7}{1992}{723.}}
\REF\ZW {E.Witten and B.Zwiebach, \npb {377}{1992}{55.}}
\REF\Ve {E. Verlinde, \npb {381}{1992}{141.}}
\REF\KlPa{I.~R. Klebanov and A. Pasquinucci, \npb {393}{1993}{261.}}
\REF\Ba {J.~L.~F. Barbon, \ijmpa {7}{1992}{7579;}
Y. Kazama and H. Nicolai, {\it ``On the exact operator formalism of
two-dimensional Liouville quantum gravity in Minkowski space-time,''}
DESY-93-043, \bb{9305023};
V.~S. Dotsenko, {\it ``Remarks on the physical states and the chiral
algebra of 2D gravity coupled to $c\le 1$ matter,''}
PAR-LPTHE-92-4, \bb{9201077}; \mpla {7}{1992}{2505.}}
\REF\GrKl {D.~J. Gross and I.~R. Klebanov, \npb {359} {1991} {3.}}
\REF\MoPl {G. Moore and R. Plesser, \prd {46} {1992} {1730.}}
\REF\MaSh {E. Martinec and S. Shatashvili, \npb {368}{1992}{338.}}
\REF\JeYo {A. Jevicki and T. Yoneya, {\it ``A deformed matrix model and
the black hole background in two-dimensional string theory'',}
NSF-ITP-93-67, BROWN-HEP-904, UT-KOMABA/93-10, \bb{9305109}.}
\REF\DDMW {S.~R. Das, \mpla {8} {1993} {69;}
A. Dhar, G. Mandal and S. Wadia, \mpla {7} {1992} {370.}}
\REF\MuVa {S. Mukhi and C. Vafa, {\it ``Two-dimensional black hole
as a topological coset model of $c=1$ string theory,''}
HUTP-93/A002, TIRF/TH/93-01, \bb{9301083}.}
\REF\DVV {R. Dijkgraaf, H. Verlinde and E. Verlinde,
\npb {371} {1992} {269.}}
\REF\new{M. Bershadsky and D. Kutasov, \plb {266}{1991}{345;}
T. Eguchi, H. Kanno and S.-K. Yang, \plb {298}{1993}{73;}
H. Ishikawa and M. Kato, \plb {302}{1993}{209.}}
\REF\AJcmp {J. Avan and A. Jevicki, \cmp {150}{1992}{149.}}
\REF\DeRo {K. Demeterfi and J.~P. Rodrigues, {\it ``States and
quantum effects in the collective field theory of a deformed matrix
model'',} PUPT-1407, CNLS-93-06, \bb {9306141};
K. Demeterfi, I.~R. Klebanov and J.~P. Rodrigues,
{\it The exact $S$-matrix of the deformed $c=1$ matrix model,''}
PUPT-1416, CNLS-93-09, \bb {9308036}.}
\REF\Da{U. Danielsson, {\it ``A matrix-model black hole,''}
CERN-TH.6916/93, \bb{9306063}.}
\REF\VV {E. Verlinde and H. Verlinde, {\it ``A unitary $S$-matrix
and 2D black hole formation and evaporation'',} IASSNS-HEP-93/18,
PUPT-1380, \bb {9302022}.}
\REF\Eguchi{T. Eguchi, {\it ``$c=1$ Liouville theory perturbed by the
black-hole mass operator,''} UT 650, \bb{9307185}.}

\singlespace
\hsize=6.0in
\vsize=8.5in
\voffset=0.0in
\hoffset=0.0in
\overfullrule=0pt

\line{}
\line{\hfill HET-918}
\line{\hfill TA-502}
\vskip .75in
\centerline{{\bf DEVELOPMENTS IN 2D STRING THEORY}}
\vskip .40in
\centerline{ANTAL JEVICKI}
\smallskip
\centerline{{\it Physics Department, Brown University}}
\centerline{{\it Providence, Rhode Island 02912, USA}}
\vskip .25in
\centerline{(Lectures presented at the Spring School }
\centerline{in String Theory, Trieste, Italy, April, 1993)}

\vskip 0.70in

{\chapter{Introduction}}
\medskip

Recent years have witnessed a remarkable progress in 2d  string theory
and quantum gravity. Beginning with matrix models one found a new and
computationally powerful description of the theory, free of mathematical
complexities. The relevance of these models to
string theory comes through a $1/N$ expansion where $1/N$ plays the role of a
bare string coupling constant $g_{\rm st}^0 = 1/N$.  This classifies
Feynman diagrams according to their topology; for a fixed topology the
sum of all graphs in a dual picture becomes a sum of triangulated
surfaces.  The continuum theory is then approached by sending
the value of lattice spacing to zero.

This heuristic picture was completely carried out in one dimension giving
an exactly solvable theory of two-dimensional strings
(for an earlier review see [\Klebreview]).
It lead first to a series of explicit results including
the computation of free energy and correlation functions at any order in
the loop expansion. The new formulation also offered a framework for
non-perturbative investigations. It provided  a new fundamental insight into
the origin of metric fluctuations and the physical nature of the Liouville
mode. Through a critical scaling limit a two-dimensional theory is
generated where the logarithmic scaling violation is seen to be the origin of
the extra dimension.

Most of the interesting features of 2d strings were
clearly exhibited in the field-theoretic description achieved in
terms of  collective field theory. Starting from matrix models one
builds a field theory describing the dynamics of observable (Wilson) loop
variables. The collective Hamiltonian describes the processes of
joining and splitting of loops, giving
A cubic interaction and a linear (tadpole) term were shown to
successfully produce all tree and loop diagrams. The theory is naturally
integrable and exactly solvable. Its
integrable nature  leads to understanding of a
$w_\infty$ algebra as a space-time symmetry of the theory. This algebra acts
in a nonlinear way on the basic collective field representing the tachyon.
It is interpreted as a spectrum-generating algebra allowing to build
an infinite sequence of discrete imaginary energy states which turn
out to be remnants of higher string modes in two dimensions.
The presence and interplay of discrete modes with the scalar tachyon are
particularly interesting. The $w_\infty $ symmetry is seen to serve as
an organizational principle specifying the dynamics.

Two-dimensional physics is made even richer by the existence of other
nontrivial backgrounds. Most interesting is the black hole type classical
solution described by an exact $SL(2,\IR)/U(1)$ sigma model. Its quantum
mechanical interpretation is of major interest and was the object of
various recent studies.

Even though there is a wealth of results coming from detailed studies of
matrix models and conformal field theories a full understanding of the theory
and its dynamics is still not available. In particular, a clear correspondence
between the two fundamentally different methods is lacking. One has an
(excellent) comparison of results and a pattern of similarities and analogies
hinting at a more unified  framework. Prospects for such a framework are
particularly exciting since this would eventually represent a
new formulation of string field theory.

A need for such a general framework is most clear already when
addressing the question of the black hole. In general one would like
to command sufficient insight to be able to go from one solution to
another. This, at present, is also one of the fundamental
challenges of string theory.

In this series of lectures we describe the progress already achieved.
The emphasis is on a unified understanding of the subject.
We will try to bridge the two major approaches: the matrix model
and conformal field theory, as much as possible describing analogies
and similarities that one has between them.  In this process a
dictionary  emerges; it is most visible in the discussion
of the infinite $w_\infty $ symmetry and the associated Ward identities.
The question of incorporating the black hole background is then addressed
and some preliminary results in this direction are described.

The selection of topics covered is as follows: In sect.~2 we give a
summary of basic two-dimensional string theory (for a more detailed
review see [\Kutreview]). In sect.~3 we describe the matrix model
and a transition to field theory. We discuss the integrability of
the theory and the construction of exact states
and their string interpretation. In sect.~4 the corresponding $w_\infty$
symmetry  is described.  A detailed comparison of Ward identities and a
description of the agreement between matrix
model and conformal field constructions is given. Sect.~5 contains
the discussion of the $S$-matrix of the theory. The latter is described by
an exact generating function, connection of which to matrix model
harmonic oscillator states we emphasize. In sect.~6 we discuss the
black hole background.

\bigskip
\chapter {String Theory in Two Dimensions}
\medskip

The conceptually simplest way to discuss the dynamics of strings
is through a $\beta$-function approach which provides
effective equations for low-lying fields . In the case of a closed
string in two dimensions these are the $m^2 = 0$ scalar $T(X^{\mu}$) (the
would-be tachyon), the graviton $G_{\mu\nu}(X)$ and the dilaton $D(X)$.
The leading $\beta$-function Lagrangian reads:
$$S_{\rm eff} = {1\over 2\pi} \int d^2 X \sqrt{G} \, e^{-2D(X)}
\Bigl\{ {1\over 2} \left[ \nabla_{\mu} T \nabla^{\mu} T + 2T^2 - V\right] +
R + 4 \nabla D\cdot\nabla D + \ldots\Bigr\}\,\,. \eqno\eq$$
The tachyon potential $V(T)$ is not so well known and neither are the
couplings to possibly higher--spin fields.  But this effective Lagrangian
exhibits several simple solutions which can serve as classical
configurations of two-dimensional string theory.

Denoting $X^{\mu} \equiv (X^0 = t , X^1= \varphi)$ one
has the {\it linear dilaton vacuum} solution
$$\eqalign{&T(X)  = 0\,\,,\cr
\noalign{\vskip 0.1cm}
&G_{\mu\nu}(X)  = \eta_{\mu\nu}\,\,,\cr
\noalign{\vskip 0.1cm}
&D(X) = - \sqrt{2}\, \varphi\,\,. }\eqno\eq$$
The scalar (tachyon) effective Lagrangian in this linear dilaton background
reads
$$S_{\rm eff} (T)={1\over 2}\int d^2 X\,e^{2\sqrt{2}\varphi}\,\,\Bigl\{
{1\over 2} \,T \left( - \partial_t^2 + \partial_{\varphi}^2 + 2\sqrt{2}
\partial_{\varphi} + 2 \right) T-V\Bigr\}\,\,.$$
Rescaling the scalar fields
$$e^{\sqrt{2}\varphi}\,\,T (t,\varphi )=\tilde{T} (t,\varphi)\eqno\eq$$
yields a massless theory
$$S = {1\over 2} \int dt d\varphi \Bigl\{ {1\over 2}\,\tilde{T}
\left( - \partial_t^2 + \partial_{\varphi}^2 \right) \tilde{T} -
{e^{-\sqrt{2}\varphi}\over 3!}\,\tilde{T}^3 + \ldots
\Bigr\}\,\,,\eqno\eq$$
with a spatially dependent string coupling constant
$$g_{\rm st} (\varphi) = e^{-\sqrt{2} \varphi}\eqno\eq$$
(we have taken for simplicity a cubic interaction).

This coupling grows and becomes infinite at $\varphi \rightarrow -
\infty$. This is usually taken as a signal that the linear dilaton
vacuum should be modified (at least in the region $\varphi \rightarrow
- \infty$).  Indeed the linearized static tachyon equation
$$\left( \partial_{\varphi}^2 + 2 \sqrt{2} \partial_{\varphi}
+ 2\right) T_0 (\varphi ) = 0\eqno\eq$$
already has two linearly independent solutions $T_0 (\varphi)=
e^{- \sqrt{2} \varphi}, \varphi e^{-\sqrt{2}\varphi}$.
This would imply that the correct vacuum is given by a tachyon
condensate [\Polone].  An (incomplete) analysis indicates that this
vacuum is then described by a $c=1$ conformal field theory coupled
to a Liouville field:
$${\cal L}={1\over 8\pi}\int d^2z\,\Bigl(\partial X \bar{\partial} X
+ \partial \varphi \bar{\partial} \varphi - 2 \sqrt{2} \varphi
(z,\bar{z}) R^{(2)}+\mu \, e^{-\sqrt{2} \varphi (z,\bar z)}\Bigl)\,\,.
\eqno\eq$$
Here the central charge $c_{X} = 1$ refers to the (matter) coordinate
$X(z,\bar{z})$ while the Liouville field with $Q = 2\sqrt{2}$ carries a
central charge $c_{\varphi} = 1 + 3Q^2 = 25$ leading to the required total of
$c=c_{X}+c_{\varphi}=26$. It is very interesting that in two dimensions
one  has another conformally invariant background, the WZW
$SL(2,\IR)/U(1)$ sigma model representing a black hole
(BH) [\WBH--\El].
Its physical properties are of major
interest as is the general question of describing different string theory
backgrounds in a single field-theoretic framework.

The presence of the cosmological term in the Liouville  theory (and of
the mass term in the black hole conformal field theory) leads to
computational difficulties when evaluating the correlation
functions (these actually become quite untractable
for the BH case).  It is a remarkable fact that the matrix model
formulation succeeds in handling  the first problem with ease and has some
promise for addressing the second as well.

The spectrum of states is usually obtained by neglecting the nonlinear
terms $\mu = 0$ (or $M=0$ for the black hole) in which case one has a
free field representation for the Virasoro generators.  In the above limit
the spectra of two theories are the same.  They consist
of a massless tachyon and an infinite sequence of discrete states.

We begin with the zero mode or tachyon states:
$$\eqalign{ & (L_0 - 1) \,\, V_{k,\beta} = 0 \,\,,\cr
\noalign {\vskip 0.1cm}
& L_0 = {1\over 2} \Bigl(  {\partial^2\over\partial X^2} +
{\partial^2\over\partial\varphi^2} + Q
{\partial\over\partial\varphi}\Bigr)\,\,,}\eqno\eq$$
with two branches of solutions
$$V_{\pm}= e^{ikX + \beta_{\pm} \varphi}\,\,,\qquad\quad
\beta_{\pm} = - \sqrt{2} \pm \vert k \vert\,\,,\eqno\eq$$
following from the on-shell condition
$$k^2 - \beta (Q+\beta ) = 0\,\,.\eqno\eq$$
Here we have taken an Euclidean (space) signature for $X$ and
$\varphi$ which is a convention in conformal field theory discussions.
One can take $X$ to be the space variable and $\varphi$ to be the
(Euclidean) time variable.
It will be more physical, and from the matrix model viewpoint more
natural, to treat $\varphi$ as a space coordinate and continue $X$ to
Minkowski time:
$$X \rightarrow - it\,\,,\qquad\quad
k \rightarrow i p\,\,.\eqno\eq$$

In the context of full Liouville
theory, the second branch with $\beta_- = - \sqrt{2} -\vert k\vert$ has a
questionable meaning since the wave functions grow at $\varphi
\rightarrow - \infty$ which is the location of the infinitely high
Liouville wall $\mu e^{-\sqrt{2} \varphi}$.  These vertex operators
are termed ``wrongly" dressed.  Operators with positive Liouville
dressing have a clear meaning.  Depending on the sign of the momentum,
$\pm =$ sign $k$, these are either right- or left-moving waves.  It is
sensible to use them to compute scattering processes and denote them
as
$$T_k^{\pm} = e^{ikX+(-\sqrt{2} \pm k)\varphi}\,,\qquad
\pm = {\rm sign} \, \, k\,\,. \eqno\eq$$
The Minkowskian continuation is $k = \pm ip$ and
$$\eqalign { T_p^+ &=e^{i p(t + \varphi )} e^{-\sqrt{2} \varphi}\,\,,\cr
\noalign {\vskip 0.1cm}
T_p^- &= e^{-ip(t-\varphi )} e^{-\sqrt{2}\varphi}\,\,, }\eqno\eq$$
for $p>0$ describe left-  and right-moving waves, respectively.

In addition one has an infinite sequence of nontrivial discrete
states specified by discrete (imaginary) values of energy and Liouville
momenta [\Po]:
$$ip_{\varphi}= -\sqrt{2} (1-j)\,,\qquad
ip = \sqrt{2} m \,\,,\eqno\eq$$
with $j = 0, {1\over 2} , 1, \ldots$ and $m = -j, \ldots , j$.
Clearly the states with $m = j$ and $m=-j$ are just special tachyon
states.  The simplest way then to reach the other states is to use the
$SU(2)$ generators as raising and lowering operators on the $m=\pm j$
tachyon states.  The $SU(2)$ generators are given by
$$\eqalign{ t_+ & = e^{i\sqrt{2} X (z)}\,\,,\cr
t_- & = e^{-i\sqrt{2} X(z)} \,\,,\cr
t_3 & = i\sqrt{2} \, \partial X (z)\,\,.}\eqno\eq$$
Denoting now the highest weight state as
$$W_{jj}^{(+)}= e^{i\sqrt{2}\,jX (z)}\,e^{-\sqrt{2}\,(1-j) \varphi
(z)}\eqno\eq$$
one gets the vertex operator for  general discrete states
$$W_{jm}^{(+)} = \left( \oint d\omega e^{-i\sqrt{2} X (\omega ) }
\right)^{j-m} W_{jj}^{(+)}\,\,,\quad -j\leq m\leq + j\,\,.
\eqno\eq$$
These can also be found in the Fock  space where they solve the Virasoro
conditions of the $c=1$ theory
$$\eqalign{ &(L_0 - 1) \vert jm \rangle =0\,\,,\cr
\noalign {\vskip 0.1cm}
&L_n \vert jm \rangle =0\,\,,\cr
&\vert jm \rangle = \int dz \, W_{jm}(z)\vert 0\rangle\,\,.}\eqno\eq$$
One also has operators with the opposite (negative) Liouville dressing
$$W_{jm}=V_{jm}(X) \,e^{-\sqrt{2} (1+j) \varphi (z)}\,\,,\eqno\eq$$
whose physical meaning is again questionable in the full Liouville
theory.  These states, however, turn out to play an important role as black
hole mass perturbations.

Evaluation of correlation functions in the continuum approach is
rather nontrivial and often relies on a number of educated guesses
involving various analytic continuations.  The problem lies
in the nontrivial Liouville potential term.  By separating and
integrating out the zero mode $\varphi (z,\bar{z} ) = \varphi_0 +
\tilde{\varphi}$ one finds, through a functional integral formulation,
the representation
$$ \langle\,\prod_{i=1}^{N} \, T_i \,\rangle =
\left( {\mu\over\pi}\right)^s \Gamma (-s)\,\bigl\langle
\Bigl( \prod_i T_i \Bigr)
\Bigl( \int d^2 z \,e^{\sqrt{2}\tilde{\varphi}}\Bigr)^s\,
\bigr\rangle_{\mu=0}\,\,.\eqno\eq$$
Here the $\Gamma$-function is a result of $\varphi_0$ integration
$$\int d\varphi_0 \, e^{Q\varphi_{0}} \Bigl( \prod_i
e^{\beta_{i}\varphi_{0}} \Bigr)\,\, {\rm exp}\,
\Bigl(-{\mu\over\hbar}\,e^{- \sqrt{2} \varphi_0} \int e^{-\sqrt{2}
\tilde{\varphi}}\Bigr)\,\,,\eqno\eq$$
and
$$-\sqrt{2} s \equiv \sum_i \, \beta_i + Q\,\,.\eqno\eq$$
The remaining correlation function is at $\mu=0$ but has a nontrivial
power of the Liouville term given by $s$. It can be evaluated only for
$s$ = integer with the full result  to be obtained by some analytic
continuation [\GoLi]. The $s=0$ amplitude is termed a ``bulk" amplitude
since the condition $s=0$ coincides with a Liouville momentum conservation.
Nontrivial computation involving major cancellation between matter and
Liouville contributions gives the simple result
$$T(k_1, k_2 , \ldots ,k_N ) = (N-3)!\,\prod_{i=1}^{N} \,\,
{\Gamma(-\sqrt{2}\,\vert k_i\vert )\over\Gamma(\sqrt{2}\,\vert k_i\vert )}
\,\,. \eqno\eq$$
At $s=0$ one has both energy and momentum conservation laws:
$$\sum_{i=1}^{N} k_i = 0\,,\qquad\quad
\sum_{i=1}^N \vert k_i \vert = -2 \sqrt{2}\,\,.\eqno\eq$$
Choosing $k_1 , k_2, \ldots ,k_{N-1} > 0$, one finds that the
$N$-th particle momentum is totally determined
$$k_N = - {N-2\over \sqrt{2}}\,\,, \eqno\eq$$
implying that the $N$-th leg factor diverges
$${\Gamma (-N+2)\over \Gamma (N-2)} \sim {1\over 0}\,\,.\eqno\eq$$
This is  in agreement with the previous $\Gamma (0) \sim {1\over 0}$
divergence.  This divergence is related to the length of the Liouville
line $\int d\varphi_0$ and is only fully understood in the matrix
description.

The final result for these $s=0$ bulk amplitudes is that they consist
of purely external leg factors
$\Delta = \Gamma (-\sqrt{2}\,\vert k\vert )/
\Gamma (\sqrt{2}\,\vert k \vert )$ and that only
$T_{++\ldots +-}$ and $T_{-\ldots - +}$ amplitudes contain a diverging
factor playing the role  of the Liouville volume.  These bulk
amplitudes can then lead to the full $s\not= 0$ amplitudes by an
appropriate continuation  [\DiFrKu]. This had to await  developments
given by the matrix model formalism. Concerning the full treatment
of Liouville theory one has the interesting algebraic approach
of [\Bab,\Ge].

\bigskip
\chapter{Matrix Model and Field Theory}
\medskip
The manner in which a simple matrix dynamics gives rise to nonlinear
two-dimensional string theory is rather interesting and is related
to collective phenomena.  The major tool employed is a field-theoretic
representation given by collective field theory [\JeSa]. We shall now
give the main features of the field-theoretic approach and describe
its significance to string theory [\DaJe]. The field theory
turns out to correctly describe interactions of strings, it therefore
represents a
very simple string field theory. It provides some major insight into
the physics of noncritical strings allowing the computation of
scattering processes [\DJR] and giving the exact $S$-matrix [\Poltwo].
New higher space-time symmetries are seen to emerge [\AvJe] with
further implications on  general string field theory being likely.

The simple model that one considers is a Hermitian matrix
$M^\dagger (t) = M(t)$ in one time dimension ($X^0 = t$) with a Lagrangian
$$L = {1\over 2} \Tr \left(\dot{M}^2 - u(M)\right)\,\,.\eqno\eq$$
It has an associated $U(N)$ conserved (matrix) charge $J = i [M,\dot{M}]$.
Restricting oneself to the singlet subspace $\hat{J}\vert\,\,\,\rangle=0$
turns this model into a gauge theory. The matrix can be diagonalized:
$M(t) \to {\rm diag}\,\, (\lambda_i (t) )$ with the eigenvalues
describing a system of nonrelativistic fermions.

The collective variables of the model are the gauge invariant (Wilson)
loop operators
$$\phi_k (t) = \Tr \left( e^{ikM(t)}\right) = \sum_{i=1}^N \,
e^{ik\lambda_{i}(t)}\,\,,\eqno\eq$$
which, after a Fourier transform
$$\phi (x,t) = \int {dk\over 2\pi} \, e^{-ikx} \, \phi_k (t) =
\sum_{i=1}^N \, \delta \left( x-\lambda_i (t)\right)\,\,,\eqno\eq$$
have a physical interpretation of a density field (of
fermions).  Introduction of a conjugate field $\Pi (x,t)$ with
Poisson brackets
$$\left\{ \phi (x), \Pi (y)\right\} = \delta (x-y)\eqno\eq$$
gives a canonical phase space.

The dynamics of this field theory is directly induced from the simple
dynamics of the matrix model variables
$M(t)$ and  $P(t) = \dot{M} (t)$. It is found to be given by the
Hamiltonian
$$H_{\rm coll}=\int dx\,\,\Bigl\{ {1\over 2}\,\Pi_{,x}\,\phi\,\Pi_{,x}+
{\pi^2\over6} \phi^3 + u (x) \phi \Bigr\}\,\,,\eqno\eq$$
where the first two terms come from the kinetic term of the matrix model
$\Tr (P^2/2 )$ while the last term represents the potential
(the latter can be easily seen through the density representation):
$${1\over 2}\,\Tr P^2 \rightarrow {1\over 2}\,\Pi_{,x}\,\phi\,\Pi_{,x}
+ {\pi^2\over 6} \phi^3\,\,, \qquad
\Tr u(M) \rightarrow \int dx \, u (x) \phi(x,t)\,\,.\eqno\eq$$
The Hamiltonian constructed in this way consists of a cubic (interaction)
term and a linear (tadpole) term.  In terms of basic loops (and strings)
the cubic interaction has the effect of splitting and joining  strings.
The linear tadpole term represents a process of string annihilation
into the vacuum. It contains the classical background potential. This
potential is tuned to get a particular string theory background;
the  noncritical $c=1$ string theory is obtained for example with
an inverted oscillator potential.  Two relevant facts are immediate
in this transition to collective
field theory:

(1)  The field $\phi (x,t)$ is two-dimensional with
the extra spatial dimension $x$ being related to the eigenvalue space
$\lambda_i$.  The appearance of an extra dimension is the first
sign that this theory will be describing $D=2$ strings.

(2)  The equations of motion for the induced fields are nonlinear while
the matrix equations (in particular for the physically relevant
oscillator potential $u(M)= -M^2/2$) are  linear
$$\ddot{M} (t) - M(t) = 0\,\,.\eqno\eq$$
Through a nonlinear transformation, $\phi (x,t) = \Tr \delta
(x-M(t))$ the matrix model provides an exact solution to the
nonlinear field theory.
 The feature of integrability and the collective transformation
itself is very similar to the well known inverse
scattering transformation in integrable field theories.  Actually
introduction of left- and right-moving chiral components
$\alpha_{\pm} (x,t) = \Pi_{,x} \pm \pi\phi (x,t)$ with Poisson
brackets
$$\left\{ \alpha_{\pm} (x) , \alpha_{\pm} (y) \right\} = \pm 2\pi
\partial_x \delta (x-y)\eqno\eq$$
brings the Hamiltonian to the form
$$H_{{\rm coll}} = {1\over 2} \int \, {dx\over 2\pi}\,\Bigl\{
{1\over 3} \left( \alpha_+^3 - \alpha_-^3\right) - \left( x^2 - \mu
\right) \left( \alpha_+ - \alpha_- \right) \Bigr\}\,\,.\eqno\eq$$
The equations of motion
$$\partial_t \alpha_{\pm} + \alpha_{\pm} \partial_x \alpha_{\pm} - x =
0\eqno\eq$$
are then seen to be two copies of a large-wavelength KdV type equation
with an external $(-x^2 )$ potential.
Collective field theory shares with some other field theories in
two dimensions the feature of exact solvability.
One can indeed write down an infinite sequence of conserved commuting
quantities (Hamiltonians).  They are simply given by [\AvJe]:
$$H_n = {1\over 2\pi} \int dx \int_{\alpha_- (x,t)}^{\alpha_+ (x,t)}
d\alpha \,\, \left( \alpha^2 - x^2 \right)^n\,\,,\eqno\eq$$
and are related to the matrix model quantities $\Tr (P^2 - M^2 )^n$.
In fact one has a simple set of transition rules between the two
descriptions. These are useful when constructing
exact eigenstates and symmetry generators of the theory.

One  easily checks that the Poisson brackets vanish
$$\left\{ H_n , H_m \right\} = 0\,\,, \eqno\eq$$
and that these charges are formally conserved
$${d\over dt}\, H_n = \int dx\,\partial_x \left( \alpha^2 - x^2 \right)
\left( \alpha^2 - x^2 \right)^n = 0\,\,. \eqno\eq$$
This naturally is correct only up to surface terms which are present and
will allow particle production.

Before continuing with the integrability features of
the theory one can study perturbation theory and small fluctuations to
clarify at this simple level the connection to string theory.
The static (ground state) equation reads
$${1\over 2} \left( \pi \phi_0 (x) \right) ^2 + u (x) =
\mu_{F}\,\,,\eqno\eq$$
where $\mu_{F}$ is the Fermi energy introduced as a linear term in the
Hamiltonian
$$\Delta H = - \int dx\,\,\mu_{F}\, \phi (x,t)\,\,.\eqno\eq$$
Denoting $\pi\phi_0 = p_0$ we see this as being simply the equation
specifying the Fermi surface:  ${1\over 2} p_0^2 + u(x)=\mu_{F}$ with
the solution
$$\pi \phi_0 = p_0 (x) = \sqrt{2(\mu_{F} - u(x) )}\,\,.\eqno\eq$$
Introducing small fluctuations with a shift $\phi (x,t) = \phi_0 (x) +
{1\over\sqrt{{\pi}}}
\partial_x \eta (x,t)$ the Hamiltonian becomes
$$H = \int dx\,\, \Bigl\{(\pi\phi_0) \Bigl( {1\over 2}\,
\Pi^2 + {1\over 2}\, \eta_{,x}^2\Bigr) + {\pi^2\over 6}\,
(\eta_{,x})^3 {\pi\over 2} \Pi^2 \eta_{,x}\Bigr\}\,\,, \eqno\eq$$
with the quadratic term (in the Lagrangian form):
$$L_2 = \int dt \int dx\,\, {1\over 2}\,\Bigl( {\dot\eta^{2}\over
\pi\phi_{0} (x) } \, - \left(\pi \phi_0 \right) \eta_{,x}^2\Bigr)\,\,.
\eqno\eq$$
This is a free massless particle in an external gravitational
background
$$g_{\mu \nu}^0=\Bigl( 1/\pi \phi_0 (x)\,,\,\,\pi\phi_0(x)\,\Bigr)
\eqno\eq$$
specified by our potential $u(x)$.  However, this
metric is removable by a coordinate transformation.
In terms of the time-of-flight coordinate
$$\tau = \int^x \, {dx\over\pi\phi_0} \qquad {\rm or} \qquad
{dx(\tau)\over d\tau} = p_0 \eqno\eq$$
one has
$$H = \int d\tau\,\,\Bigl\{ {1\over 2}\,\Bigl(\Pi^2 + (\partial_{\tau}
\eta )^2 \Bigr) + {1\over 6 p_0^2}\,\Bigl(
(\partial_{\tau} \eta ) ^3 +
3\Pi^2(\partial_\tau \eta)\Bigr) \Bigr\} \,\,,\eqno\eq$$
which describes a massless theory with a spatially dependent coupling
constant
$$g_{\rm st} (\tau) = {1\over p^2_0(\tau )}\,\,.\eqn\stringcc$$
The continuum $c=1$ string theory is approached for a special choice
of the potential $v(x)= -x^2/2$.  In this case one has a
critical theory near $\mu_F = - \mu \rightarrow 0$.
For the oscillator we have
$$ \eqalign { x(\tau) & = \sqrt{2\mu}\, \cosh\,\tau\,\,,\cr
\noalign {\vskip 0.1cm}
p_0 (\tau) & = \sqrt{2\mu} \, \sinh \, \tau\,\,.}\eqno\eq$$
The length of the (physical) $\tau$-space diverges at the turning
point $x_0 = \sqrt{2\mu}$.  The string coupling constant \stringcc\
is now
$$g_{\rm st}(\tau)={1\over 2\mu \, \sinh^2 \tau}\,\,.\eqno\eq$$
It depends on the Fermi level as $g_{\rm st} \sim 1/\mu$.
This is in parallel with the dependence of the string coupling on the
cosmological constant of the $c=1$ string theory. We also see that
asymptotically $ g_{\rm st} \sim {1\over \mu}\,e^{-2\tau}$
as  $\tau \rightarrow + \infty$.
Comparing it to the expected behavior in $c=1$ string
theory  $g_{\rm st} \sim {1\over\mu}\,e^{-\sqrt{2} \varphi}$ one has
the (asymptotic) identification
$$\tau \leftrightarrow {1\over \sqrt{2}}\,\varphi\,,\qquad\quad
t_M \leftrightarrow {1\over\sqrt{2}}\,t_{c=1} \,\,.\eqno\eq$$
One can  now identify $\eta (\tau,t)$ with the tachyon
field $T(\varphi, t)$.  Remembering that $e^{\sqrt{2}\varphi}\,
T(\varphi, t)$ was the field satisfying the massless Klein-Gordon
equation, one  also has the identification of the
energy-momenta:
$$ip_{\tau}\leftrightarrow 2 + i\sqrt{2} \, p_\varphi\,,\qquad\quad
i\epsilon \leftrightarrow i \sqrt{2}\,p \,\,,\eqno\eq$$
where $\epsilon$ is the energy in the matrix model picture.

The above identification of the collective field
$\eta(\tau,t)$ was only done asymptotically when the
$\mu e^{- \sqrt{2}\varphi}$ term in the Liouville equation is ignored.
A much more precise identification can be performed with the cosmological
term also present.

In the matrix model the time-of-flight coordinate is
introduced to bring the quadratic mass operator of the
collective field into a Klein-Gordon form:
$$\Bigl[\,\partial_t^2 -\sqrt{x^2-2\mu}\,\,\partial_x\,
\sqrt{x^2-2\mu}\,\,\partial_x\Bigr]\,\eta =
(\partial_t^2 -\partial_\tau^2)\,\eta(t,\tau)\,\,, \eqno\eq$$
with $x=\sqrt{2\mu}\,\cosh\tau$. If we use a basis
conjugate to $x$: $p=-i\,(\partial/\partial x)$ the spatial
operator reads
$$\omega^2 =p^2x^2-2\mu p^2\,\,,\eqno\eq$$
and after a change of variables
$p=\sqrt{2}\,e^{-\varphi/\sqrt{2}}$ this gives the Liouville
operator
$$\hat\omega^2 =-{1\over 2}\,(\partial_\varphi)^2
-4\mu\,e^{-\varphi/\sqrt{2}}\equiv {\cal H}_L\,\,.\eqno\eq$$
We see that the Liouville coordinate is to be identified more
precisely [\MoSe] with the variable $p$ conjugate to the matrix
eigenvalue $\lambda$. The conjugate basis is not unnatural
in collective theory, it is associated with the
(Wilson) loop operator itself
$$W(\ell,t)= \Tr\,(e^{-\ell M}) =
\int dx\,e^{-\ell x}\,\phi(x,t) \,\,.\eqno\eq$$
which at the linearized level
$$W(\ell,t)=
\int_0^\infty  d\tau\,e^{-\sqrt{2\mu}\,\ell \cosh\tau}\,
\partial_\tau \eta\,\,, \eqno\eq$$
is  seen to obey
$$(\partial_t^2 -\hat\omega^2)\,\hat W(\ell,t)=0\,\,,\eqno\eq$$
with
$$\hat\omega^2=\partial_\tau^2 \quad
\Rightarrow\quad  -(\ell\,\partial_\ell)^2 +
2\mu\,\ell^2\,\,.\eqno\eq$$
After a change $\ell=2e^{-\varphi/\sqrt{2}}$
one has the Liouville operator ${\cal H}_L$.
In this conjugate momentum basis the connection to Liouville
theory is therefore manifest. One could obviously write all
equations in this representation  but formulae are
much simpler in terms of time-of-flight coordinate $\tau$.
The (Wilson) loop field and its natural connection to the
Liouville picture will be relevant for defining the string
theory $S$-matrix.

To further clarify the identification of the Liouville mode
let us write the transformation between the matrix
eigenvalue $x=\lambda$ and the time-of-flight coordinate
$\tau$ as a point canonical transformation:
$$\eqalign{x&=\sqrt{2\mu}\,\cosh\tau\,\,,\cr
\noalign{\vskip 0.2cm}
p&={1\over \sqrt{2\mu}\,\sinh\tau}\,\,p_\tau\,\,,}\eqno\eq$$
where $p$ and $p_\tau$ are the conjugates:
$\{x,p\}=\{\tau,p_\tau\} =1$.
Introducing $p=\sqrt{2}\,e^{-\varphi/\sqrt{2}}$
we have
$$\eqalign{&p_\varphi=\sqrt{2\mu}\, e^{-\varphi/\sqrt{2}}\,
\cosh\tau\,\,,\cr
\noalign{\vskip 0.2cm}
&p_\tau=\sqrt{2}\,\sqrt{2\mu}\, e^{-\varphi/\sqrt{2}}\,
\sinh\tau \,\,,}\eqno\eq$$
as a canonical transformation between the Liouville and
time-of-flight coordinates. The property of this
transformation is that
$${1\over 2}\,\omega^2 ={1\over 2}\,p_\tau^2 =
p_\varphi^2 -2\mu\, e^{-\varphi/\sqrt{2}}\,\,.$$
Now in Liouville theory one also usually deals
with two alternate descriptions and two different fields:
the original Liouville field $\varphi(z,\bar z)$
and a free field $\psi(z,\bar z)$.
They are related by a canonical (B\"acklund) transformation
$$\eqalign{\dot\varphi &= \psi'+\sqrt{2\mu}\,
e^{-\varphi/\sqrt{2}}\, \cosh (\psi/\sqrt{2})\,\,,\cr
\noalign{\vskip 0.2cm}
\dot\psi &=\varphi'+\sqrt{2\mu}\,e^{-\varphi/\sqrt{2}}\,
\sinh (\psi/\sqrt{2})\,\,,}\eqno\eq$$
where the two derivatives correspond to the two-dimensional space
$z=\sigma+i\xi$. The above transformation relates the
Liouville action to the action of a free field
$\psi(z,\bar z)$. Clearly for the center of mass mode
$(\varphi' = \psi' =0)$ one sees
$$\eqalign{\Pi_\varphi &= \sqrt{2\mu}\,
e^{-\varphi/\sqrt{2}}\, \cosh (\psi/\sqrt{2})\,\,,\cr
\noalign{\vskip 0.2cm}
\Pi_\psi &=\sqrt{2\mu}\,e^{-\varphi/\sqrt{2}}\,
\sinh (\psi/\sqrt{2})\,\,. }\eqno\eq$$

The transformation between $\varphi$ and $\psi$ is
identical to the one in the matrix model. We have then the
fact that the time-of-flight coordinate $\tau$ is to be
identified with the free field zero mode $\psi_0:\,\psi_0=
\sqrt{2}\,\tau$. In most of the vertex operator construction
it is the free field which is used.

We shall now continue and discuss the exact classical solution
of the theory and exhibit its integrability.
Consider first  the physical meaning of the
component fields $\alpha_{\pm}$ and the nature of boundary conditions
at the turning point or wall $\tau = 0$.  Shifting by the classical
solution, $\alpha_{\pm} = \pm p_0 + \epsilon_{\pm}\,$, the equations
of motion linearize to
$$\partial_{\pm} \epsilon_{\pm} \pm \left( p_0 \partial_x + \partial_x
p_0 \right) \epsilon_{\pm} = 0\,\,. \eqno\eq$$
Denoting $\epsilon_{\pm} \equiv \mp {1\over p_0} \psi_{\mp}$ we have
$$\left( \partial_t \pm \partial_{\tau} \right) \psi_{\mp}=0\,\,.
\eqno\eq$$
So indeed, $\psi_{\pm} = \psi_{\pm} ( t\pm \tau)$ are left- and
right-moving waves, respectively.  There is however a nontrivial
boundary condition in the theory which comes in as follows:  The
eigenvalue density $\phi = {1\over 2\pi} \left( \alpha_+ - \alpha_-
\right)$ gives the conserved fermion number
$$\dot{N} = \int dx \,\dot{\phi} = {1\over 2\pi} \int dx \left(
\alpha_+^2 - \alpha_-^2 \right)=0\,\,.\eqno\eq$$
At the boundary point for $x$ (or $\tau = 0)$, this implies
$$\left( \alpha_+^2 - \alpha_-^2 \right)\Bigl\vert_{{\rm boundary}} =
0\,\,,\eqno\eq$$
so that there is no leakage into the region under the barrier (this
may have to be given up in nonperturbative discussion [\LM]).  For the
small fluctuations we then have
$$\psi_+ (x) = \psi_- (x)\,\,,\eqno\eq$$
implying Dirichlet boundary conditions. In terms of Fourier modes
$$\psi_{\pm} = \int_{-\infty}^{\infty} dk\,\,\alpha_k^{\pm}\,\,
e^{ik (t\pm \tau)}\eqno\eq$$
with $\alpha_{-k} = \alpha_k^+$ our boundary condition implies that one
has only one set of oscillators with positive momenta
$$\alpha_k^+ = \alpha_k^- = \alpha_k\,,\qquad k>0\,\,.\eqno\eq$$
This is appropriate for a theory defined on a half-line
$\tau \in [0,\infty)$.

A very simple form for the exact solution of the collective equations
was given by Polchinski [\Poltwo].
At  the classical level one has a  phase space picture of
the eigenvalues $\lambda (\sigma,t)$ and their momenta $p(\sigma,t) =
\dot{\lambda}$.  They obey the classical equations of motion
$$\dot{p}(\sigma,t) = - u'\left(\lambda(\sigma,t)\right)\,\,.
\eqno\eq$$
The information that the particles are fermions is contained in the
statement that the equation $x = \lambda (\sigma,t)$ is invertible: $\sigma
= \sigma (x,t)$ so that for each $\sigma$ there is only one particle
(actually there is a degeneracy corresponding to the upper and lower
Fermi surface).  Consider in particular the inverted oscillator: the
solution is immediately written as
$$\eqalign{ x & = a(\sigma) \cosh (t - \sigma )\,\,,\cr
p & = a(\sigma ) \sinh(t - \sigma )\,\,. }\eqn\parsol$$
Here $a(\sigma )$ is an arbitrary function giving an arbitrary initial
condition.  The simplest configuration is obtained for
$a (\sigma ) = \sqrt{\mu} = {\rm const.}$ and we have
$$p_{\pm}=p(\sigma_{\pm},t)=\pm\sqrt{x^2 -\mu}\,\,.\eqno\eq$$
This is recognized as the static ground state collective field
configuration $\pi \phi_0 (x)$. It is easy to see that the general
configuration leads to the solution of collective
equations. The collective field is identified with the Fermi momentum
densities
$$\alpha_{\pm}(x,t)\equiv p_{\pm}=p (\sigma_{\pm}(x,t),t)\,\,.\eqno\eq$$
Conversely, $p(\sigma,t)=\alpha(x(\sigma,t),t)$. Using the chain rule
$${\partial p\over \partial t} = {\partial\alpha\over \partial t} +
{\partial\alpha\over \partial x}\,\, {\partial x\over\partial t} = -
u'(x)\,\,,\eqno\eq$$
and the equation of motion obeyed by $p(\sigma,t)$, there follows the
equation
$${\partial\alpha\over\partial t}= -u'(x)-\alpha\,\partial_x \alpha
\,\,.\eqno\eq$$
These are the decoupled quadratic equations for the collective fields
$\alpha_{\pm} (x,t)$ associated with the cubic Hamiltonian.

Knowledge of the exact solution can be directly  used to determine
scattering amplitudes. One considers and follows the time evolution of
an incoming left-moving lump.
A point parametrized by $\sigma$ which passes through $x$
at some (early) time $t$ will reflect on the boundary and pass through
the same point $x$ at some later time $t'$ as a right-moving lump.
The time evolution of the particle coordinates is known
explicitly~\parsol\
so one can determine the relationship between the two times $t$ and
$t'$. Consider the exact solution given by Eq.~\parsol,
at a distance $\tau$ large enough one has
$$x=e^{\tau} = \cases{a(\sigma)\, e^{-(t-\sigma )}\,, \quad &
$\quad t\rightarrow - \infty$\cr
a(\sigma)\, e^{+(t' + \sigma )}\,, \quad & $\quad t' \rightarrow +
\infty$\cr }$$
from where
$$ t' - \tau = t + \tau + \ln a^2 (\sigma)\,\,.\eqno\eq$$
On the other hand $a^2 (\sigma ) $ is related to $\alpha_{\pm}$:
$$a^2 = x^2 - p^2 = x^2 - \alpha_{\pm}^2 \approx 1 +
\psi_{\mp}\,\,.\eqno\eq$$
The outgoing particle momentum $p_+ (t', \sigma)$ is equal in magnitude
(but opposite in sign) to the incoming momentum of the particle
$p_- (t, \sigma)$:
$$p_+ (t', \sigma ) = - p_- (t,\sigma )\,\,.\eqno\eq$$
This elementary relationship provides a relationship between the
incoming and outgoing wave and therefore yields the $S$-matrix.
Collecting the above formulas we have
$$\psi_-(z)=\psi_+ \Bigl(z-\ln(1+\psi_-(z))\Bigr)\,\,.\eqno\eq$$
This functional equation determines the relation between the left- and
right-moving (incoming and outgoing fields). It represents a nonlinear
version of our Dirichlet boundary conditions and is characteristic of
scattering problems involving a wall. An expansion in power series can
be performed determining explicitly the outgoing modes in terms of the
incoming ones.  This is then sufficient to give the
scattering amplitudes. We shall return to this subject in sect.~5.

In general, all features of the exactly solvable matrix model
translate into string theory.  More precisely there is a direct
translation of matrix model quantities into the collective field
theory which itself is then completely integrable as we have
emphasized. We  end this section by summarizing the  set of
translation rules between the matrix model and collective field
theory representations.

At the  classical level one thinks of matrix variables
as coordinates in a fermionic phase space $M\rightarrow \lambda ,
P\rightarrow p$. Collective field theory
represents a second quantization according to $p\rightarrow
\alpha (x,t)$.  So we have the correspondences:
$$\eqalign{ M & \leftrightarrow \lambda \leftrightarrow x\,\,,\cr
P &\leftrightarrow p \leftrightarrow \alpha (x,t)\,\,.}\eqno\eq$$
The $U(N)$ trace becomes a phase space integration in the fermionic
picture and in  the collective representation:
$$\Tr \left\{ \,\,\right\}\quad \rightarrow \quad\int{dx\over 2\pi}
\int_{\alpha_{-}(x,t)}^{\alpha_{+}(x,t)} d\alpha \,\, \left\{ \,\,
\right\}\,\,, \eqno\eq$$
where $\alpha_{\pm} (x,t)$ are the chiral components of the scalar
field density. For example the collective Hamiltonian comes out
as follows:
$${1\over 2}\,\Tr \left( P^2 - M^2 \right)\rightarrow{1\over 2}
(p^2 - x^2) \rightarrow \int {dx\over 2\pi} \int d\alpha \,
{1\over 2}(\alpha^2- x^2) ={1\over 2} \int {dx\over 2\pi}
\Bigl[ {\alpha\over 3}^3 - x^2 \alpha \Bigr]_-^+ \,\,.\eqno\eq$$
The above transition rules summarize the statement that
the Poisson brackets of single particle quantities in the Fermi (or
matrix) phase space
$$\left\{ f_1 (x,p), f_2 (x,p)\right\}_{\rm P.B.}\eqno\eq$$
remain preserved in the field theory.  For example, the field-theoretic
operator inferred from the  oscillator states is
$$B_n^{\pm}=\int{dx\over 2\pi}\int d\alpha \,\,(\alpha \pm x)^n\,\,.
\eqno\eq$$
We can now use the $\alpha$-field Poisson brackets
$\left\{ \alpha_{\pm} (x), \alpha_{\pm} (y) \right\} = \mp 2\pi i
\delta '(x-y)$  to verify that indeed
$$\left\{ H_{{\rm coll}} , B_n^{\pm} \right\} = \pm n \,
B_n^{\pm}\,\,.\eqno\eq$$
This represents an eigenstate of the collective field theory Hamiltonian.
At the quantum level a normal ordering prescription is used to
completely define the operators.

The outlined string field theory gives a systematic perturbation theory
in the string coupling constant. The Feynman rules that are constructed
are characterized by a nontrivial cubic vertex exhibiting discrete poles
in the momenta. Most importantly a fully quantized Hamiltonian is achieved
through normal ordering with the counter-terms being supplied by the original
collective formalism. So what one has is a totally finite string field
theory capable of reproducing string theory diagrams to all orders. It
works at loop level without further counter-terms giving a single covering of
modular space. This, as is well known, has always been quite nontrivial in a
string-theoretic framework. For more details of the quantum theory and
explicit calculations at the loop level the reader should
consult [\DJR].

\bigskip
\chapter{$w_{\infty}$ Symmetry}
\medskip

The matrix model description has the virtue of great simplicity:  it
is linear and trivially exactly solvable.  For the matrix Hamiltonian
$$H = {1\over 2}\, \Tr\left(P^2-M^2\right)\eqno\eq$$
one can write down exact creation--annihilation operators
$$B_n^{\pm} = \Tr \left( P \pm M\right)^n\,,
\quad\quad n = 0,1,2,\ldots\eqno\eq$$
creating imaginary energy eigenstates
$$\left[\,H, B_n^{\pm}\,\right]= \mp in B_n^{\pm}\,,\qquad
\epsilon_n = \pm in \,\,.\eqno\eq$$
The whole point here is to be able to translate this exact information
into physical results which, as we have emphasized, is achieved through
collective field theory. The direct connection of the space-time string
field theory with the matrix model leads then further insight. The simple
oscillator structure with its creation--annihilation basis implies the
presence of a similar structure in the field theory and therefore string
theory.

To understand the physical meaning of the (oscillator) states
recall that in the collective field theoretic description we have another
spatial quantum number in addition to the energy.
This feature arose as a consequence of scaling invariance.
The coordinate and the fields transform as
$$x\rightarrow ax\,,\qquad\quad
\alpha (x,t)\rightarrow {1\over a}\,\alpha(ax,t)\,\,,\eqno\eq$$
and the Hamiltonian, without the chemical potential term,
$-\mu \alpha$, scales as
$$H \rightarrow {1\over a^4}\,H\,\,.\eqno\eq$$
The classical equations of motion are consequently scale invariant.
One then defines the scaling momentum as
$$i p_s = - 4 + s\,\,,\eqno\eq$$
where $s$ is the naive scaling dimension $s[x] = s[\alpha] = 1$.  The
creation--annihilation operators
$$\tilde T_n^{\pm} ={1\over n}\,\int {dx \over 2\pi}\,\,
{(\alpha\pm x)^{n+1}\over n+1}\,\,\Big\vert_{\alpha_{-}}^{\alpha_{+}}
\eqno\eq$$
consequently have the following energy-momentum:
$$i \epsilon = n \,,\qquad\quad
i p_s = -2 + n \,\,.\eqno\eq$$
We find these to be in precise agreement with the discrete tachyon
vertex operator states since there
$$i \sqrt{2}\, p  = \pm 2 j\,,\qquad\quad
i \sqrt{2}\, p_{\varphi} = -2 + 2j \,\,,\eqno\eq$$
and we have already noted the relations $\sqrt{2}\, p = \epsilon,\,\,
\sqrt{2}\,p_{\varphi}=p_s$. We then have a one-to-one correspondence
between oscillator states of the matrix model and discrete tachyon
vertex operators of the conformal description of $c=1$ string theory
$$B_n^{\pm}=\Tr\left(P\pm M\right)^n \quad \leftrightarrow \quad
T_{p}^{(\pm)}=e^{\pm i\sqrt{2} jX}\,e^{-\sqrt{2}(1-j)\varphi}\eqno\eq$$
with $n=2j$ or $j=n/2$.

An analytic continuation of discrete imaginary momenta to real
values $(n = i\kappa,\, p_s = 2i - \kappa )$
gives the scattering operators
$$\eqalign{ B_{-i\kappa}^- & = \Tr \left( P - M\right)^{-i\kappa}
\sim e^{-i\kappa (t+\tau )}\,\,,\cr
B_{-i\kappa}^+ & = \Tr \left( P + M\right)^{-i\kappa}
\sim e^{-i\kappa (t-\tau )}\,\,, }\eqno\eq$$
describing left- and right-moving waves, respectively. These operators
can be used to construct the in- and out-states of scattering theory
$$\eqalign{\Tr \left( P-M \right)^{-i\kappa} \vert 0 \rangle & = \vert
\kappa;\,{\rm in} \rangle \,\,,\cr
\Tr\left( P + M \right)^{+i\kappa}\vert 0\rangle & = \vert\kappa;\,
{\rm out} \rangle \,\,. }\eqno\eq$$
Namely, for an in-state, one needs a {\it left}-moving wave while the
out-state is necessarily given by a {\it right}-moving one. Here we
have used the picture where the wall is at $\tau = -\infty$
corresponding to the physical space being defined on the right semi-axis
$x = e^{\tau} \geq 0$. Had we chosen to define the theory on the
other side of the barrier, the states
$$\eqalign{\Tr\left(P+M\right)^{i\kappa}&=e^{i\kappa(t+\tau)}\,\,,\cr
\Tr\left(P-M\right)^{i\kappa}&=e^{i\kappa(t-\tau)}\,\,, }\eqno\eq$$
would be physical since they have the meaning of a right-moving
{\it in}-wave and a left-moving {\it out}-wave.
Hence there is a one-to-one correspondence between the scattering
operators in the matrix model and the string theory vertex operators
$$\Tr\,(P\pm M)^{-i \kappa}\quad\leftrightarrow \quad
T^{\pm}=e^{\pm {\kappa\over\sqrt{2}} X}\,e^{-\sqrt{2}+i{\kappa\over\sqrt{2}}
\varphi}\,\,.\eqno\eq$$
A typical transition amplitude reads
$$S = \langle {\rm out} \vert {\rm in}\rangle=\langle 0\vert
\Tr (P+M)^{{\kappa\over i}}\Tr(P-M)^{{\kappa\over i}}\vert 0\rangle\,\,.
\eqno\eq$$
It only contains operators with the same (Liouville)-exponential
dressing. This is in total agreement with the continuum string theory
situation.

In addition to the tachyon states, the matrix oscillator description
immediately allows a construction of an infinite sequence of discrete
states [\GKN,\AvJe].
They are created by the operators
$$B_{n,\bar n}=\Tr\Bigl((P+M)^n (P-M)^{\bar{n}}\Bigr)\,\,,\eqno\eq$$
with energies and momenta given by
$$i\epsilon=n-\bar{n}\,,\qquad\quad
i p_s =-2+(n+ \bar{n})\,\,.\eqno\eq$$
Comparing this with the discrete spectrum of the string theory given by
$$i \sqrt{2} p = 2m\,,\qquad\quad
i \sqrt{2} p_{\varphi} =-2+2j\,\,,\eqno\eq$$
we find the correspondence
$$m ={n-\bar{n}\over 2}\,,\qquad\quad
j ={n + \bar{n}\over 2} \,\,.\eqno\eq$$
These are indeed half-integers once $n,\bar{n}$ are integers.  The
field theory operators
$$B_{jm}=\int {dx\over 2\pi}\int_{\alpha_-}^{\alpha_+} d\alpha\,
(\alpha + x )^{j+m}\,(\alpha - x )^{j-m}\eqno\eq$$
can be shown (again by using the Poisson brackets or the commutators)
to generate discrete imaginary energy eigenstates of the Hamiltonian
$$\left[ H, B_{jm} \right] = -2i m B_{jm}\,\,.\eqno\eq$$
This commutator shows that the operators $B_{jm}$ are
spectrum-generating operators for the Hamiltonian $H$;
but it also signals the existence of a large symmetry algebra which
operates in this theory [\AvJe,\MoSe,\winf,\WiGR,\KlPo,].

First we had the  sequence of conserved quantities
$$H_l =\Tr\left( P^2 - M^2 \right)^{l+1}\eqno\eq$$
commuting among themselves
$$\left[ H_l , H_{l'} \right] = 0\,\,.\eqno\eq$$
These are  particular cases of the spectrum-generating operators
$B_{jm}$. One is then lead to consider the complete algebra of all the
operators. Introducing the more standard notation
$$O_{JM} = (p+x)^{J+M+1} (p-x)^{J-M+1}\,\,,\eqno\eq$$
with the associated collective field realization
$$O_{JM} = \int {dx\over 2\pi} \int_{\alpha_{-}}^{\alpha_{+}} d\alpha
\,\, (\alpha + x )^{J+M+1} (\alpha - x )^{J-M+1}\,\,,\eqno\eq$$
one checks that they obey the $w_{\infty}$ commutation relations
$$\left[ O_{J_{1}M_{1}} , O_{J_{2} M_{2}} \right] = 4i
\Bigl((J_2+1)M_1-(J_1+1)M_2\Bigr)\,O_{J_{1}+J_{2},M_{1}+M_{2}}\,\,.
\eqn\winfalg$$
We note that this commutator results if no special ordering is taken
for the noncommuting factors.  At the full operator
quantization level, field theory requires special normal ordering.
It is likely that this modifies the simple $w_{\infty}$ algebra to a
$W_{1+\infty}$ algebra.

Recalling the form of tachyon operators $T_n^{\pm}=\Tr(P\pm M )^n$ one sees
a special relationship between the tachyon operators and the
$w_{\infty}$ generators.  A simple computation shows that
$$O_{JM} = {1\over 2i} \, {1\over (J+M+2) (J-M+2)} \, \bigl[
T_{J+M+2}^+ , T_{J-M+2}^- \bigr] \,\,.\eqno\eq$$
This first sheds some light on the nature of higher discrete modes in
the collective formalism:  they are composite states of the tachyon.
More importantly, one then expects a simple realization of the
$w_{\infty}$ algebra on the tachyon sector.

To understand the role played by the scalar collective field with
respect to the $w_\infty$ algebra one can first look at the
following Virasoro subalgebra:
$$O_{l} \equiv O_{{l\over 2},{l\over 2}} =
\int {dx \over 2\pi}\,\int d\alpha\,\, (\alpha+x)^{l+1}\,
(\alpha -x)\,\,,\eqno\eq$$
with
$$[ O_{l},  O_{l'} ]= 2i(l-l')\,O_{l+l'} \,\,.  \eqno\eq$$
We can then determine the transformation property of the tachyon
field under this subalgebra. Actually, the exact tachyon
creation operator $T_n$ can itself be written as an
extension of the whole algebra
$$\tilde T_n^{+} ={1\over n}\,\int {dx \over 2\pi}\,
{(\alpha+x)^{n+1}\over n+1} =
{1\over n}\, O_{{n\over 2}-1, {n\over 2}}\,\,,\eqno\eq$$
and this determines the commutator (with the indices extended  outside
the standard range $|m| \leq j$). Alternatively one  can also directly
use the basic commutation relations to find
$$[O_l, \tilde T_n^{+}] = 2i (n+l)\,\tilde T_{n+l}^{+}\,\,,\eqno\eq$$
which shows that the tachyon transforms as a field of conformal weight 1.
This is understood to be a space-time and not a world sheet feature.
The fact that an infinite space-time  symmetry appears
in the collective field theory explains many similarities that it
has with conformal field theory.
 The $w_\infty$ generators act in a
nonlinear way  on the tachyon field.
This implies that this symmetry can be used to write down Ward
identities for correlation functions and the $S$-matrix.

Let us study in more detail the nonlinearity involved in
the collective representation(here we summarize the results
achieved in [\JRvT]).
One is in general interested in comparison with similar
nonlinearities (and Ward identities) obtainable in the
world sheet conformal field theory analysis.
The latter is only performed in the approximation neglecting
the cosmological constant term $(\mu \to 0)$ which  represents the
strong coupling regime of the field theory
$(g_{\rm st} = 1/\mu \to \infty)$.
In this limit one simply expands
$$\alpha_{\pm}(x,t) =\pm x+{1\over 2x}\,\hat\alpha_{\pm}\,\,,\eqno\eq$$
which is an approximate form when
$$\pi\phi_0(x) =\sqrt{x^2-\mu} \to x = {e^{\tau}\over 2}\,\,.
\eqno\eq$$
The exact tachyon operators reduce
in the leading (linear) approximation to
$$\tilde T_n^{\pm} ={1\over n}\,\int {d\tau \over 2\pi}\,\,
e^{n\tau}\,\hat\alpha_\pm \,\,,\eqno\eq$$
This is as it should be since they are to describe left- and right-moving
waves,
respectively. Consider now the $w_\infty$ generators
in the same approximation. With the
above background shift one easily finds that they reduce to
$$O_{JM}={1\over J-M+2}\,\int {d\tau\over 2\pi}\,\,
e^{2M\tau}\,\hat\alpha_{+}^{J-M+2}
+{(-1)^{J-M}\over J+M+2}\,\int {d\tau\over 2\pi}\,\,
e^{-2M\tau}\,\hat\alpha_{-}^{J+M+2}\,\,.\eqno\eq$$
Here we see that the operator $O_{JM}$ behaves as the
$(J-M+2)$th power of the right-moving tachyon $\alpha_+$ and
also the $(J+M+2)$th power of the left-moving tachyon $\alpha_-$.
These are the leading polynomial powers in the left- and
right-moving components of the tachyon; even in this
strong coupling limit the theory is nonlinear and one has
further higher order terms. Concerning these one can go to the
in (out) fields (where it is likely that only leading terms
remain). The in (out) fields are simply limits of the component fields
$\alpha_\pm$:
$$\alpha_{\rm out}(t-\tau) = \lim_{t\to +\infty} \alpha_+\,,
\qquad\quad \alpha_{\rm in}(t+\tau) = \lim_{t\to -\infty} \alpha_-\,\,,
\eqn\inoutfields$$
Since the operators $O_{JM}$ are conserved (up to a phase),
looking at the $t\to \pm\infty$ limit of $e^{2Mt}\,O_{JM}$ we
obtain an identity (between the in- and out-representation):
$$\eqalign{O_{JM}&={1\over J-M+2}\,\int {dz\over 2\pi}\,\,
\alpha_{\rm out}^{J-M+2} (z)\cr
\noalign{\vskip 0.2cm}
&={(-1)^{J-M}\over J+M+2}\,\int {dz\over 2\pi}\,\,
\alpha_{\rm in}^{J+M+2} (z) \,\,.}\eqno\eq$$
Introducing creation--annihilation operators
$$\alpha_{\rm in}(z)=\int dz\,e^{-ikz}\,\alpha(k)\,,
\qquad\quad \alpha_{\rm out}(z)= \int dz\,e^{-ikz}\,\beta(k)\,\,,
\eqno\eq$$
with $\alpha(k)=a(k)$ and $\alpha(-k)=ka(k)^\dagger$;$\beta(k)=b(k)$ and
$\beta(-k)=kb(k)^\dagger$ we have the expressions
after a continuation $k \to ik$:
$$\eqalign{O_{J,-M}&={1\over J-M+2}\,\int dk_1\ldots dk_{J-M+2}\,\,
\alpha(k_1)\ldots\alpha(k_{J-M+2})
\delta(\sum k_i +2M)\cr
\noalign{\vskip 0.2cm}
&={(-1)^{J-M}\over J+M+2}\,\int dp_1\ldots dp_{J+M+2}\,\,
\beta(p_1)\ldots\beta(p_{J+M+2})
\delta(\sum p_i +2M)\,\,.}\eqno\eq$$
These representations can be compared with analogue expressions
found in conformal field theory [\KlPo].

The Ward identities essentially follow from the in--out
representations of the generators in terms of the tachyon
field. A typical $S$-matrix element is given by
$$S(\{k_i\};\{p_j\})=\langle 0| \prod_j \,\beta(p_j)\,\prod_i \, \alpha(-k_i),
|0\rangle \,\,.\eqno\eq$$
Consider a general matrix element of the $w_\infty$
generator $O_{JM}$,
$$\langle \,\,0|\beta\, O_{JM}\, \alpha |0\,\,\rangle \,\,.$$
It can be evaluated by commuting to the left or to the right
using alternatively the above in--out representations. The
two different evaluations give the identity
$$\langle 0| [\beta, O_{JM}]\, \alpha|0\rangle =
\langle 0| \beta \,[O_{JM}, \alpha]|0\rangle\,\,,\eqno\eq$$
which summarizes the general Ward identities. These, when
written out explicitly using the representations for
$O_{JM}$, have the form of recursion relations reducing the
$N$-point amplitude to lower point ones. Specifically, the
one creation operator term of the $\alpha$-representation
for $O_{JM}$ gives:
$$O_{M+1,-M}\,a^\dagger(k_1)\,a^\dagger(k_2) |0\rangle =
4\pi (k_1+k_2+2M)\,a^\dagger(k_1+k_2+2M) |0\rangle\,\,,\eqno\eq$$
which turns a two-particle state into one-particle state.
Generally,
$$O_{M+N,-M}\,a^\dagger(k_1)\ldots a^\dagger(k_{N+1}) |0\rangle =
2\pi^N\,(N+1)! (\sum k_i+2M)\,a^\dagger(\sum k_i+2M)
|0\rangle\,\,,\eqno\eq$$
showing a reduction of the $(N+1)$-particle state into a
single-particle state.

As an example let us calculate the $3\to 1$ amplitude
$$S_{3,1}=\langle 0|b(p)\,a^\dagger(k_1)\,
a^\dagger(k_2)\,a^\dagger(k_3) |0\rangle\,\,.\eqno\eq$$
The energy-momentum conservation laws imply
(recall that for $B^\pm_n$, $\epsilon=\pm n,
p_s =-2+n$):
$$\eqalign{&k_1+k_2+k_3-p=0\,\,,\cr
&(-2+k_1)+(-2+k_2)+(-2+k_3)+(-2+p)=-4\,\,.}$$
The latter is a specific case of the general
Liouville conservation (or bulk condition):
$$\sum_{i=1}^N = p_s^i = -4\,\,.\eqno\eq$$
The energy-momentum relations specify the momentum of
the 4th particle
$$p=2 \qquad {\rm or} \qquad k_1+k_2+k_3=2\,\,.$$
Use now the operator
$$\eqalign{O_{{1\over 2},-{1\over 2}} &= {1\over \sqrt{\mu}}\,
\int dk_1 dk_2 dk_3\,\, k_1 \,a^\dagger(k_1)\,
a(k_2)\,a(k_3)\,\delta(k_1-k_2-k_3+1) \cr
\noalign{\vskip 0.2cm}
&= {\sqrt{\mu}}\, \int dp_1 dp_2 \,\, p_1\,
b^\dagger(p_1)\,b(p_2)\,\delta(p_1-p_2+1) }$$
to deduce
$$\eqalign{S_{3,1}&=\langle \,b(2)\,a^\dagger(k_1)\,
a^\dagger(k_2)\,a^\dagger(k_3)\,\rangle\cr
\noalign{\vskip 0.2cm}
&={\pi\over \mu}\,(k_1+k_2-1)\,
\langle \,b(1)\,a^\dagger(k_1+k_2-1)\,
a^\dagger(k_3)\,\rangle\cr
\noalign{\vskip 0.2cm}
&+{\pi\over \mu}\,(k_2+k_3-1)\,
\langle \,b(1)\,a^\dagger(k_2+k_3-1)\,
a^\dagger(k_1)\,\rangle\cr
\noalign{\vskip 0.2cm}
&+{\pi\over \mu}\,(k_1+k_3-1)\,
\langle \,b(1)\,a^\dagger(k_1+k_3-1)\,
a^\dagger(k_2)\,\rangle \,\,.}$$
Taking the normalized three-point function to be
$S_3 = 1/\mu$, the result
$$S_{3,1} ={\pi\over \mu^2}\,\bigl(2(k_1+k_2+k_3)-3\bigr)=
{\pi\over \mu^2} $$
follows. One can iteratively repeat the same reduction for higher
point amplitudes and find
$$S_{N,1}= {\pi^{N-1}\over (N-2)!}\,\,\mu^{-N+1}\,\,,\eqno\eq$$
which is the $(N+1)$-point  ``bulk'' scattering amplitude.

In describing the infinite symmetry  we have followed the
matrix model approach were the appearance of the symmetry structure
is most natural. The features described arise also in the continuum
conformal field theory language where the Ward identities take a
particularly elegant form.

Of crucial importance in establishing continuum
quantities that are analogous with those of the matrix model is Witten's
identification of the ground ring [\WiGR]. This consist of ghost
number zero, conformal spin zero operators ${\cal O}_{JM}$
which are closed under operator products
${\cal O'}\cdot {\cal O''} \sim{\cal O''}$ (up to BRST
commutators). The basic generators are
$${\cal O}_{0,0}=1\,,\qquad
{\cal O}_{{1\over 2},\pm{1\over 2}} =
\Bigl[ c\,b \pm {i\over\sqrt{2}}\,\partial X -
{1\over\sqrt{2}}\,\partial\varphi\Bigr]\,
e^{(\pm iX+\varphi)/\sqrt{2}}\,\,.\eqno\eq$$
The suggestion (of Witten) was that
${\cal O}_{{1\over 2},\pm{1\over 2}}$ are the
variables which correspond to the phase space coordinates of the
matrix model
$$\eqalign{
{\cal O}_{{1\over 2},+{1\over 2}} &=a_+ \equiv p+x\,\,,\cr
\noalign{\vskip 0.2cm}
{\cal O}_{{1\over 2},-{1\over 2}} &=a_- \equiv p-x\,\,,}\eqno\eq$$
with ${\cal O}_{0,0}=1$ being the cosmological constant
operator. Once the (fermionic) matrix eigenvalue coordinates have
been identified one could study the action of discrete states
vertex operators upon them. They turn out to act as vector
fields on the scalar ring
$$\Psi_{JM}={\partial h\over \partial a_+}\,\,
{\partial\over\partial a_-} -
{\partial h\over \partial a_-}\,\,
{\partial\over\partial a_+} \,\,,\eqno\eq$$
with the familiar matrix model form
$h_{JM}=a_+^{J+M}\,a_-^{J-M}$. In the continuum
approach the $w_\infty$ generators are integrals of
conserved currents which are for closed string theory
constructed as
$$\eqalign{
&Q_{JM}=\oint {dz\over 2\pi i}\,\,W_{JM}\,(z,\bar z)\,\,,\cr
\noalign{\vskip 0.2cm}
&W_{JM}\,(z,\bar z)=\Psi^+_{J+1,M}(z)\,{\cal O}_{JM}(\bar z)\,\,.}
\eqno\eq$$
One can study the action of these operators on the
the tachyon vertex operators.
A formula derived by Klebanov [\Kl] reads
$$Q_{M+N-1}\,T^+_{k_1}(0)\int T^+_{k_2}\ldots\int T^+_{k_N}=
F_{N,M}(k_1,\ldots,k_N)\, T^+_{-\sum k_i +M}\,\,,\eqno\eq$$
where
$$F_{N,M}(k_1,\ldots,k_N)=2\pi^{N-1}\,N!\,k\,
{\Gamma(2k)\over \Gamma(1-2k)}\,
\prod_{i=1}^N\,{\Gamma(1-2k_i)\over \Gamma(2k_i)}\,\,,\eqno\eq$$
with a similar formula for the action of
$Q_{-M+N-1}$ on $N$ oppositely moving $T^-$ tachyons.
These representations of the $w_\infty$ generators on
tachyon vertex operators are clearly comparable to the
direct representation obtained in the matrix model
(or more precisely collective field formalism). The
comparison and agreement of these representations
 is the closest one comes in being
able to identify the two approaches.

The conformal (vertex operator) formalism gives a very
elegant summary of Ward identities in the form of general
(master) equation. We end this section with a short description
of this equation [\ZW,\Ve,\KlPa]. It  follows from BRST
invariance of the discrete state vertex operators
$$\{ Q_{\rm BRST}, c(z) W_{JM}(z)\}=0\,\,,\eqno\eq$$
which implies that for general tachyon correlation function
$$\big\langle \{ Q_{\rm BRST}, c\, W_{JM}\}\,
V^\pm_{k_1} \ldots V^\pm_{k_n}\,\big\rangle =0\,\,.\eqno\eq$$
Changing to operator formalism
$$\sum_{{\rm perm}} \big\langle V^\pm_{k_1}\,
\{ Q_{\rm BRST}, c\, W_{JM}\}\Delta V^\pm_{1} \ldots
\Delta V^\pm_{1}\,V^\pm_{k_n} \big\rangle =0\,\,,\eqno\eq$$
allows one to eliminate $ Q_{\rm BRST}$. The vertex operators are all
BRST invariant while the propagator is essentially the
inverse of $Q$:
$$[Q,\Delta]=\Pi_{L_0 -\bar L_0}\,b^-_0\,\,,\eqno\eq$$
where the $\Pi$ projects on the subspace
$(L_0 -\bar L_0)|\Phi_i \rangle =0$. The final form of the
Ward identity then follows
$$\sum_{{\rm partitions}\atop {m+m'=n}}\,
\langle V_{i_1}\ldots  V_{i_m}\,\Phi \rangle
\langle \Phi  V_{j_1}\ldots  V_{j_{m'}}\,c W_{JM} \rangle =0\,\,.
\eqno\eq$$

Most of the considerations of this section and most of the studies
of the $w_\infty$ symmetry are performed in the extreme limit where the
cosmological term is ignored. This is particularly the case for the
continuum, conformal field theory approach. Some attempts to
extensions and inclusion of the nontrivial cosmological constant
effect were made however. In the matrix model the cosmological
term is introduced in a simple and elegant way corresponding
 to nonzero Fermi energy
$$h_0={1\over 2}\,(p^2 -x^2)\quad \to \quad
h_\mu={1\over 2}\,(p^2 -x^2)+\mu\,\,.\eqno\eq$$
Since the ground ring generators $a_\pm$ were identified
to be analogues to $p\pm x$ it is then expected that an
equivalent deformation from $a_+ a_- =0$ can be established in
the continuum conformal field theory approach. This is seen
by considering the action of a ground ring on tachyons. For
$\mu =0$ it reads
$$\eqalign{&a_+\,c\,\bar c \,\tilde T^+_k = c\,\bar c \,
\tilde T^+_{k+1}\,\,,\cr
\noalign{\vskip 0.2cm}
&a_-\,c\,\bar c \,\tilde T^+_k = 0\,\,.}\eqno\eq$$
The effect of cosmological perturbation
$\mu\,T^+_{k=0}$ is found by evaluating the first order
perturbation theory contribution
$$a_-\,c\,\bar c \,\tilde T^+_k =
-a_-\,c\,\bar c \,\tilde T^+_k(0)\,
\Bigl( \mu\int d^2 z\,,\tilde T^+_{k=0}(z) \Bigr)\,\,.\eqno\eq$$
On the right-hand side $a_-$ essentially fuses the two
tachyon operators into one giving (to first order)
$$a_- \,\tilde T^+_k = -\mu \,\tilde T^+_{k-1}\,\,.\eqno\eq$$
This replaces the second relation above and now the nonzero
Fermi level condition $a_+ a_- =-\mu$ results. To first
order [\WiGR,\Ba] one then has an agreement with the matrix model.
This is encouraging and one would clearly like to establish the complete
agreement at an exact level.

\bigskip
\chapter {$S$-matrix}
\medskip

Let us now describe the complete tree-level $S$-matrix of the
$c=1$ theory. In the previous sections we have seen the ``bulk''
scattering amplitudes which follow from the Ward identities or
are computed in conformal field theory. The complete
$N$-point scattering amplitude
$S_N=\langle T_{p_1}  T_{p_2}\ldots  T_{p_N} \rangle$
takes the (factorized) form
$$S_N=\prod_{i=1}^N (-)\mu^{i p_i}\,
{\Gamma(-ip_i)\over \Gamma(+ip_i)}\,\,A_{\rm coll}(p_1,\ldots,p_N)\,\,.
\eqno\eq$$
The external leg factors are associated with a field
redefinition [\GrKl] of vertex operators
$$T_k^\pm = {\Gamma(\mp k)\over \Gamma(\pm k)}\,\,
\tilde T_k^\pm\,\,.\eqno\eq$$
It is the redefined tachyon vertex operator $\tilde T$
that found its natural role in collective field theory.

The external leg factors of the full $S$-matrix have a
very relevant physical meaning which we now discuss. In
Minkowski space-time, $(k=ip)$, one has
$$\Delta =\mu^{\mp ip}\,\,
{\Gamma(\pm ip)\over \Gamma(\mp ip)}\,\,.\eqno\eq$$
So, the factors $\Delta =e^{i\theta_p}$ are pure phases.
As such they give no contributions to the actual
transition amplitudes and could be ignored.
The fact is however that they carry physical information
on the nature of tachyon background. The factors exhibit
poles at discrete imaginary energy
$$p\sqrt{2}=in\,,\quad n=1,2,3,\ldots \eqno\eq$$
If we consider a process with an incoming tachyon and
$N$ outgoing ones, the discrete imaginary value of the
incoming momenta signifies the resonant on-shell
process in which a certain number  $r$ of Liouville
exponentials participate
$$\langle T_-\,(\mu\,e^{-\sqrt{2}\varphi})^r\, T_+ \ldots
t_+ \rangle\,\,.\eqno\eq$$
The on-shell condition in this case indeed gives
$$i\sqrt{2}\,p =-(r+N-1) \,\,,\eqno\eq$$
in agreement with the discrete imaginary energy poles noted above.

In collective field theory the
external leg factors are associated with
a field redefinition given by an integral transformation. The transformation
comes from the change of coordinates between the Liouville  and the
time-of-flight variables. It is the later that appears naturally in the
collective field formalism  and as we have seen provides a simple description
of the theory.
Let us recall the basic  (Wilson) loop operator of the matrix model
with its Laplace transform
$$\hat W(\ell,t)\equiv \Tr\,(e^{-\ell M}) =
W_0 +\int dx\,e^{-\ell x}\,\partial_x \eta \,\,.\eqno\eq$$
After the change to the time of flight coordinate
$x=\sqrt{2\mu}\,\cosh\tau$ and the explicit identification of the Liouville
$\ell =2e^{-\varphi/\sqrt{2}}$ the integral
transformation results
$$\hat W(\ell,t)=\int_0^\infty d\tau\,\,
{\rm exp}\,\Bigl[-2\sqrt{2\mu}\,e^{-\varphi/\sqrt{2}}\,
\cosh \tau\Bigr]\,\partial_\tau\,\eta(\tau,t)\,\,,\eqno\eq$$
We have seen in our earlier study of the linearized theory that
this integral transformation takes the
Liouville operator into a Klein-Gordon operator
$$(\partial_t^2 -\partial_\tau^2)\,\eta
\quad \Longleftrightarrow \quad
\bigl(\partial_t^2 -{1\over 2}\,\partial_\varphi^2
+4\mu e^{-\sqrt{2}\,\varphi}\,\bigr) \hat W\,\,.\eqno\eq$$

The integral transformation therefore expresses the tachyon field in
terms of a simple Klein-Gordon field $\eta(\tau,t)$:
$$T(\varphi,X)\equiv e^{-\sqrt{2}\,\varphi}\,
\hat W(\ell,t) = \int_0^\infty d\tau\,\,
{\rm exp}\,\,\Bigl[-2\sqrt{2\mu}\,e^{-\varphi/\sqrt{2}}\,
\cosh \tau\Bigr]\,\partial_\tau\,\eta \,\,,\eqno\eq$$
with the expected relation between the matrix model and
string theory times $X=\sqrt{2}\,t$.
The correlation functions of the
tachyon field $T$  are then expressible in terms of correlation
functions of the collective field
$\eta$. The transformation described takes plane wave solutions of the
Klein-Gordon equation
$$\eta(\tau,t)=\int_{-\infty}^\infty\,{dp\over p}\,\,
\tilde\eta (p)\,e^{-ipt}\,\sin (p\tau) \eqno\eq$$
into Liouville solutions
$$T(\varphi,t)=\int dp\,e^{-ipt}\,\gamma(p)\,
K_{ip}(2\sqrt{\mu}\,e^{-\varphi/\sqrt{2}}\,)\,
\tilde\eta(p)\,\,.\eqno\eq$$
The above redefinition of the
in--out fields does have an effect in supplying external leg
factors. The asymptotic behavior of $T(\varphi,t)$ reads
$$T\sim \int dp\,e^{-ipt}\,\Bigl(\Gamma(ip)\mu^{-ip/2}\,
e^{ip\varphi/\sqrt{2}} +
\Gamma(-ip)\mu^{ip/2}\, e^{-ip\varphi/\sqrt{2}} \Bigr)\,\,,\eqno\eq$$
giving the reflection coefficient
$$R(p)=-\mu^{ip}\,\,{\Gamma(-ip)\over \Gamma(ip)} \eqno\eq$$
for each external leg of the $S$-matrix.

After this redefinition the problem is reduced to
calculating amplitudes in collective field theory:
$ A_{\rm coll}\,(p_1,\ldots,p_N)$. There, as we have already seen,
one can derive an exact relationship between the in- and
out-field which contains the complete information about the $S$-matrix.
The solution to the scattering problem
can also be directly deduced from the exact oscillator
states. It is this procedure that turns out to be the most
straightforward and we now describe it in detail.

Consider the exact (tachyon) creation--annihilation operators
in collective field theory
$$B_{\pm ip}^\pm = \int {dx\over 2\pi}\,
\left\{ {(\alpha_+ \pm x)^{1\pm ip}\over 1\pm ip} -
{ (\alpha_- \pm x)^{1\pm ip}\over 1\pm ip} \right\}\,\,.
\eqno\eq$$
Previously we have seen that at fixed time these operators serve
as exact creation--annihilation operators of the
nonlinear collective Hamiltonian. Let us now follow the
time-dependent formalism (we describe here the derivation given
in [\JRvT]). The exact creation--annihilation operators have a simple
time evolution
$$B_{\pm ip}^\pm (t)= e^{-ipt}\, B_{\pm ip}^\pm (0)\,\,.
\eqno\eq$$
Consequently the quantity
$$\hat B_{\pm ip}^\pm \equiv e^{ipt}\, B_{\pm ip}^\pm (t)
\eqno\eq$$
is time-independent. We can simply look at the operator
$\hat B_{\pm ip}^\pm$ at asymptotic times $t= \pm\infty$
and obtain a relationship between the in and out fields~\inoutfields.
The  operator
$$B_{\pm ip}^\pm = \int {dx\over 2\pi}\,
\left\{ {(\alpha_+ \pm x)^{1\pm ip}\over 1\pm ip} -
{ (\alpha_- \pm x)^{1\pm ip}\over 1\pm ip} \right\}
\eqno\eq$$
contains contributions from both $\alpha_+$ and
$ \alpha_-$. At $t=\pm \infty$ only one of the terms
survives and we have an identity
(recall that $\hat B$ is time-independent),
$\hat B(+\infty) =\hat B(-\infty)$, which reads
$$\int {dx\over 2\pi}\,(\alpha_\pm \pm x)^{1\pm ip} =
\int {dx\over 2\pi}\,(\alpha_\mp \pm x)^{1\pm ip} \,\,.
\eqn\scatteq$$
It relates in and out fields ($\alpha_\pm$ now represent the
asymptotic fields). This is the scattering equation. It contains the
full specification of the $S$-matrix.

One can evaluate
and expand the left- and right-hand side of the Eq.~\scatteq.
Shifting by the static background
$$\alpha_\pm(t,x) \approx \pm \bigl(x-{1\over 2x}\bigr)
+{1\over 2x}\,\hat\alpha_\pm(t\mp \tau)\,\,,\eqno\eq$$
the left-hand side becomes
$$L=\int dx\,(-2x)^{1\pm ip}\,\Bigl\{
1-{(1\pm ip)\over 4x^2}\,(1\pm \hat\alpha_\pm) +
{\cal O}({1\over x^4})\Bigr\}\,\,.\eqno\eq$$
After a change of integration variable
$x=\cosh \tau \approx e^\tau/2$ the
${\cal O}(1/x^4)$ terms are seen to decay away
exponentially and what remains is only the term linear in
$\hat\alpha_\pm$. For the right-hand side we
simply find
$$R=\int dx\,\Bigl(-{1\over 2x}\Bigr)^{1\pm ip}\,
(1\pm \hat\alpha_\mp)^{1\pm ip}\,\,.\eqno\eq$$
The scattering equation then becomes
$$\int dz\,e^{-ipz}\,{1\over \mu}\,\alpha_\pm (z)=
{1\over 1\pm ip}\,\int dz\, e^{-ipz}\,
\left\{ \Bigl(1\pm {1\over\mu}\,\alpha_\mp \Bigr)^{1\mp ip}
-1 \right\}\,\,,\eqno\eq$$
giving the solution for the in-field as a function of the out-field
and vice versa. We have also explicitly restated the string
coupling constant $g_{\rm st}= 1/\mu$. This solution was
originally obtained [\MoPl] by explicitly solving the functional
relationships between the left and right collective
field components $\alpha_+$ and $\alpha_-$ given
earlier. We see here that it directly follows from the
exact oscillator states.

Before proceeding with the consideration of the $S$-matrix
let us note that the solution found has a reasonable
strong coupling limit. Indeed, for $\mu \to 0$ one is
lead to choose $ip$ to be an integer, $ip=N$, and the strong coupling
relation
$$\int dz\,e^{-Nz}\,\alpha_+ (z)=
{1\over 1+N}\,\int dz\, e^{-Nz}\,\Bigl({1\over\mu}\,
\alpha_- \Bigr)^{N+1}\eqno\eq$$
results. It is recognized as a statement specifying the bulk
amplitude where the $S_{1,N}$ and $S_{N,1}$ amplitudes
were nonzero with the momenta of the first (last) particle
being equal to $-1+N$. We have already used
relations of the above type (and their $w_\infty$ generalizations)
in our discussion of the Ward identities at (strong coupling)
$\mu=0$.

One can explicitly perform the series expansion in
$g_{\rm st}=1/\mu$, it reads
$$\hat\alpha_\pm (z) =\sum_{l=1}^{\infty}\,
{(-g_{\rm st})^{l-1}\over l!}\,\,
{\Gamma(\mp\partial +1)\over \Gamma(\mp\partial +2-l)}\,\,
\hat\alpha_\mp^l (z)\,\,.\eqno\eq$$
The $S$-matrix is defined in terms of momentum space
creation--annihilation operators
$$\pm\alpha_\pm (z) =\int {dp\over 2\pi}\,e^{-ipz}\,
\tilde\alpha_\pm (p)\,,\qquad\quad
[\tilde\alpha_\pm (p),\tilde\alpha_\pm (p')]=p\,\delta(p+p')\,\,,
\eqno\eq$$
with $\tilde\alpha_- (-p)$ and  $\tilde\alpha_+ (-p)$
being the in--out creation operators, respectively
($\alpha(-p)$ and $\beta(-p)$ in our earlier notation).
In momentum space
$$\tilde\alpha_\mp (p)= \sum_{l=1}^{\infty}\,
{(-g_{\rm st})^{l-1}\over l!}\,\,
{\Gamma(1 \pm ip)\over \Gamma(2\pm ip -l)}\,\,
\int dp_i\,\,\delta(p-\sum p_i)\,
\tilde\alpha_\pm(p_1)\ldots \tilde\alpha_\pm(p_l)\,\,,\eqn\momsp$$
and the $n\to m$ $S$-matrix element is defined by
$$A_{\rm coll} (\{p_i\}\to\{p'_j\}) =
\langle 0|\,\prod_{j=1}^m\,\tilde\alpha_+ (p'_j)\,\,
\prod_{i=1}^n\,\tilde\alpha_- (p_i)\,|0\rangle\,\,.\eqno\eq$$
Consider for example $n=1, m=3$ which is the four-point amplitude
$$A_{1,3}=\langle 0| \alpha_+(p'_1)\, \alpha_+(p'_2)\,
\alpha_+(p'_3)\,\alpha_-(-p_1)\,|0\rangle\,\,.\eqno\eq$$
It is given by the cubic term ${\cal O}(g_{\rm st}^2)$
in the expansion of \momsp\ which equals
$$\alpha_- (-p_1)= {1\over 3!}\,\,g_{\rm st}^2\,\,
{\Gamma(1-ip_1)\over \Gamma(-1-ip_1)}\, \int dp'_i\,
\delta(p_1-\sum p'_i)\,
\alpha_+(p'_1)\, \alpha_+(p'_2)\, \alpha_+(p'_3)\eqno\eq$$
and
$$A_{1,3}(p_1;p'_1,p'_2,p'_3)=
i\, g_{\rm st}^2 \,p_1 p'_1 p'_2 p'_3\,(1+ip_1)\,\,.
\eqno\eq$$
In general, an arbitrary amplitude is given in [\MoPl] to read
$$A_{n,m}=i\,(- g_{\rm st})^{n+m-2}\,
\Bigl( \prod_{i=1}^n p_i \Bigr) \Bigl( \prod_{j=1}^m p'_j \Bigr)\,
{\Gamma(-ip_n)\over \Gamma(1-m-ip_n)}\,\,
{\Gamma(1-m-i\Omega)\over \Gamma(-3-n-m-i\Omega)}\eqno\eq$$
where $\Omega=\sum_{i=1}^n\, p_i$ and the result is valid
in the kinematic region $p_n> p'_k > \sum_{j=1}^{n-1}\,p_j$.
This completes the derivation of the collective tree-level
amplitudes.

\bigskip
\chapter {Black Hole}
\medskip

Two-dimensional string theory possesses another interesting curved
space solution taking the form of a black hole. It is
described exactly by the $SL(2,\IR)/U(1)$
nonlinear $\sigma$-model
$$S_{\rm WZW} = {k\over 8\pi}\int d^2 z\,\Tr\left( g^{-1}
\partial g g^{- 1} \bar{\partial} g\right) -
ik \Gamma_{\rm WZW} + {\rm Gauge}\,\,,\eqno\eq$$
with $k = {9\over 4}$. This then gives the required central charge
$c = {3k\over k-2} -1 = 26$.
As such the model should be thought of as a different classical solution of the
same theory. We have in the earlier lectures seen that the flat
space-time string theory is very nicely and very completely described by a
matrix model. The black hole solution is however markedly different from the
$c=1$ theory. It is characterized by the absence of tachyon
condensation and a nontrivial metric and dilaton field:
$$\eqalign{&T(X) = 0\,\,,\cr
\noalign{\vskip 0.1cm}
&(ds)^2= - {k\over 2} \, {dudv\over M-uv}\,\,\cr
\noalign{\vskip 0.1cm}
&D = \log (M-uv)\,\,. }\eqno\eq$$
Here a particular $SL(2,\IR)$ parametrization is chosen:
$g=\left( {\alpha\atop -v} \,\,{u\atop \beta}\right) $,
$\alpha\beta +uv=1$ and $M$
is the black hole mass.  There is actually a parametrization (related
to the $c=1$ theory) in which the $\sigma$-model Lagrangian reads
$$\eqalign{S_{\rm eff}={1\over 8\pi} \int d^2 z\,\, \Bigl\{
&(\partial X')^2 + (\partial\varphi')^2 - 2\sqrt{2} \phi' R^{(2)} \cr
&+ M \vert{1\over 2\sqrt{2}} \partial\varphi' + i \sqrt{{k\over2}}
\partial X'\vert^2\, e^{-2\sqrt{2} \varphi'} \Bigr\}\,\,.}\eqno\eq$$
This parametrization corresponds to a linear dilaton but
in contrast to the $c=1$ theory, one has a black hole mass term
perturbation represented by a gravitational vertex operator instead of
the cosmological constant term given by a
tachyon operator $e^{-\sqrt{2} \varphi}$.
One of the surprising facts is however that there exists a classical
duality transformation that can be used to relate the two sigma models to
each other [\MaSh]. From this there arises a hope that one could
possibly be able to describe the black hole by a matrix model also.
More generally from a string field theory viewpoint one would hope to
be able to describe different classical solutions in the same setting.
In what follows we will present some joint work done with T. Yoneya
on this subject [\JeYo]. For other different attempts see
[\DDMW, \MuVa].

First insight into the black hole problem is gained by
considering the linearized tachyon [\DVV] field in the external background.
In the conformal field theory  this is given by the zero mode Virasoro
condition. The Virasoro operator $L_0(u,v)$ consists of
two parts,
$$L_0= -\Delta_0 + {1\over 4}\,(u\partial_u-v\partial_v)^2\,\,,
\eqno\eq$$
where $\Delta_0$ is the Casimir operator of $SL(2,\IR)$.
The Virasoro condition for the linear tachyon field (vertex operator)
reads:
$$L_0(u,v)T\equiv {1 \over k-2}\Bigl[(1-uv)\partial_u \partial_v -
{1 \over 2}(u\partial_u + v\partial_v)-{1 \over 2k}
(u\partial_u-v\partial_v)^2 \Bigr]\,T=T\,\,.\eqn\Vircond$$

The on-shell  tachyon corresponds to the continuous
representation of $SL(2,\IR)$ which has eigenvalues
$\Delta_0= -\lambda^2 - {1 \over 4} \, \quad (\lambda = {\rm  real})$
and  $-i\partial_t= 2i\omega$ with the on-shell condition
$\lambda^2 = 9\omega^2$ at $k=9/4$.
The above equation can be interpreted as corresponding to a covariant
Laplacian $L_0=-{1 \over 2e^{D}\sqrt{G}}\,\partial^{\mu}\,e^{D}\,
\sqrt{G}\,G_{\mu\nu}\,\partial_{\nu}$ in the  background space-time metric
$G_{\mu\nu}$ and dilaton $D$,
which can be read off  from Eq.~\Vircond
$$\eqalign{
&ds^2={k-2 \over 2}\,[\,dr^2 - \beta^2(r)\,d\bar t^2\,]\,\,,\cr
\noalign{\vskip 0.15cm}
&D = \log \,\bigl(\sinh {r\over\beta(r)}\,\bigr) + a\,\,,\cr
\noalign{\vskip 0.15cm}
&\beta(r)= 2\,(\coth^2{r \over 2} -{2 \over k})^{-1/2}\,\,.}\eqno\eq$$
These are  then candidates for the
``exact'' background.
Here the new coordinate $r$ and time $\bar t$ are defined by
$$u= \sinh{r \over 2} \,e^{\bar t}\,,\qquad\quad
v= -\sinh{r \over 2} \,e^{-\bar t}\,\,.\eqno\eq$$
These variables describe the static exterior region outside the
event horizon located  at $r=0$. The constant $a$
determines the  mass of the black hole
$$M_{{\rm bh}}= \sqrt{{2 \over k-2}}\,\,e^a\,\,.\eqno\eq$$

The  exact metric can be shown to be free of curvature singularity.
However, one still has a ``dilaton singularity'' at $uv=1$
where the string coupling $g_{\rm st}\sim e^{-D/2}$
diverges. In terms of the variables $u$ and $v$, the dilaton reads
$$D=\log\,\Bigl[4\Bigl(-uv(1-uv)(-{(1-uv) \over uv}-{2 \over k})
\Bigr)^{1/2}\Bigr]+a\,\,,\eqno\eq$$
and the region $uv >1$ corresponds to a disjoint
region with a naked singularity.

The free parameter $a$ can be eliminated by a scale transformation
$$u \rightarrow M^{-1/2}\,u\,,\qquad v \rightarrow M^{-1/2}\,v\,,
\qquad M\equiv e^a\,\,.\eqno\eq$$
This introduces the black hole mass parameter in more explicit way,
where one replaces $(1-uv)$ by $(M-uv)$ in the expressions for the
dilaton and the metric. An important relation is the connection of the
string coupling constant with the parameter
$a$, or rather the black hole mass. In general, the dilaton field
determines the string coupling constant and in the present case one
obtains
$$g_{\rm st}\,(r=0)  \propto e^{-a/2}= M^{-1/2}\,\,.\eqno\eq$$
This is to be compared  with the dependence of
$g_{\rm st} \propto \mu^{-1}$
on the cosmological constant in flat space-time. One notes the
different power which comes from the different scaling dimensions
of the two parameters.
The two backgrounds become identical in the asymptotic region.
Consider the asymptotic  behavior of the Virasoro operator
and the dilaton  when $r\rightarrow \infty$. Using
$u\sim e^{{r\over 2}+\bar t}\,, \,\,
v \sim e^{{r\over 2}-\bar t}$ one finds
$$\eqalign{&L_0\sim {1 \over 4(k-2)}\,(\partial_r^2+\partial_r)+
{1 \over 4k} \,\partial_{\bar t}^2\,\,,\cr
\noalign{\vskip 0.1 cm}
&D\sim r + a - \log 4\,\,.}\eqno\eq$$
This is the form of Virasoro operator in the linear dilaton case,
the parameters $r, \bar t$ are identified asymptotically with the
$\varphi$ and $t$ for the linear dilaton background as
$$\eqalign{\bar t &\leftrightarrow\sqrt{{1 \over 2k}}\,\,t=
{\sqrt{2} \over 3}\,\,t\,\,,\cr
\noalign{\vskip 0.1 cm}
r &\leftrightarrow\sqrt{{2 \over k-2}}\,\,\varphi=2\sqrt{2}\,\varphi\,\,.}
\eqno\eq$$
For the conjugate  momentum and energy, the correspondence is then
$$\eqalign{&ip_{\varphi}= -\sqrt{2}+i2\sqrt{2}\,\lambda = -\sqrt{2} +
{i \over \sqrt{2}}\,\,p_{\tau}\,\,,\cr
\noalign{\vskip 0.1 cm}
&ip = i{2\sqrt{2} \over 3}\,\omega={i \over\sqrt{2}}\,p_{t}\,\,.}
\eqn\enmom$$
This implies a  one-to-one correspondence of tachyon states in the
black hole and linear dilaton backgrounds. There is also a correspondence
between the discrete states spectra in the two theories.

In the Minkowski metric, the spectrum of the
discrete states  for the black hole is isomorphic to that in the linear
dilaton background. In particular,
the first nontrivial discrete state with zero energy
($j=1, m=0$ or $ip_{\varphi}= -2\sqrt{2}, \, p =0$) is
identified with the operator associated with the mass
of black hole, as can be seen from  the first correction to the
asymptotic behavior of the exact space-time metric
$$ds^2 \sim {k-2 \over 2}\,\Bigl[dr^2 - {4k \over k-2}\,\Bigl(1-
{4k \over k-2}\,e^{-r} + {\cal O}(e^{-2r}) \Bigr)dt^2\Bigr]\,\,.
\eqno\eq$$
It is important to note that the $\varphi$ momentum
is twice that of the operator corresponding to
tachyon condensation.

The solutions of the tachyon Virasoro conditions describe the scattering of
a single tachyon on the black hole. It represents
one of the few quantities that has been rigorously computed
in black hole string theory [\DVV]. The amplitude provides some nontrivial
physical insight and is obtained as follows. One writes an  integral
representation for the solution with definite energy $\omega$ and momentum
$\lambda$ as
$$\int_C {dx\over x}\, x^{-2i\omega}\,\bigl(\sqrt{M-uv}
+{u \over x}\bigr)^{-\nu_-}\, \bigl(\sqrt{M-uv} -vx\bigr)^{-\nu_+}\,\,,
\eqn\intrep$$
with $\nu_{\pm}={1 \over 2}-i(\lambda \pm \omega)$. In general, one has
four different contours of integration with two linearly independent
solutions corresponding, for example, to the coutours
$C_2\equiv[-u\sqrt{M-uv}, 0\,],\,
C_4\equiv(-\infty,\nu^{-1}\sqrt{M-uv}\,]$ as
($y\equiv uv =-\sinh^2 {r \over 2}$):
$$\eqalign{&T_{C_2}= U_{\omega}^{\lambda}=e^{-2i\omega \bar t}\,
F_{\omega}^{\lambda} (y)\,\,,\cr
\noalign{\vskip 0.2cm}
&T_{C_4}= V_{\omega}^{\lambda}=e^{-2i\omega \bar t}\,
F_{-\omega}^{\lambda} (y)\,\,,}\eqno\eq$$
where
$$F_{\omega}^{\lambda}(y)= (-y)^{-i\omega}B(\nu_+, \nu_-)
F(\nu_+, \nu_-, 1-2i\omega, y)\,\,.\eqno\eq$$
The asymptotic behaviors of the solutions are,
for $r \rightarrow 0 \, ({\rm horizon})$:
$$\eqalign{&U_{\omega}^{\lambda} \sim \beta(\lambda, \omega)
({u \over \sqrt{M}})^{-2i \omega}\,\,,\cr
\noalign{\vskip 0.1cm}
&V_{\omega}^{\lambda} \sim \beta(\lambda, -\omega)
(-{v \over \sqrt{M}})^{-2i \omega}\,\,}\eqno\eq$$
while for null-infinity $r \rightarrow \infty$:
$$F_{\omega}^{\lambda}\sim \alpha (\lambda, \omega)(-y)
^{-{1 \over 2}+ i\lambda} + \alpha (-\lambda,\omega)(-y)
^{-{1 \over 2}- i\lambda}\,\,,\eqn\asymp$$
where
$$\eqalign{\alpha(\lambda, \omega) &={\Gamma(\nu_+)\,
\Gamma(\bar\nu_- - \nu_+)\over \Gamma(\bar\nu_-)}\,\,,\cr
\noalign{\vskip 0.1cm}
\beta(\lambda, \omega) &= B(\nu_+, \bar \nu_-)\,\,.}\eqno\eq$$
We  see  that $U_{\omega}^{\lambda}$ describes
a wave coming from past null-infinity scattering on
the black hole, while $V_{\omega}^{\lambda}$ describes a wave emitted
by the white hole crossing the past event horizon.
The solution $U_{\omega}^{\lambda}$ gives the $S$-matrix elements of
tachyons, incoming from the asymptotic flat region at past null-infinity and
scattered out to future null-infinity. On-shell
$\omega = 3\lambda \,(>0)$, and this solution gives the reflection
and transmission coefficients as ratios of the coefficients appearing
in the above asymptotic forms:
$$\eqalign{R_B(\lambda)&= {\alpha(\lambda, \omega)\over
\alpha (-\lambda, \omega)}\,\,,\cr
\noalign{\vskip 0.1cm}
T_B(\lambda)&={\beta(\lambda, \omega)\over\alpha (-\lambda,\omega)}\,\,.}
\eqn\rtcoeff$$
The reflection and abosorbtion coefficients satisfy the unitarity relation
$$|R_B|^2 + {\omega \over \lambda} |T_B|^2 = 1\,\,.\eqno\eq$$
This describes the two-point correlation function and it
is of major interest to formulate a full quantum field theory in
the presence of a black hole which would be capable of giving
general $N$-point scattering amplitudes and correlation functions.
One is also very interested in being able to evaluate loop effects
and even to discuss formation and evaporation of black holes in
the general field theoretic framework.
In the absence of a general theory one can try to follow the analogy
with the $c=1$ theory and attempt to guess the structure required for the
black hole. This is what was done in [\JeYo]. Persuing the above
analogy we can postulate again that string theory in the black hole
background is described by
a factorized $S$-matrix. It is then reasonable to expect that the
external leg factors of the full $S$-matrix
are again determined through a non-local field redefinition whose role
is to connect
the Virasoro equations in the black hole background with the free
massless Klein-Gordon equation. The main part of the $S$-matrix is then
to be determined. The suggestion based on the analogy with the $c=1$
theory is that one again has a description  in terms of a
matrix model and the associated  collective field theory.
To simulate the black hole background the matrix model is expected to
include a deformation from the standard inverted oscillator potential.
 There as yet exist no general principles for constructing the theory but
one can make certain concrete suggestions on the eventual form of the
matrix model.
Let us describe first  the expected form for the external leg factors.
These are supplied by a field redefinition whose
purpose is to reduce the black hole background Virasoro condition
to the scalar free field equation. We have summarized the  black hole
Virasoro equation and its solutions in detail, so let us
consider the integral representation \intrep\
with the contour $C_2$ which is appropriate for
the scattering problem in the exterior region $(u>0, v<0)$:
$$U_{\omega}^{\lambda}(u,v)=
\int_{C_2}{dx \over x} x^{-2i\omega}(\sqrt{M-uv}+ {u \over x})^{-\nu_-}
(\sqrt{M-uv}-vx)^{-\nu_+}\,\,.\eqno\eq$$
Since the spectrum of the on-shell solution has a one-to-one
correspondence through \enmom\ with that of the free Klein-Gordon
equation, it is natural to make the following change of integration
variable:
$$\eqalign{&(\sqrt{M-uv} + {u \over x})^{-1}\,(\sqrt{M-uv} - vx)
= e^{-4t /3} \,\,,\cr
\noalign{\vskip 0.1cm}
&(\sqrt{M-uv} + {u \over x})\,(\sqrt{M-uv} - vx)
=e^{-4\tau} \,\,.}\eqno\eq$$
The integral formula for the solution takes  now the form
$$U_{\omega}^{\lambda}
=\int_{-\infty}^{\infty}dt \int_0^{\infty} d\tau\,
\delta({u\,e^{-2t/3}+v\,e^{2t/3} \over 2}-\sqrt{M}\cosh 2\tau)\,
e^{-4i\omega t/3}\cos 4\lambda \tau\,\,.\eqno\eq$$
This is seen to be an integral transform of a  Klein-Gordon plane wave with
momentum and energy
$$p_{\tau}=4\lambda\,,\qquad\quad p_{t}={4\over 3}\,\omega\,\,.
\eqno\eq$$
Since the plane waves are recognized as natural eigenstates of the
linearied collective field  terms of time-of-flight variable
we have the candidate for the non-local field redefinition
$$T(u,v) =
\int_{-\infty}^{\infty}dt \int_0^{\infty} d\tau\,
\delta({u\,e^{-2t/3}+v\,e^{2t/3} \over 2} - \sqrt{M} \cosh 2\tau)
\gamma (i\partial_{t})\partial_{\tau}\eta (t, \tau)\,\,,
\eqn\bhfieldred$$
where $\gamma(i\partial_{t})^* = \gamma (-i\partial_{t})$
is an arbitrary weight function to be fixed by normalization
condition.

In terms of the Fourier decomposition
$$\eta(t,\tau)=\int_{-\infty}^\infty {dp\over p}\,\,
\tilde\eta (p)\,e^{-ipt}\,\sin\,p\tau\,\,,\eqno\eq$$
it reads
$$T(u,v) = \int_{-\infty}^{\infty}dp\,\,
\tilde\eta(p)\,\gamma (p)\,U_{\omega (p)}^{\lambda(p)}(u,v)\,\,,
\eqno\eq$$
with $\omega (p) = 3p/2, \lambda (p) = p/2$. In particular, the
asymptotic behavior for $y\rightarrow \infty$ is
$$\eqalign{T(u,v) = \int_{-\infty}^{\infty}dp\,\,
\tilde\eta(p)\,\gamma(p)
\Bigl[&(-y)^{-{1 \over 2}+i\lambda(p)}\,\alpha\bigl(\lambda (p),\omega(p)
\bigr) +\cr
&(-y)^{-{1 \over 2} - i \lambda (p)}\,\alpha\bigl(-\lambda (p),\omega(p)
\bigr)\Bigr]\,\, e^{-2i\omega(p) t}\,\,.}\eqno\eq$$
This shows that an asymptotic wave packet of $\eta$ field is
transformed into a deformed wave packet of the tachyon field.
The integral transformation that we have given suplies the leg factors of the
conjectured black hole $S$-matrix.
Even if the factorization becomes only an
approximate feature of the full theory one could expect that the
factorization holds near the poles of the $S$-matrix.

Let us now study the possible resonance poles produced by the external leg
factors. It turns out that studying the location of these poles
gives useful and nontrivial constraints on the full $S$-matrix.
 From the asymptotic behavior of \asymp\ and the associated
reflection coefficient \rtcoeff, we see that the positions
of the resonance poles are
$$i4\lambda = i{4 \over 3}\omega=i\sqrt{2}\,p_t =-2,-4,-6,\ldots\,\,.
\eqno\eq$$
This contrasts with the case of the usual $c=1$ model where we
have poles at all negative integers of the corresponding energy.
On the other hand, if we consider an amplitude for
an incoming tachyon with producing $N-1$ outgoing
tachyons, the energy and momentum conservation laws
are satisfied when the energy of incoming tachyon obeys
$$i\sqrt{2}p_t = -(2r + N-2)\,\,,\eqno\eq$$
where $r$ now counts the number of insertions of the
black hole mass operator. The factor $2$ multiplying  $r$
comes about because the momentum carried by the
black hole mass is twice that of the tachyon condensation.
Comparing next the two expressions for the location of the poles
we seee that these are consistent only if $N$ is even. More precisely,
only the even $N=2k$ point amplitudes are to be nonzero while the odd
$N=2k+1$ point amplitudes should vanish. This represents a strong requirement
on the form of the complete theory.

We are than lead to the main problem of specifying the full dynamics
in the form of a generalized matrix model.
In the limit of vanishing black hole mass, the black hole
background reduces to the linear dilaton vacuum.
This is a singular limit in the sense that
the string coupling diverges, corresponding to the $c=1$ matrix
model with vanishing 2d cosmological constant $\mu = 0$,
or zero Fermi energy. Since, according to our
hypothesis, the deformation corresponding to non-vanishing black hole
mass cannot be described by the usual matrix model,
we have to seek for other possible
deformations than the one given by  the
Fermi energy.  We  assume that the Fermi energy is
kept exactly at zero, while the Hamiltonian itself is modified.

{}From the earlier analysis we have several hints or constraints which the
modified Hamiltonian should obey. The first is that there is a double scaling
limit and that the resulting string coupling constant squared should be given
by the black hole mass $M$. The second constraint is the required vanishing
of all odd $N$-point amplitudes. Finally, in agreement with the world sheet
description of the black hole string theory one should have a natural
$SL(2,\IR)$ symmetry.

Consider a general modification of the inverted oscillator Hamiltonian
$$h(p,x) \quad\rightarrow \quad h_M (p, x)={1\over 2}\,(p^2 -  x^2)
+ M \delta h (p, x)\,\,.\eqno\eq$$
We have assumed that the deformation is
described by a term linear in $M$.
The first requirement for $\delta h$ is a scaling property
to ensure that the string coupling is proportional to $M^{-1/2}$.
In collective field theory after a shift by the classical ground state one has
that the string coupling  generally
proportional to $({dx \over d\tau})^{-2}$. Thus the above
requirement is satisfied if the deformation operator
$\delta h$ scales as
$\delta h(p, x) \rightarrow \rho^{-2}\delta h(p,x)$
under scale transformations $(p, x) \rightarrow
(\rho p, \rho x)$. This leads to
$$\delta h(p,x)={1\over 2 x^2}\,\,f({p \over x})\,\,.\eqno\eq$$

To further specify the  general function $f(p/x)$,
one invokes the requirement of $SL(2,\IR)$ symmetry. We have seen
in sect.~4, that the usual $c=1$
Hamiltonian $h= (p^2 - x^2)/2$ allows a set of
eigenoperators $O_{j,m}$ satisfying  the
the $w_{\infty}$ algebra~\winfalg.
The origin of this  algebraic structure, which is supposed to encode
the extended nature of strings, can be traced to  existence
of an $SL(2,\IR)$ algebra consisting of
$$\eqalign{&L_1={1 \over 4}\,(p^2-x^2)= h(p,x)\,\,,\cr
&L_2= -{1 \over 4}\,(px + xp)\,\,,\cr
&L_3= {1 \over 4}\,(p^2+x^2)\,\,.}\eqno\eq$$
The  eigenoperators satisfying the $w_{\infty}$
algebra are constructed in terms of the  $SL(2,\IR)$ operators according to
$$O_{j,m}=L_+^{{j+m \over 2}}\,L_-^{{j-m \over 2}}\,,
\qquad L_{\pm}=L_3 \pm L_2\,\,,\eqno\eq$$
which close under the Poisson bracket since the Casimir invariant has
a fixed value
$$L_1^2 + L_2^2 - L_3^2= {3\hbar \over 16}\,\,.\eqno\eq$$
(the Planck constant indicates the effect of operator ordering).

Since the spectrum of  discrete states in the
black hole background is expected to be the same
as that of the usual $c=1$ model in the Minkowski metric,
it is natural to require that the deformed model
should also share a similar algebraic structure.

There is actually a very simple model with the above structure. It is given
for $f=1$ in which case one has the extra term represented by a well known
singular potential. One has  the $SL(2,\IR)$ generators of the form:
$$\eqalign{&L_1(M)= {1\over 2}\,h_M (p,x)
={1 \over 4}\,(p^2-x^2+{M \over x^2})\,\,,\cr
&L_2(M)= -{1 \over 4}\,(px + xp)\,\,,\cr
&L_3(M)= {1 \over 4}\,(p^2 + x^2 + {M \over x^2})\,\,,}\eqno\eq$$
which satisfy
$$L_1(M)+L^2_2(M)-L^2_3(M)= -{M\over 2}+{3\hbar\over 16}\,\,.\eqno\eq$$
We  note that because of  different constraint
for the Casimir invariant  the  algebra of eigenoperators is now
modified in an $M$-dependent way. The algebraic properties of the
model with the singular potential have been investigated in detail
in [\AJcmp].
The model is exactly solvable and possesses some features characteristic of
black hole background.

Let us proceed to describe the properties of the  deformed model:
$$h_M(p,x) = {1 \over 2}(p^2 - x^2) + {M \over 2 x^2}\,\,.
\eqn\defham$$
We assume here that $M>0$. Then the genus zero free energy in the limit of
vanishing scaling parameter, $\bar M \rightarrow 0$, behaves like
$F\sim {N^2\over 8\pi\sqrt{2}}\,\bar M \log {\bar M \over \sqrt{2}}$.
The double scaling  limit is thus the limit $\bar M \rightarrow 0,
\,N\rightarrow \infty$
with $M \equiv N^2 \bar M$ being kept fixed. After  the
usual rescaling, $\,x\equiv \sqrt{N}\times$  matrix eigenvalue,
the system is reduced to the free fermion system with the one-body
potential $-{1 \over 2}x^2 + {M \over 2x^2}$.
Note that in the limit $M \rightarrow 0$ the potential
approaches the usual inverted harmonic oscillator potential with
a repulsive $\delta$-function-like singularity.

The solution of the classical equations
with energy $\epsilon$ reads
$$x^2(t) =-\epsilon +  \sqrt{M + \epsilon^2} \cosh 2t\,\,.\eqno\eq$$
The ground state corresponding to zero Fermi energy is obtained
by setting $\epsilon =0$ and replacing the time variable $t$
by the time-of-flight coordinate $\tau$, $x^2=\sqrt{M}\cosh 2\tau$.
This is recognized as  precisely the quantity appearing in
the integral transformation \bhfieldred. The
$\delta$-function present in the transformation gives a relation
between the black hole and the matrix model variables.
It serves to identify the matrix eigenvalue as
$$x^2=(u e^{-2t/3}+v e^{2t/3})/2\,\,.$$
The string coupling is now space dependent
$$g(\tau) \equiv {\sqrt{\pi} \over 12}\,\Bigl({dx \over
d\tau}\Bigr)^{-2} = {1 \over 48}\sqrt{{\pi \over M}}\,
\Bigl({1 \over \sinh^2\tau } +{1 \over \cosh^2 \tau}\Bigr)\,\,,
\eqno\eq$$
with the required relation with the black hole mass and
the asymptotic behavior at large $\tau$.

The tree  level scattering amplitudes are generally obtained
from the exact solution of the classical equations. The exact
solution to the collective  equations  has the following
parametrized  form
$$\eqalign{&x(t,\sigma)=\Bigl[-a(\sigma)+\sqrt{M+a^2(\sigma)}\,
\cosh 2(\sigma-t)\Bigr]^{1/2}\,\,,\cr
&\alpha(t,\sigma)={1 \over x(t,\sigma)}\,\sqrt{M+a^2(\sigma)}\,
\sinh 2(\sigma-t)\,\,.}\eqn\bhparasol$$
It contains an arbitrary function
$a(\sigma)$ describing the deviation of the
Fermi surface from its ground state form.
The asymptotic behavior for large $x$, of the profile
function reads
$$\alpha_{\pm}(t,\tau)=\pm x(\tau)\Bigl(1- {\psi_{\pm}
(t\pm \tau) \over x^2(\tau)}\Bigr) + {\cal O}({1 \over x^2})
\,\,.\eqno\eq$$
The functions $\psi_{\pm}(t\pm \tau)$
represent incoming and outgoing waves, respectively.
In terms of the $\eta$ field, we have
$$(\partial_{t}\pm \partial_{\tau})\eta =
\pm {1 \over \sqrt{\pi}}\,\,\psi_{\pm}(t\pm \tau)\eqno\eq$$
for $t \rightarrow \mp \infty$.

A nonlinear relation between incoming and outgoing fields
$\psi_+$ and $\psi_-$ can be established
by studying the time delay. Take the times at which a parametrized
point $\sigma$ is passed by the incoming and outgoing waves at a fixed
value of large $\tau$ be $t_1 (\rightarrow -\infty)$ and
$t_2 (\rightarrow \infty)$, respectively.
{}From \bhparasol\ we have then
$$\eqalign{\bigl(&M+a^2(\sigma)\bigr)^{1/4}\,e^{\sigma-t_1}=
M^{1/4}\, e^{\tau}\,\,,\cr
\big(&M+a^2(\sigma)\bigr)^{1/4}\,e^{t_2-\sigma}=M^{1/4}\,e^{\tau}\,\,.}
\eqno\eq$$
This implies
$$t_1+\tau=t_2 -\tau +{1\over 2}\log\Bigl(1+{a^2(\sigma)\over M}
\Bigr)\,\,,\eqno\eq$$
and hence
$$a(\sigma) = \psi_+(t_1 +\tau) = \psi_-(t_2 -\tau)\,\,.\eqno\eq$$
This then gives  functional scattering equations connecting
the incoming and outgoing waves
$$\psi_{\pm}(z)= \psi_{\mp}\Bigl(z \mp {1 \over 2}\log\,
(1+ {1 \over M}\,\psi^2_{\pm}(z))\Bigr)\,\,. \eqn\scatt$$
The result is  similar  in form to that of the usual $c=1$ model.
However, one notes a crucial difference that
Eq.~\scatt\ is even, i.e. it is invariant under
the change of sign of $\psi_{\pm}\rightarrow -\psi_{\pm}$.
This  ensures that the number of particles participating in
the scattering is even. All the odd point amplitudes do vanish in
the deformed model.

The explicit power series solution of \scatt\ is
$$\psi_{\pm}(z) = \sum_{p=0}^{\infty}{M^{-p} \over p!\,(2p+1)}\,\,
{\Gamma(1\pm {1 \over 2}\partial_z) \over \Gamma(1-p \pm
{1 \over 2}\partial_z)}\,\,\psi^{2p+1}_{\mp}(z)\,\,,\eqno\eq$$
which shows that the amplitudes are essentially polynomial with respect
to the momenta without any singularity.

The scattering equation \scatt\ can also be derived using directly the
exact states [\DeRo,\Da], as was done in sect.~5 for the $c=1$ model.
First, one recalls the symmetry structure of the collective
theory with Hamiltonian~\defham\ given in [\AJcmp]:
$$\eqalign{
\bigl[ O_{j_1,m_1}^{a_1},  O_{j_2,m_2}^{a_2} \bigr] =
&-4i(j_1 m_2 - m_1 j_2)\,O_{j_1+j_2-2, m_1+m_2}^{a_1+a_2+1}\cr
\noalign{\vskip 0.2cm}
&-4i(a_1 m_2 - m_1 a_2)\,O_{j_1+j_2, m_1+m_2}^{a_1+a_2-1}\,\,,}\eqno\eq$$
where
$$\eqalign{
O_{j,m}^{a} \equiv \int {dx\over 2\pi} \int_{\alpha_-}^{\alpha_+}
d\alpha &\Bigl( \alpha^2 - x^2 +{M\over x^2} \Bigr)^a
\Bigl( (\alpha + x)^2 +{M\over x^2} \Bigr)^{{j+m\over 2}}\cr
\noalign{\vskip 0.1cm}
&\Bigl( (\alpha - x)^2 +{M\over x^2} \Bigr)^{{j-m\over 2}}\,\,.}\eqn\op$$
The operators $T^{(-)}_{-ip}\, (T^{(+)}_{ip})$
which create exact tachyon in (out) states
are obtained by analytic continuation
$j\to \pm ip/2$ of some special operators \op:
$$\eqalign{
&O_{j,j}^{a=0} = \int  {dx\over 2\pi} \int_{\alpha_-}^{\alpha_+}
d\alpha
\Bigl[ (\alpha +x)^2 +{M\over x^2}\Bigr]^j \equiv T_{2j}^{(+)}\,\,, \cr
\noalign{\vskip 0.2truecm}
&O_{j,-j}^{a=0} = \int  {dx\over 2\pi} \int_{\alpha_-}^{\alpha_+}
d\alpha \Bigl[ (\alpha - x)^2 + {M\over x^2}\Bigr]^j
\equiv T_{2j}^{(-)}\,\,.}\eqno\eq$$
Eq.~\scatt\ then easily follows from an asymptotic expansion of
$$T^{(+)}_{ip,+} = -T^{(+)}_{ip,-}\,\,,$$
for large $\tau$, where $T^{(+)}_{ip,+}$ and $T^{(+)}_{ip,-}$
are defined by
$$T^{(+)}_{ip}=T^{(+)}_{ip,+}-T^{(+)}_{ip,-}\,\,.\eqno\eq$$

The scattering equation
can also be rewritten in terms of  energy-momentum tensor
$$T_{\pm\pm}(z) = {1 \over 2\pi}\, \psi^2_{\pm}(z)\,\,,\eqno\eq$$
as
$$\int dz\,\,e^{i\omega z} \,T_{\pm\pm}(z)
={M \over 2\pi}\int dz \,\,e^{i\omega z}
{1 \over 1 \pm {i\omega \over 2}}\,
\Bigl[\Bigl(1+{2\pi\over M}\,T_{\mp\mp}(z)\Bigr)^{1\pm{i\omega \over 2}}
-1\Bigr]\,\,.\eqno\eq$$
One can easily check that this defines a canonical
transformation by confirming that the Virasoro algebra
(at the level of Poisson bracket) is preserved by this transformation.
This relation for  the energy momentum tensor is very
similar to the one obtained recently by Verlinde and Verlinde
[\VV] for the $S$-matrix of the $N=24$ dilaton gravity.A slight difference is
that in the case of dilaton gravity one has the derivative of the energy
momentum tensor participating in the equation.

In conclusion, the framework presented above gives some initial
picture of a black hole in the matrix model. It contains some basic
requirements for a consistent formalism. In particular, the scaling
properties of the black hole mass deformation are in agreement with
the corresponding vertex operators (see also [\Eguchi]). The
particular singular matrix model studied has an interesting double
scaling limit with an $SL(2,\IR)$ algebraic structure. This clearly
is not enough to completely describe black hole and further
generalizations and studies are likely to lead to further interesting
results.

\noindent
{\bf Acknowledgement}

These notes were written while the author was visiting LPTHE, Paris 6,
Paris, France.
He is grateful to the members of the high energy group for
their hospitality.

\refout

\end

======================================================================== 2200
Return-Path: <@BROWNVM.BROWN.EDU:kresimir@PUHEP1.PRINCETON.EDU>
Received: from BROWNVM (NJE origin SMTP@BROWNVM) by BROWNVM.BROWN.EDU (LMail
          V1.1d/1.7f) with BSMTP id 1733 for <maryannr@BROWNVM>; Mon,
          20 Sep 1993 12:40:58 -0400
Received: from Princeton.EDU by BROWNVM.brown.edu (IBM VM SMTP V2R2) with TCP;
   Mon, 20 Sep 93 12:40:16 EDT
Received: from puhep1.Princeton.EDU by Princeton.EDU (5.65b/2.98/princeton)
        id AA05296; Mon, 20 Sep 93 12:42:02 -0400
Received: by puhep1.Princeton.EDU (5.52/1.113)
        id AA03431; Mon, 20 Sep 93 12:42:01 EDT
Date: Mon, 20 Sep 93 12:42:01 EDT
{}From: "Kresimir Demeterfi" <kresimir@puhep1.Princeton.EDU>
Message-Id: <9309201642.AA03431@puhep1.Princeton.EDU>
To: maryannr@brownvm.brown.edu

\input phyzzx
\nopagenumbers

%
\def\refout{\par\penalty-400\vskip\chapterskip
   \spacecheck\referenceminspace
   \ifreferenceopen \Closeout\referencewrite \referenceopenfalse \fi
   \noindent{\bf References\hfil}\vskip\headskip
   \input \jobname.refs }
\def\chapter#1{\par \penalty-300
   \chapterreset \noindent{\bf \chapterlabel.~~#1}
   \nobreak \penalty 30000 }

\def\IR{\relax{\rm I\kern-.18em R}}

\def\npb#1#2#3{{\it Nucl. Phys.} {\bf B#1} (#2) #3 }
\def\plb#1#2#3{{\it Phys. Lett.} {\bf B#1} (#2) #3 }
\def\prd#1#2#3{{\it Phys. Rev. } {\bf D#1} (#2) #3 }
\def\prl#1#2#3{{\it Phys. Rev. Lett.} {\bf #1} (#2) #3 }
\def\mpla#1#2#3{{\it Mod. Phys. Lett.} {\bf A#1} (#2) #3 }
\def\ijmpa#1#2#3{{\it Int. J. Mod. Phys.} {\bf A#1} (#2) #3 }

\def\cmp#1#2#3{{\it Commun. Math. Phys.} {\bf #1} (#2) #3 }

\def\ptp#1#2#3{{\it Prog. Theor. Phys.} {\bf #1} (#2) #3 }
\def\bb#1{{\tt hep-th/#1}}
\REF\Klebreview{I. R. Klebanov, {\it ``String theory in two
dimensions'',} in ``String Theory and Quantum Gravity'',
Proceedings of the Trieste Spring School 1991, eds. J. Harvey et al.,
(World Scientific, Singapore, 1992).}
\REF\Kutreview {D. Kutasov, {\it ``Some properties of (non) critical
strings'',} in ``String Theory and Quantum Gravity'',  Proceedings of
the Trieste Spring School 1991, eds. J. Harvey et al., (World Scientific,
Singapore, 1992).}
\REF\Polone {J. Polchinski, \npb {346}{1990}{253.}}
\REF\WBH {E. Witten, \prd {44} {1991} {314.}}
\REF\BaNe {I. Bars and B. Nemeschansky, \npb {348} {1991} {89.}}
\REF\RSS {M. Ro\v cek, K. Schoutens and A. Sevrin, \plb {265} {1991}
{303.}}
\REF\MaWa {G. Mandal, A. Sengupta and S. Wadia, \mpla {6}{1991}{1685.}}
\REF\El {S. Elizur, A. Forge and E. Rabinovici, \npb {359}{1991}{581.}}
\REF\Po{A.~M. Polyakov, \mpla {6}{1991}{635.}}
\REF\GoLi {M. Goulian and M. Li, \prl {66}{1990}{2051.}}
\REF\DiFrKu{P. DiFrancesco and D. Kutasov, \plb {261}{1991}{385;}
\npb {375}{1991}{119;} Y. Kitazawa, \plb {265}{1991}{262;}
Y. Tanii, \ptp {86}{1991}{547;}
V.~S. Dotsenko, \mpla {6}{1991}{3601.}}
\REF\Bab{O. Babelon, \plb {215}{1988}{523.}}
\REF\Ge {J.-L. Gervais, \cmp {130}{1990}{257;} \npb {391}{1993}{287.}}
\REF\JeSa {A. Jevicki and B. Sakita, \npb {165} {1980} {511.}}
\REF\DaJe {S.~R. Das and A. Jevicki, \mpla {5} {1990} {1639.}}
\REF\DJR {K. Demeterfi, A. Jevicki and J.~P. Rodrigues, \npb {362}{1991}
{173;}  \npb {365} {1991} {499.}}
\REF\Poltwo { J. Polchinski, \npb {362} {1991} {125.}}

\REF\AvJe {J. Avan and A. Jevicki, \plb {266}{1991}{35;}
\plb {272}{1991}{17.}}
\REF\MoSe {G. Moore and N. Seiberg, \ijmpa{7}{1992}{2601.}}
\REF\GKN {D.~J. Gross, I.~R. Klebanov and M. Newman,
\npb {350} {1991} {671.}}
\REF\LM {J. Lee and P.~F. Mende, \plb {312} {1993} {433.}}
\REF\winf {D. Minic, J. Polchinski and Z. Yang, \npb {369}{1992}{324;}}
\REF\WiGR {E. Witten, \npb {373}{1992}{187.}}
\REF\KlPo {I.~R. Klebanov and A.~M. Polyakov, \mpla {6}{1991}{3273.}}
\REF\JRvT {A. Jevicki, J.~P. Rodrigues and A. van Tonder,
\npb {404} {1993} {91.}}
\REF\Kl {I.~R. Klebanov, \mpla {7}{1992}{723.}}
\REF\ZW {E.Witten and B.Zwiebach, \npb {377}{1992}{55.}}
\REF\Ve {E. Verlinde, \npb {381}{1992}{141.}}
\REF\KlPa{I.~R. Klebanov and A. Pasquinucci, \npb {393}{1993}{261.}}
\REF\Ba {J.~L.~F. Barbon, \ijmpa {7}{1992}{7579;}
Y. Kazama and H. Nicolai, {\it ``On the exact operator formalism of
two-dimensional Liouville quantum gravity in Minkowski space-time,''}
DESY-93-043, \bb{9305023};
V.~S. Dotsenko, {\it ``Remarks on the physical states and the chiral
algebra of 2D gravity coupled to $c\le 1$ matter,''}
PAR-LPTHE-92-4, \bb{9201077}; \mpla {7}{1992}{2505.}}
\REF\GrKl {D.~J. Gross and I.~R. Klebanov, \npb {359} {1991} {3.}}
\REF\MoPl {G. Moore and R. Plesser, \prd {46} {1992} {1730.}}
\REF\MaSh {E. Martinec and S. Shatashvili, \npb {368}{1992}{338.}}
\REF\JeYo {A. Jevicki and T. Yoneya, {\it ``A deformed matrix model and
the black hole background in two-dimensional string theory'',}
NSF-ITP-93-67, BROWN-HEP-904, UT-KOMABA/93-10, \bb{9305109}.}
\REF\DDMW {S.~R. Das, \mpla {8} {1993} {69;}
A. Dhar, G. Mandal and S. Wadia, \mpla {7} {1992} {370.}}
\REF\MuVa {S. Mukhi and C. Vafa, {\it ``Two-dimensional black hole
as a topological coset model of $c=1$ string theory,''}
HUTP-93/A002, TIRF/TH/93-01, \bb{9301083}.}
\REF\DVV {R. Dijkgraaf, H. Verlinde and E. Verlinde,
\npb {371} {1992} {269.}}
\REF\new{M. Bershadsky and D. Kutasov, \plb {266}{1991}{345;}
T. Eguchi, H. Kanno and S.-K. Yang, \plb {298}{1993}{73;}
H. Ishikawa and M. Kato, \plb {302}{1993}{209.}}
\REF\AJcmp {J. Avan and A. Jevicki, \cmp {150}{1992}{149.}}
\REF\DeRo {K. Demeterfi and J.~P. Rodrigues, {\it ``States and
quantum effects in the collective field theory of a deformed matrix
model'',} PUPT-1407, CNLS-93-06, \bb {9306141};
K. Demeterfi, I.~R. Klebanov and J.~P. Rodrigues,
{\it The exact $S$-matrix of the deformed $c=1$ matrix model,''}
PUPT-1416, CNLS-93-09, \bb {9308036}.}
\REF\Da{U. Danielsson, {\it ``A matrix-model black hole,''}
CERN-TH.6916/93, \bb{9306063}.}
\REF\VV {E. Verlinde and H. Verlinde, {\it ``A unitary $S$-matrix
and 2D black hole formation and evaporation'',} IASSNS-HEP-93/18,
PUPT-1380, \bb {9302022}.}
\REF\Eguchi{T. Eguchi, {\it ``$c=1$ Liouville theory perturbed by the
black-hole mass operator,''} UT 650, \bb{9307185}.}

\singlespace
\hsize=6.0in
\vsize=8.5in
\voffset=0.0in
\hoffset=0.0in
\overfullrule=0pt

\line{}
\line{\hfill HET-918}
\line{\hfill TA-502}
\vskip .75in
\centerline{{\bf DEVELOPMENTS IN 2D STRING THEORY}}
\vskip .40in
\centerline{ANTAL JEVICKI}
\smallskip
\centerline{{\it Physics Department, Brown University}}
\centerline{{\it Providence, Rhode Island 02912, USA}}
\vskip .25in
\centerline{(Lectures presented at the Spring School }
\centerline{in String Theory, Trieste, Italy, April, 1993)}

\vskip 0.70in

{\chapter{Introduction}}
\medskip

Recent years have witnessed a remarkable progress in 2d  string theory
and quantum gravity. Beginning with matrix models one found a new and
computationally powerful description of the theory, free of mathematical
complexities. The relevance of these models to
string theory comes through a $1/N$ expansion where $1/N$ plays the role of a
bare string coupling constant $g_{\rm st}^0 = 1/N$.  This classifies
Feynman diagrams according to their topology; for a fixed topology the
sum of all graphs in a dual picture becomes a sum of triangulated
surfaces.  The continuum theory is then approached by sending
the value of lattice spacing to zero.

This heuristic picture was completely carried out in one dimension giving
an exactly solvable theory of two-dimensional strings
(for an earlier review see [\Klebreview]).
It lead first to a series of explicit results including
the computation of free energy and correlation functions at any order in
the loop expansion. The new formulation also offered a framework for
non-perturbative investigations. It provided  a new fundamental insight into
the origin of metric fluctuations and the physical nature of the Liouville
mode. Through a critical scaling limit a two-dimensional theory is
generated where the logarithmic scaling violation is seen to be the origin of
the extra dimension.

Most of the interesting features of 2d strings were
clearly exhibited in the field-theoretic description achieved in
terms of  collective field theory. Starting from matrix models one
builds a field theory describing the dynamics of observable (Wilson) loop
variables. The collective Hamiltonian describes the processes of
joining and splitting of loops, giving
A cubic interaction and a linear (tadpole) term were shown to
successfully produce all tree and loop diagrams. The theory is naturally
integrable and exactly solvable. Its
integrable nature  leads to understanding of a
$w_\infty$ algebra as a space-time symmetry of the theory. This algebra acts
in a nonlinear way on the basic collective field representing the tachyon.
It is interpreted as a spectrum-generating algebra allowing to build
an infinite sequence of discrete imaginary energy states which turn
out to be remnants of higher string modes in two dimensions.
The presence and interplay of discrete modes with the scalar tachyon are
particularly interesting. The $w_\infty $ symmetry is seen to serve as
an organizational principle specifying the dynamics.

Two-dimensional physics is made even richer by the existence of other
nontrivial backgrounds. Most interesting is the black hole type classical
solution described by an exact $SL(2,\IR)/U(1)$ sigma model. Its quantum
mechanical interpretation is of major interest and was the object of
various recent studies.

Even though there is a wealth of results coming from detailed studies of
matrix models and conformal field theories a full understanding of the theory
and its dynamics is still not available. In particular, a clear correspondence
between the two fundamentally different methods is lacking. One has an
(excellent) comparison of results and a pattern of similarities and analogies
hinting at a more unified  framework. Prospects for such a framework are
particularly exciting since this would eventually represent a
new formulation of string field theory.

A need for such a general framework is most clear already when
addressing the question of the black hole. In general one would like
to command sufficient insight to be able to go from one solution to
another. This, at present, is also one of the fundamental
challenges of string theory.

In this series of lectures we describe the progress already achieved.
The emphasis is on a unified understanding of the subject.
We will try to bridge the two major approaches: the matrix model
and conformal field theory, as much as possible describing analogies
and similarities that one has between them.  In this process a
dictionary  emerges; it is most visible in the discussion
of the infinite $w_\infty $ symmetry and the associated Ward identities.
The question of incorporating the black hole background is then addressed
and some preliminary results in this direction are described.

The selection of topics covered is as follows: In sect.~2 we give a
summary of basic two-dimensional string theory (for a more detailed
review see [\Kutreview]). In sect.~3 we describe the matrix model
and a transition to field theory. We discuss the integrability of
the theory and the construction of exact states
and their string interpretation. In sect.~4 the corresponding $w_\infty$
symmetry  is described.  A detailed comparison of Ward identities and a
description of the agreement between matrix
model and conformal field constructions is given. Sect.~5 contains
the discussion of the $S$-matrix of the theory. The latter is described by
an exact generating function, connection of which to matrix model
harmonic oscillator states we emphasize. In sect.~6 we discuss the
black hole background.

\bigskip
\chapter {String Theory in Two Dimensions}
\medskip

The conceptually simplest way to discuss the dynamics of strings
is through a $\beta$-function approach which provides
effective equations for low-lying fields . In the case of a closed
string in two dimensions these are the $m^2 = 0$ scalar $T(X^{\mu}$) (the
would-be tachyon), the graviton $G_{\mu\nu}(X)$ and the dilaton $D(X)$.
The leading $\beta$-function Lagrangian reads:
$$S_{\rm eff} = {1\over 2\pi} \int d^2 X \sqrt{G} \, e^{-2D(X)}
\Bigl\{ {1\over 2} \left[ \nabla_{\mu} T \nabla^{\mu} T + 2T^2 - V\right] +
R + 4 \nabla D\cdot\nabla D + \ldots\Bigr\}\,\,. \eqno\eq$$
The tachyon potential $V(T)$ is not so well known and neither are the
couplings to possibly higher--spin fields.  But this effective Lagrangian
exhibits several simple solutions which can serve as classical
configurations of two-dimensional string theory.

Denoting $X^{\mu} \equiv (X^0 = t , X^1= \varphi)$ one
has the {\it linear dilaton vacuum} solution
$$\eqalign{&T(X)  = 0\,\,,\cr
\noalign{\vskip 0.1cm}
&G_{\mu\nu}(X)  = \eta_{\mu\nu}\,\,,\cr
\noalign{\vskip 0.1cm}
&D(X) = - \sqrt{2}\, \varphi\,\,. }\eqno\eq$$
The scalar (tachyon) effective Lagrangian in this linear dilaton background
reads
$$S_{\rm eff} (T)={1\over 2}\int d^2 X\,e^{2\sqrt{2}\varphi}\,\,\Bigl\{
{1\over 2} \,T \left( - \partial_t^2 + \partial_{\varphi}^2 + 2\sqrt{2}
\partial_{\varphi} + 2 \right) T-V\Bigr\}\,\,.$$
Rescaling the scalar fields
$$e^{\sqrt{2}\varphi}\,\,T (t,\varphi )=\tilde{T} (t,\varphi)\eqno\eq$$
yields a massless theory
$$S = {1\over 2} \int dt d\varphi \Bigl\{ {1\over 2}\,\tilde{T}
\left( - \partial_t^2 + \partial_{\varphi}^2 \right) \tilde{T} -
{e^{-\sqrt{2}\varphi}\over 3!}\,\tilde{T}^3 + \ldots
\Bigr\}\,\,,\eqno\eq$$
with a spatially dependent string coupling constant
$$g_{\rm st} (\varphi) = e^{-\sqrt{2} \varphi}\eqno\eq$$
(we have taken for simplicity a cubic interaction).

This coupling grows and becomes infinite at $\varphi \rightarrow -
\infty$. This is usually taken as a signal that the linear dilaton
vacuum should be modified (at least in the region $\varphi \rightarrow
- \infty$).  Indeed the linearized static tachyon equation
$$\left( \partial_{\varphi}^2 + 2 \sqrt{2} \partial_{\varphi}
+ 2\right) T_0 (\varphi ) = 0\eqno\eq$$
already has two linearly independent solutions $T_0 (\varphi)=
e^{- \sqrt{2} \varphi}, \varphi e^{-\sqrt{2}\varphi}$.
This would imply that the correct vacuum is given by a tachyon
condensate [\Polone].  An (incomplete) analysis indicates that this
vacuum is then described by a $c=1$ conformal field theory coupled
to a Liouville field:
$${\cal L}={1\over 8\pi}\int d^2z\,\Bigl(\partial X \bar{\partial} X
+ \partial \varphi \bar{\partial} \varphi - 2 \sqrt{2} \varphi
(z,\bar{z}) R^{(2)}+\mu \, e^{-\sqrt{2} \varphi (z,\bar z)}\Bigl)\,\,.
\eqno\eq$$
Here the central charge $c_{X} = 1$ refers to the (matter) coordinate
$X(z,\bar{z})$ while the Liouville field with $Q = 2\sqrt{2}$ carries a
central charge $c_{\varphi} = 1 + 3Q^2 = 25$ leading to the required total of
$c=c_{X}+c_{\varphi}=26$. It is very interesting that in two dimensions
one  has another conformally invariant background, the WZW
$SL(2,\IR)/U(1)$ sigma model representing a black hole
(BH) [\WBH--\El].
Its physical properties are of major
interest as is the general question of describing different string theory
backgrounds in a single field-theoretic framework.

The presence of the cosmological term in the Liouville  theory (and of
the mass term in the black hole conformal field theory) leads to
computational difficulties when evaluating the correlation
functions (these actually become quite untractable
for the BH case).  It is a remarkable fact that the matrix model
formulation succeeds in handling  the first problem with ease and has some
promise for addressing the second as well.

The spectrum of states is usually obtained by neglecting the nonlinear
terms $\mu = 0$ (or $M=0$ for the black hole) in which case one has a
free field representation for the Virasoro generators.  In the above limit
the spectra of two theories are the same.  They consist
of a massless tachyon and an infinite sequence of discrete states.

We begin with the zero mode or tachyon states:
$$\eqalign{ & (L_0 - 1) \,\, V_{k,\beta} = 0 \,\,,\cr
\noalign {\vskip 0.1cm}
& L_0 = {1\over 2} \Bigl(  {\partial^2\over\partial X^2} +
{\partial^2\over\partial\varphi^2} + Q
{\partial\over\partial\varphi}\Bigr)\,\,,}\eqno\eq$$
with two branches of solutions
$$V_{\pm}= e^{ikX + \beta_{\pm} \varphi}\,\,,\qquad\quad
\beta_{\pm} = - \sqrt{2} \pm \vert k \vert\,\,,\eqno\eq$$
following from the on-shell condition
$$k^2 - \beta (Q+\beta ) = 0\,\,.\eqno\eq$$
Here we have taken an Euclidean (space) signature for $X$ and
$\varphi$ which is a convention in conformal field theory discussions.
One can take $X$ to be the space variable and $\varphi$ to be the
(Euclidean) time variable.
It will be more physical, and from the matrix model viewpoint more
natural, to treat $\varphi$ as a space coordinate and continue $X$ to
Minkowski time:
$$X \rightarrow - it\,\,,\qquad\quad
k \rightarrow i p\,\,.\eqno\eq$$

In the context of full Liouville
theory, the second branch with $\beta_- = - \sqrt{2} -\vert k\vert$ has a
questionable meaning since the wave functions grow at $\varphi
\rightarrow - \infty$ which is the location of the infinitely high
Liouville wall $\mu e^{-\sqrt{2} \varphi}$.  These vertex operators
are termed ``wrongly" dressed.  Operators with positive Liouville
dressing have a clear meaning.  Depending on the sign of the momentum,
$\pm =$ sign $k$, these are either right- or left-moving waves.  It is
sensible to use them to compute scattering processes and denote them
as
$$T_k^{\pm} = e^{ikX+(-\sqrt{2} \pm k)\varphi}\,,\qquad
\pm = {\rm sign} \, \, k\,\,. \eqno\eq$$
The Minkowskian continuation is $k = \pm ip$ and
$$\eqalign { T_p^+ &=e^{i p(t + \varphi )} e^{-\sqrt{2} \varphi}\,\,,\cr
\noalign {\vskip 0.1cm}
T_p^- &= e^{-ip(t-\varphi )} e^{-\sqrt{2}\varphi}\,\,, }\eqno\eq$$
for $p>0$ describe left-  and right-moving waves, respectively.

In addition one has an infinite sequence of nontrivial discrete
states specified by discrete (imaginary) values of energy and Liouville
momenta [\Po]:
$$ip_{\varphi}= -\sqrt{2} (1-j)\,,\qquad
ip = \sqrt{2} m \,\,,\eqno\eq$$
with $j = 0, {1\over 2} , 1, \ldots$ and $m = -j, \ldots , j$.
Clearly the states with $m = j$ and $m=-j$ are just special tachyon
states.  The simplest way then to reach the other states is to use the
$SU(2)$ generators as raising and lowering operators on the $m=\pm j$
tachyon states.  The $SU(2)$ generators are given by
$$\eqalign{ t_+ & = e^{i\sqrt{2} X (z)}\,\,,\cr
t_- & = e^{-i\sqrt{2} X(z)} \,\,,\cr
t_3 & = i\sqrt{2} \, \partial X (z)\,\,.}\eqno\eq$$
Denoting now the highest weight state as
$$W_{jj}^{(+)}= e^{i\sqrt{2}\,jX (z)}\,e^{-\sqrt{2}\,(1-j) \varphi
(z)}\eqno\eq$$
one gets the vertex operator for  general discrete states
$$W_{jm}^{(+)} = \left( \oint d\omega e^{-i\sqrt{2} X (\omega ) }
\right)^{j-m} W_{jj}^{(+)}\,\,,\quad -j\leq m\leq + j\,\,.
\eqno\eq$$
These can also be found in the Fock  space where they solve the Virasoro
conditions of the $c=1$ theory
$$\eqalign{ &(L_0 - 1) \vert jm \rangle =0\,\,,\cr
\noalign {\vskip 0.1cm}
&L_n \vert jm \rangle =0\,\,,\cr
&\vert jm \rangle = \int dz \, W_{jm}(z)\vert 0\rangle\,\,.}\eqno\eq$$
One also has operators with the opposite (negative) Liouville dressing
$$W_{jm}=V_{jm}(X) \,e^{-\sqrt{2} (1+j) \varphi (z)}\,\,,\eqno\eq$$
whose physical meaning is again questionable in the full Liouville
theory.  These states, however, turn out to play an important role as black
hole mass perturbations.

Evaluation of correlation functions in the continuum approach is
rather nontrivial and often relies on a number of educated guesses
involving various analytic continuations.  The problem lies
in the nontrivial Liouville potential term.  By separating and
integrating out the zero mode $\varphi (z,\bar{z} ) = \varphi_0 +
\tilde{\varphi}$ one finds, through a functional integral formulation,
the representation
$$ \langle\,\prod_{i=1}^{N} \, T_i \,\rangle =
\left( {\mu\over\pi}\right)^s \Gamma (-s)\,\bigl\langle
\Bigl( \prod_i T_i \Bigr)
\Bigl( \int d^2 z \,e^{\sqrt{2}\tilde{\varphi}}\Bigr)^s\,
\bigr\rangle_{\mu=0}\,\,.\eqno\eq$$
Here the $\Gamma$-function is a result of $\varphi_0$ integration
$$\int d\varphi_0 \, e^{Q\varphi_{0}} \Bigl( \prod_i
e^{\beta_{i}\varphi_{0}} \Bigr)\,\, {\rm exp}\,
\Bigl(-{\mu\over\hbar}\,e^{- \sqrt{2} \varphi_0} \int e^{-\sqrt{2}
\tilde{\varphi}}\Bigr)\,\,,\eqno\eq$$
and
$$-\sqrt{2} s \equiv \sum_i \, \beta_i + Q\,\,.\eqno\eq$$
The remaining correlation function is at $\mu=0$ but has a nontrivial
power of the Liouville term given by $s$. It can be evaluated only for
$s$ = integer with the full result  to be obtained by some analytic
continuation [\GoLi]. The $s=0$ amplitude is termed a ``bulk" amplitude
since the condition $s=0$ coincides with a Liouville momentum conservation.
Nontrivial computation involving major cancellation between matter and
Liouville contributions gives the simple result
$$T(k_1, k_2 , \ldots ,k_N ) = (N-3)!\,\prod_{i=1}^{N} \,\,
{\Gamma(-\sqrt{2}\,\vert k_i\vert )\over\Gamma(\sqrt{2}\,\vert k_i\vert )}
\,\,. \eqno\eq$$
At $s=0$ one has both energy and momentum conservation laws:
$$\sum_{i=1}^{N} k_i = 0\,,\qquad\quad
\sum_{i=1}^N \vert k_i \vert = -2 \sqrt{2}\,\,.\eqno\eq$$
Choosing $k_1 , k_2, \ldots ,k_{N-1} > 0$, one finds that the
$N$-th particle momentum is totally determined
$$k_N = - {N-2\over \sqrt{2}}\,\,, \eqno\eq$$
implying that the $N$-th leg factor diverges
$${\Gamma (-N+2)\over \Gamma (N-2)} \sim {1\over 0}\,\,.\eqno\eq$$
This is  in agreement with the previous $\Gamma (0) \sim {1\over 0}$
divergence.  This divergence is related to the length of the Liouville
line $\int d\varphi_0$ and is only fully understood in the matrix
description.

The final result for these $s=0$ bulk amplitudes is that they consist
of purely external leg factors
$\Delta = \Gamma (-\sqrt{2}\,\vert k\vert )/
\Gamma (\sqrt{2}\,\vert k \vert )$ and that only
$T_{++\ldots +-}$ and $T_{-\ldots - +}$ amplitudes contain a diverging
factor playing the role  of the Liouville volume.  These bulk
amplitudes can then lead to the full $s\not= 0$ amplitudes by an
appropriate continuation  [\DiFrKu]. This had to await  developments
given by the matrix model formalism. Concerning the full treatment
of Liouville theory one has the interesting algebraic approach
of [\Bab,\Ge].

\bigskip
\chapter{Matrix Model and Field Theory}
\medskip
The manner in which a simple matrix dynamics gives rise to nonlinear
two-dimensional string theory is rather interesting and is related
to collective phenomena.  The major tool employed is a field-theoretic
representation given by collective field theory [\JeSa]. We shall now
give the main features of the field-theoretic approach and describe
its significance to string theory [\DaJe]. The field theory
turns out to correctly describe interactions of strings, it therefore
represents a
very simple string field theory. It provides some major insight into
the physics of noncritical strings allowing the computation of
scattering processes [\DJR] and giving the exact $S$-matrix [\Poltwo].
New higher space-time symmetries are seen to emerge [\AvJe] with
further implications on  general string field theory being likely.

The simple model that one considers is a Hermitian matrix
$M^\dagger (t) = M(t)$ in one time dimension ($X^0 = t$) with a Lagrangian
$$L = {1\over 2} \Tr \left(\dot{M}^2 - u(M)\right)\,\,.\eqno\eq$$
It has an associated $U(N)$ conserved (matrix) charge $J = i [M,\dot{M}]$.
Restricting oneself to the singlet subspace $\hat{J}\vert\,\,\,\rangle=0$
turns this model into a gauge theory. The matrix can be diagonalized:
$M(t) \to {\rm diag}\,\, (\lambda_i (t) )$ with the eigenvalues
describing a system of nonrelativistic fermions.

The collective variables of the model are the gauge invariant (Wilson)
loop operators
$$\phi_k (t) = \Tr \left( e^{ikM(t)}\right) = \sum_{i=1}^N \,
e^{ik\lambda_{i}(t)}\,\,,\eqno\eq$$
which, after a Fourier transform
$$\phi (x,t) = \int {dk\over 2\pi} \, e^{-ikx} \, \phi_k (t) =
\sum_{i=1}^N \, \delta \left( x-\lambda_i (t)\right)\,\,,\eqno\eq$$
have a physical interpretation of a density field (of
fermions).  Introduction of a conjugate field $\Pi (x,t)$ with
Poisson brackets
$$\left\{ \phi (x), \Pi (y)\right\} = \delta (x-y)\eqno\eq$$
gives a canonical phase space.

The dynamics of this field theory is directly induced from the simple
dynamics of the matrix model variables
$M(t)$ and  $P(t) = \dot{M} (t)$. It is found to be given by the
Hamiltonian
$$H_{\rm coll}=\int dx\,\,\Bigl\{ {1\over 2}\,\Pi_{,x}\,\phi\,\Pi_{,x}+
{\pi^2\over6} \phi^3 + u (x) \phi \Bigr\}\,\,,\eqno\eq$$
where the first two terms come from the kinetic term of the matrix model
$\Tr (P^2/2 )$ while the last term represents the potential
(the latter can be easily seen through the density representation):
$${1\over 2}\,\Tr P^2 \rightarrow {1\over 2}\,\Pi_{,x}\,\phi\,\Pi_{,x}
+ {\pi^2\over 6} \phi^3\,\,, \qquad
\Tr u(M) \rightarrow \int dx \, u (x) \phi(x,t)\,\,.\eqno\eq$$
The Hamiltonian constructed in this way consists of a cubic (interaction)
term and a linear (tadpole) term.  In terms of basic loops (and strings)
the cubic interaction has the effect of splitting and joining  strings.
The linear tadpole term represents a process of string annihilation
into the vacuum. It contains the classical background potential. This
potential is tuned to get a particular string theory background;
the  noncritical $c=1$ string theory is obtained for example with
an inverted oscillator potential.  Two relevant facts are immediate
in this transition to collective
field theory:

(1)  The field $\phi (x,t)$ is two-dimensional with
the extra spatial dimension $x$ being related to the eigenvalue space
$\lambda_i$.  The appearance of an extra dimension is the first
sign that this theory will be describing $D=2$ strings.

(2)  The equations of motion for the induced fields are nonlinear while
the matrix equations (in particular for the physically relevant
oscillator potential $u(M)= -M^2/2$) are  linear
$$\ddot{M} (t) - M(t) = 0\,\,.\eqno\eq$$
Through a nonlinear transformation, $\phi (x,t) = \Tr \delta
(x-M(t))$ the matrix model provides an exact solution to the
nonlinear field theory.
 The feature of integrability and the collective transformation
itself is very similar to the well known inverse
scattering transformation in integrable field theories.  Actually
introduction of left- and right-moving chiral components
$\alpha_{\pm} (x,t) = \Pi_{,x} \pm \pi\phi (x,t)$ with Poisson
brackets
$$\left\{ \alpha_{\pm} (x) , \alpha_{\pm} (y) \right\} = \pm 2\pi
\partial_x \delta (x-y)\eqno\eq$$
brings the Hamiltonian to the form
$$H_{{\rm coll}} = {1\over 2} \int \, {dx\over 2\pi}\,\Bigl\{
{1\over 3} \left( \alpha_+^3 - \alpha_-^3\right) - \left( x^2 - \mu
\right) \left( \alpha_+ - \alpha_- \right) \Bigr\}\,\,.\eqno\eq$$
The equations of motion
$$\partial_t \alpha_{\pm} + \alpha_{\pm} \partial_x \alpha_{\pm} - x =
0\eqno\eq$$
are then seen to be two copies of a large-wavelength KdV type equation
with an external $(-x^2 )$ potential.
Collective field theory shares with some other field theories in
two dimensions the feature of exact solvability.
One can indeed write down an infinite sequence of conserved commuting
quantities (Hamiltonians).  They are simply given by [\AvJe]:
$$H_n = {1\over 2\pi} \int dx \int_{\alpha_- (x,t)}^{\alpha_+ (x,t)}
d\alpha \,\, \left( \alpha^2 - x^2 \right)^n\,\,,\eqno\eq$$
and are related to the matrix model quantities $\Tr (P^2 - M^2 )^n$.
In fact one has a simple set of transition rules between the two
descriptions. These are useful when constructing
exact eigenstates and symmetry generators of the theory.

One  easily checks that the Poisson brackets vanish
$$\left\{ H_n , H_m \right\} = 0\,\,, \eqno\eq$$
and that these charges are formally conserved
$${d\over dt}\, H_n = \int dx\,\partial_x \left( \alpha^2 - x^2 \right)
\left( \alpha^2 - x^2 \right)^n = 0\,\,. \eqno\eq$$
This naturally is correct only up to surface terms which are present and
will allow particle production.

Before continuing with the integrability features of
the theory one can study perturbation theory and small fluctuations to
clarify at this simple level the connection to string theory.
The static (ground state) equation reads
$${1\over 2} \left( \pi \phi_0 (x) \right) ^2 + u (x) =
\mu_{F}\,\,,\eqno\eq$$
where $\mu_{F}$ is the Fermi energy introduced as a linear term in the
Hamiltonian
$$\Delta H = - \int dx\,\,\mu_{F}\, \phi (x,t)\,\,.\eqno\eq$$
Denoting $\pi\phi_0 = p_0$ we see this as being simply the equation
specifying the Fermi surface:  ${1\over 2} p_0^2 + u(x)=\mu_{F}$ with
the solution
$$\pi \phi_0 = p_0 (x) = \sqrt{2(\mu_{F} - u(x) )}\,\,.\eqno\eq$$
Introducing small fluctuations with a shift $\phi (x,t) = \phi_0 (x) +
{1\over\sqrt{{\pi}}}
\partial_x \eta (x,t)$ the Hamiltonian becomes
$$H = \int dx\,\, \Bigl\{(\pi\phi_0) \Bigl( {1\over 2}\,
\Pi^2 + {1\over 2}\, \eta_{,x}^2\Bigr) + {\pi^2\over 6}\,
(\eta_{,x})^3 {\pi\over 2} \Pi^2 \eta_{,x}\Bigr\}\,\,, \eqno\eq$$
with the quadratic term (in the Lagrangian form):
$$L_2 = \int dt \int dx\,\, {1\over 2}\,\Bigl( {\dot\eta^{2}\over
\pi\phi_{0} (x) } \, - \left(\pi \phi_0 \right) \eta_{,x}^2\Bigr)\,\,.
\eqno\eq$$
This is a free massless particle in an external gravitational
background
$$g_{\mu \nu}^0=\Bigl( 1/\pi \phi_0 (x)\,,\,\,\pi\phi_0(x)\,\Bigr)
\eqno\eq$$
specified by our potential $u(x)$.  However, this
metric is removable by a coordinate transformation.
In terms of the time-of-flight coordinate
$$\tau = \int^x \, {dx\over\pi\phi_0} \qquad {\rm or} \qquad
{dx(\tau)\over d\tau} = p_0 \eqno\eq$$
one has
$$H = \int d\tau\,\,\Bigl\{ {1\over 2}\,\Bigl(\Pi^2 + (\partial_{\tau}
\eta )^2 \Bigr) + {1\over 6 p_0^2}\,\Bigl(
(\partial_{\tau} \eta ) ^3 +
3\Pi^2(\partial_\tau \eta)\Bigr) \Bigr\} \,\,,\eqno\eq$$
which describes a massless theory with a spatially dependent coupling
constant
$$g_{\rm st} (\tau) = {1\over p^2_0(\tau )}\,\,.\eqn\stringcc$$
The continuum $c=1$ string theory is approached for a special choice
of the potential $v(x)= -x^2/2$.  In this case one has a
critical theory near $\mu_F = - \mu \rightarrow 0$.
For the oscillator we have
$$ \eqalign { x(\tau) & = \sqrt{2\mu}\, \cosh\,\tau\,\,,\cr
\noalign {\vskip 0.1cm}
p_0 (\tau) & = \sqrt{2\mu} \, \sinh \, \tau\,\,.}\eqno\eq$$
The length of the (physical) $\tau$-space diverges at the turning
point $x_0 = \sqrt{2\mu}$.  The string coupling constant \stringcc\
is now
$$g_{\rm st}(\tau)={1\over 2\mu \, \sinh^2 \tau}\,\,.\eqno\eq$$
It depends on the Fermi level as $g_{\rm st} \sim 1/\mu$.
This is in parallel with the dependence of the string coupling on the
cosmological constant of the $c=1$ string theory. We also see that
asymptotically $ g_{\rm st} \sim {1\over \mu}\,e^{-2\tau}$
as  $\tau \rightarrow + \infty$.
Comparing it to the expected behavior in $c=1$ string
theory  $g_{\rm st} \sim {1\over\mu}\,e^{-\sqrt{2} \varphi}$ one has
the (asymptotic) identification
$$\tau \leftrightarrow {1\over \sqrt{2}}\,\varphi\,,\qquad\quad
t_M \leftrightarrow {1\over\sqrt{2}}\,t_{c=1} \,\,.\eqno\eq$$
One can  now identify $\eta (\tau,t)$ with the tachyon
field $T(\varphi, t)$.  Remembering that $e^{\sqrt{2}\varphi}\,
T(\varphi, t)$ was the field satisfying the massless Klein-Gordon
equation, one  also has the identification of the
energy-momenta:
$$ip_{\tau}\leftrightarrow 2 + i\sqrt{2} \, p_\varphi\,,\qquad\quad
i\epsilon \leftrightarrow i \sqrt{2}\,p \,\,,\eqno\eq$$
where $\epsilon$ is the energy in the matrix model picture.

The above identification of the collective field
$\eta(\tau,t)$ was only done asymptotically when the
$\mu e^{- \sqrt{2}\varphi}$ term in the Liouville equation is ignored.
A much more precise identification can be performed with the cosmological
term also present.

In the matrix model the time-of-flight coordinate is
introduced to bring the quadratic mass operator of the
collective field into a Klein-Gordon form:
$$\Bigl[\,\partial_t^2 -\sqrt{x^2-2\mu}\,\,\partial_x\,
\sqrt{x^2-2\mu}\,\,\partial_x\Bigr]\,\eta =
(\partial_t^2 -\partial_\tau^2)\,\eta(t,\tau)\,\,, \eqno\eq$$
with $x=\sqrt{2\mu}\,\cosh\tau$. If we use a basis
conjugate to $x$: $p=-i\,(\partial/\partial x)$ the spatial
operator reads
$$\omega^2 =p^2x^2-2\mu p^2\,\,,\eqno\eq$$
and after a change of variables
$p=\sqrt{2}\,e^{-\varphi/\sqrt{2}}$ this gives the Liouville
operator
$$\hat\omega^2 =-{1\over 2}\,(\partial_\varphi)^2
-4\mu\,e^{-\varphi/\sqrt{2}}\equiv {\cal H}_L\,\,.\eqno\eq$$
We see that the Liouville coordinate is to be identified more
precisely [\MoSe] with the variable $p$ conjugate to the matrix
eigenvalue $\lambda$. The conjugate basis is not unnatural
in collective theory, it is associated with the
(Wilson) loop operator itself
$$W(\ell,t)= \Tr\,(e^{-\ell M}) =
\int dx\,e^{-\ell x}\,\phi(x,t) \,\,.\eqno\eq$$
which at the linearized level
$$W(\ell,t)=
\int_0^\infty  d\tau\,e^{-\sqrt{2\mu}\,\ell \cosh\tau}\,
\partial_\tau \eta\,\,, \eqno\eq$$
is  seen to obey
$$(\partial_t^2 -\hat\omega^2)\,\hat W(\ell,t)=0\,\,,\eqno\eq$$
with
$$\hat\omega^2=\partial_\tau^2 \quad
\Rightarrow\quad  -(\ell\,\partial_\ell)^2 +
2\mu\,\ell^2\,\,.\eqno\eq$$
After a change $\ell=2e^{-\varphi/\sqrt{2}}$
one has the Liouville operator ${\cal H}_L$.
In this conjugate momentum basis the connection to Liouville
theory is therefore manifest. One could obviously write all
equations in this representation  but formulae are
much simpler in terms of time-of-flight coordinate $\tau$.
The (Wilson) loop field and its natural connection to the
Liouville picture will be relevant for defining the string
theory $S$-matrix.

To further clarify the identification of the Liouville mode
let us write the transformation between the matrix
eigenvalue $x=\lambda$ and the time-of-flight coordinate
$\tau$ as a point canonical transformation:
$$\eqalign{x&=\sqrt{2\mu}\,\cosh\tau\,\,,\cr
\noalign{\vskip 0.2cm}
p&={1\over \sqrt{2\mu}\,\sinh\tau}\,\,p_\tau\,\,,}\eqno\eq$$
where $p$ and $p_\tau$ are the conjugates:
$\{x,p\}=\{\tau,p_\tau\} =1$.
Introducing $p=\sqrt{2}\,e^{-\varphi/\sqrt{2}}$
we have
$$\eqalign{&p_\varphi=\sqrt{2\mu}\, e^{-\varphi/\sqrt{2}}\,
\cosh\tau\,\,,\cr
\noalign{\vskip 0.2cm}
&p_\tau=\sqrt{2}\,\sqrt{2\mu}\, e^{-\varphi/\sqrt{2}}\,
\sinh\tau \,\,,}\eqno\eq$$
as a canonical transformation between the Liouville and
time-of-flight coordinates. The property of this
transformation is that
$${1\over 2}\,\omega^2 ={1\over 2}\,p_\tau^2 =
p_\varphi^2 -2\mu\, e^{-\varphi/\sqrt{2}}\,\,.$$
Now in Liouville theory one also usually deals
with two alternate descriptions and two different fields:
the original Liouville field $\varphi(z,\bar z)$
and a free field $\psi(z,\bar z)$.
They are related by a canonical (B\"acklund) transformation
$$\eqalign{\dot\varphi &= \psi'+\sqrt{2\mu}\,
e^{-\varphi/\sqrt{2}}\, \cosh (\psi/\sqrt{2})\,\,,\cr
\noalign{\vskip 0.2cm}
\dot\psi &=\varphi'+\sqrt{2\mu}\,e^{-\varphi/\sqrt{2}}\,
\sinh (\psi/\sqrt{2})\,\,,}\eqno\eq$$
where the two derivatives correspond to the two-dimensional space
$z=\sigma+i\xi$. The above transformation relates the
Liouville action to the action of a free field
$\psi(z,\bar z)$. Clearly for the center of mass mode
$(\varphi' = \psi' =0)$ one sees
$$\eqalign{\Pi_\varphi &= \sqrt{2\mu}\,
e^{-\varphi/\sqrt{2}}\, \cosh (\psi/\sqrt{2})\,\,,\cr
\noalign{\vskip 0.2cm}
\Pi_\psi &=\sqrt{2\mu}\,e^{-\varphi/\sqrt{2}}\,
\sinh (\psi/\sqrt{2})\,\,. }\eqno\eq$$

The transformation between $\varphi$ and $\psi$ is
identical to the one in the matrix model. We have then the
fact that the time-of-flight coordinate $\tau$ is to be
identified with the free field zero mode $\psi_0:\,\psi_0=
\sqrt{2}\,\tau$. In most of the vertex operator construction
it is the free field which is used.

We shall now continue and discuss the exact classical solution
of the theory and exhibit its integrability.
Consider first  the physical meaning of the
component fields $\alpha_{\pm}$ and the nature of boundary conditions
at the turning point or wall $\tau = 0$.  Shifting by the classical
solution, $\alpha_{\pm} = \pm p_0 + \epsilon_{\pm}\,$, the equations
of motion linearize to
$$\partial_{\pm} \epsilon_{\pm} \pm \left( p_0 \partial_x + \partial_x
p_0 \right) \epsilon_{\pm} = 0\,\,. \eqno\eq$$
Denoting $\epsilon_{\pm} \equiv \mp {1\over p_0} \psi_{\mp}$ we have
$$\left( \partial_t \pm \partial_{\tau} \right) \psi_{\mp}=0\,\,.
\eqno\eq$$
So indeed, $\psi_{\pm} = \psi_{\pm} ( t\pm \tau)$ are left- and
right-moving waves, respectively.  There is however a nontrivial
boundary condition in the theory which comes in as follows:  The
eigenvalue density $\phi = {1\over 2\pi} \left( \alpha_+ - \alpha_-
\right)$ gives the conserved fermion number
$$\dot{N} = \int dx \,\dot{\phi} = {1\over 2\pi} \int dx \left(
\alpha_+^2 - \alpha_-^2 \right)=0\,\,.\eqno\eq$$
At the boundary point for $x$ (or $\tau = 0)$, this implies
$$\left( \alpha_+^2 - \alpha_-^2 \right)\Bigl\vert_{{\rm boundary}} =
0\,\,,\eqno\eq$$
so that there is no leakage into the region under the barrier (this
may have to be given up in nonperturbative discussion [\LM]).  For the
small fluctuations we then have
$$\psi_+ (x) = \psi_- (x)\,\,,\eqno\eq$$
implying Dirichlet boundary conditions. In terms of Fourier modes
$$\psi_{\pm} = \int_{-\infty}^{\infty} dk\,\,\alpha_k^{\pm}\,\,
e^{ik (t\pm \tau)}\eqno\eq$$
with $\alpha_{-k} = \alpha_k^+$ our boundary condition implies that one
has only one set of oscillators with positive momenta
$$\alpha_k^+ = \alpha_k^- = \alpha_k\,,\qquad k>0\,\,.\eqno\eq$$
This is appropriate for a theory defined on a half-line
$\tau \in [0,\infty)$.

A very simple form for the exact solution of the collective equations
was given by Polchinski [\Poltwo].
At  the classical level one has a  phase space picture of
the eigenvalues $\lambda (\sigma,t)$ and their momenta $p(\sigma,t) =
\dot{\lambda}$.  They obey the classical equations of motion
$$\dot{p}(\sigma,t) = - u'\left(\lambda(\sigma,t)\right)\,\,.
\eqno\eq$$
The information that the particles are fermions is contained in the
statement that the equation $x = \lambda (\sigma,t)$ is invertible: $\sigma
= \sigma (x,t)$ so that for each $\sigma$ there is only one particle
(actually there is a degeneracy corresponding to the upper and lower
Fermi surface).  Consider in particular the inverted oscillator: the
solution is immediately written as
$$\eqalign{ x & = a(\sigma) \cosh (t - \sigma )\,\,,\cr
p & = a(\sigma ) \sinh(t - \sigma )\,\,. }\eqn\parsol$$
Here $a(\sigma )$ is an arbitrary function giving an arbitrary initial
condition.  The simplest configuration is obtained for
$a (\sigma ) = \sqrt{\mu} = {\rm const.}$ and we have
$$p_{\pm}=p(\sigma_{\pm},t)=\pm\sqrt{x^2 -\mu}\,\,.\eqno\eq$$
This is recognized as the static ground state collective field
configuration $\pi \phi_0 (x)$. It is easy to see that the general
configuration leads to the solution of collective
equations. The collective field is identified with the Fermi momentum
densities
$$\alpha_{\pm}(x,t)\equiv p_{\pm}=p (\sigma_{\pm}(x,t),t)\,\,.\eqno\eq$$
Conversely, $p(\sigma,t)=\alpha(x(\sigma,t),t)$. Using the chain rule
$${\partial p\over \partial t} = {\partial\alpha\over \partial t} +
{\partial\alpha\over \partial x}\,\, {\partial x\over\partial t} = -
u'(x)\,\,,\eqno\eq$$
and the equation of motion obeyed by $p(\sigma,t)$, there follows the
equation
$${\partial\alpha\over\partial t}= -u'(x)-\alpha\,\partial_x \alpha
\,\,.\eqno\eq$$
These are the decoupled quadratic equations for the collective fields
$\alpha_{\pm} (x,t)$ associated with the cubic Hamiltonian.

Knowledge of the exact solution can be directly  used to determine
scattering amplitudes. One considers and follows the time evolution of
an incoming left-moving lump.
A point parametrized by $\sigma$ which passes through $x$
at some (early) time $t$ will reflect on the boundary and pass through
the same point $x$ at some later time $t'$ as a right-moving lump.
The time evolution of the particle coordinates is known
explicitly~\parsol\
so one can determine the relationship between the two times $t$ and
$t'$. Consider the exact solution given by Eq.~\parsol,
at a distance $\tau$ large enough one has
$$x=e^{\tau} = \cases{a(\sigma)\, e^{-(t-\sigma )}\,, \quad &
$\quad t\rightarrow - \infty$\cr
a(\sigma)\, e^{+(t' + \sigma )}\,, \quad & $\quad t' \rightarrow +
\infty$\cr }$$
from where
$$ t' - \tau = t + \tau + \ln a^2 (\sigma)\,\,.\eqno\eq$$
On the other hand $a^2 (\sigma ) $ is related to $\alpha_{\pm}$:
$$a^2 = x^2 - p^2 = x^2 - \alpha_{\pm}^2 \approx 1 +
\psi_{\mp}\,\,.\eqno\eq$$
The outgoing particle momentum $p_+ (t', \sigma)$ is equal in magnitude
(but opposite in sign) to the incoming momentum of the particle
$p_- (t, \sigma)$:
$$p_+ (t', \sigma ) = - p_- (t,\sigma )\,\,.\eqno\eq$$
This elementary relationship provides a relationship between the
incoming and outgoing wave and therefore yields the $S$-matrix.
Collecting the above formulas we have
$$\psi_-(z)=\psi_+ \Bigl(z-\ln(1+\psi_-(z))\Bigr)\,\,.\eqno\eq$$
This functional equation determines the relation between the left- and
right-moving (incoming and outgoing fields). It represents a nonlinear
version of our Dirichlet boundary conditions and is characteristic of
scattering problems involving a wall. An expansion in power series can
be performed determining explicitly the outgoing modes in terms of the
incoming ones.  This is then sufficient to give the
scattering amplitudes. We shall return to this subject in sect.~5.

In general, all features of the exactly solvable matrix model
translate into string theory.  More precisely there is a direct
translation of matrix model quantities into the collective field
theory which itself is then completely integrable as we have
emphasized. We  end this section by summarizing the  set of
translation rules between the matrix model and collective field
theory representations.

At the  classical level one thinks of matrix variables
as coordinates in a fermionic phase space $M\rightarrow \lambda ,
P\rightarrow p$. Collective field theory
represents a second quantization according to $p\rightarrow
\alpha (x,t)$.  So we have the correspondences:
$$\eqalign{ M & \leftrightarrow \lambda \leftrightarrow x\,\,,\cr
P &\leftrightarrow p \leftrightarrow \alpha (x,t)\,\,.}\eqno\eq$$
The $U(N)$ trace becomes a phase space integration in the fermionic
picture and in  the collective representation:
$$\Tr \left\{ \,\,\right\}\quad \rightarrow \quad\int{dx\over 2\pi}
\int_{\alpha_{-}(x,t)}^{\alpha_{+}(x,t)} d\alpha \,\, \left\{ \,\,
\right\}\,\,, \eqno\eq$$
where $\alpha_{\pm} (x,t)$ are the chiral components of the scalar
field density. For example the collective Hamiltonian comes out
as follows:
$${1\over 2}\,\Tr \left( P^2 - M^2 \right)\rightarrow{1\over 2}
(p^2 - x^2) \rightarrow \int {dx\over 2\pi} \int d\alpha \,
{1\over 2}(\alpha^2- x^2) ={1\over 2} \int {dx\over 2\pi}
\Bigl[ {\alpha\over 3}^3 - x^2 \alpha \Bigr]_-^+ \,\,.\eqno\eq$$
The above transition rules summarize the statement that
the Poisson brackets of single particle quantities in the Fermi (or
matrix) phase space
$$\left\{ f_1 (x,p), f_2 (x,p)\right\}_{\rm P.B.}\eqno\eq$$
remain preserved in the field theory.  For example, the field-theoretic
operator inferred from the  oscillator states is
$$B_n^{\pm}=\int{dx\over 2\pi}\int d\alpha \,\,(\alpha \pm x)^n\,\,.
\eqno\eq$$
We can now use the $\alpha$-field Poisson brackets
$\left\{ \alpha_{\pm} (x), \alpha_{\pm} (y) \right\} = \mp 2\pi i
\delta '(x-y)$  to verify that indeed
$$\left\{ H_{{\rm coll}} , B_n^{\pm} \right\} = \pm n \,
B_n^{\pm}\,\,.\eqno\eq$$
This represents an eigenstate of the collective field theory Hamiltonian.
At the quantum level a normal ordering prescription is used to
completely define the operators.

The outlined string field theory gives a systematic perturbation theory
in the string coupling constant. The Feynman rules that are constructed
are characterized by a nontrivial cubic vertex exhibiting discrete poles
in the momenta. Most importantly a fully quantized Hamiltonian is achieved
through normal ordering with the counter-terms being supplied by the original
collective formalism. So what one has is a totally finite string field
theory capable of reproducing string theory diagrams to all orders. It
works at loop level without further counter-terms giving a single covering of
modular space. This, as is well known, has always been quite nontrivial in a
string-theoretic framework. For more details of the quantum theory and
explicit calculations at the loop level the reader should
consult [\DJR].

\bigskip
\chapter{$w_{\infty}$ Symmetry}
\medskip

The matrix model description has the virtue of great simplicity:  it
is linear and trivially exactly solvable.  For the matrix Hamiltonian
$$H = {1\over 2}\, \Tr\left(P^2-M^2\right)\eqno\eq$$
one can write down exact creation--annihilation operators
$$B_n^{\pm} = \Tr \left( P \pm M\right)^n\,,
\quad\quad n = 0,1,2,\ldots\eqno\eq$$
creating imaginary energy eigenstates
$$\left[\,H, B_n^{\pm}\,\right]= \mp in B_n^{\pm}\,,\qquad
\epsilon_n = \pm in \,\,.\eqno\eq$$
The whole point here is to be able to translate this exact information
into physical results which, as we have emphasized, is achieved through
collective field theory. The direct connection of the space-time string
field theory with the matrix model leads then further insight. The simple
oscillator structure with its creation--annihilation basis implies the
presence of a similar structure in the field theory and therefore string
theory.

To understand the physical meaning of the (oscillator) states
recall that in the collective field theoretic description we have another
spatial quantum number in addition to the energy.
This feature arose as a consequence of scaling invariance.
The coordinate and the fields transform as
$$x\rightarrow ax\,,\qquad\quad
\alpha (x,t)\rightarrow {1\over a}\,\alpha(ax,t)\,\,,\eqno\eq$$
and the Hamiltonian, without the chemical potential term,
$-\mu \alpha$, scales as
$$H \rightarrow {1\over a^4}\,H\,\,.\eqno\eq$$
The classical equations of motion are consequently scale invariant.
One then defines the scaling momentum as
$$i p_s = - 4 + s\,\,,\eqno\eq$$
where $s$ is the naive scaling dimension $s[x] = s[\alpha] = 1$.  The
creation--annihilation operators
$$\tilde T_n^{\pm} ={1\over n}\,\int {dx \over 2\pi}\,\,
{(\alpha\pm x)^{n+1}\over n+1}\,\,\Big\vert_{\alpha_{-}}^{\alpha_{+}}
\eqno\eq$$
consequently have the following energy-momentum:
$$i \epsilon = n \,,\qquad\quad
i p_s = -2 + n \,\,.\eqno\eq$$
We find these to be in precise agreement with the discrete tachyon
vertex operator states since there
$$i \sqrt{2}\, p  = \pm 2 j\,,\qquad\quad
i \sqrt{2}\, p_{\varphi} = -2 + 2j \,\,,\eqno\eq$$
and we have already noted the relations $\sqrt{2}\, p = \epsilon,\,\,
\sqrt{2}\,p_{\varphi}=p_s$. We then have a one-to-one correspondence
between oscillator states of the matrix model and discrete tachyon
vertex operators of the conformal description of $c=1$ string theory
$$B_n^{\pm}=\Tr\left(P\pm M\right)^n \quad \leftrightarrow \quad
T_{p}^{(\pm)}=e^{\pm i\sqrt{2} jX}\,e^{-\sqrt{2}(1-j)\varphi}\eqno\eq$$
with $n=2j$ or $j=n/2$.

An analytic continuation of discrete imaginary momenta to real
values $(n = i\kappa,\, p_s = 2i - \kappa )$
gives the scattering operators
$$\eqalign{ B_{-i\kappa}^- & = \Tr \left( P - M\right)^{-i\kappa}
\sim e^{-i\kappa (t+\tau )}\,\,,\cr
B_{-i\kappa}^+ & = \Tr \left( P + M\right)^{-i\kappa}
\sim e^{-i\kappa (t-\tau )}\,\,, }\eqno\eq$$
describing left- and right-moving waves, respectively. These operators
can be used to construct the in- and out-states of scattering theory
$$\eqalign{\Tr \left( P-M \right)^{-i\kappa} \vert 0 \rangle & = \vert
\kappa;\,{\rm in} \rangle \,\,,\cr
\Tr\left( P + M \right)^{+i\kappa}\vert 0\rangle & = \vert\kappa;\,
{\rm out} \rangle \,\,. }\eqno\eq$$
Namely, for an in-state, one needs a {\it left}-moving wave while the
out-state is necessarily given by a {\it right}-moving one. Here we
have used the picture where the wall is at $\tau = -\infty$
corresponding to the physical space being defined on the right semi-axis
$x = e^{\tau} \geq 0$. Had we chosen to define the theory on the
other side of the barrier, the states
$$\eqalign{\Tr\left(P+M\right)^{i\kappa}&=e^{i\kappa(t+\tau)}\,\,,\cr
\Tr\left(P-M\right)^{i\kappa}&=e^{i\kappa(t-\tau)}\,\,, }\eqno\eq$$
would be physical since they have the meaning of a right-moving
{\it in}-wave and a left-moving {\it out}-wave.
Hence there is a one-to-one correspondence between the scattering
operators in the matrix model and the string theory vertex operators
$$\Tr\,(P\pm M)^{-i \kappa}\quad\leftrightarrow \quad
T^{\pm}=e^{\pm {\kappa\over\sqrt{2}} X}\,e^{-\sqrt{2}+i{\kappa\over\sqrt{2}}
\varphi}\,\,.\eqno\eq$$
A typical transition amplitude reads
$$S = \langle {\rm out} \vert {\rm in}\rangle=\langle 0\vert
\Tr (P+M)^{{\kappa\over i}}\Tr(P-M)^{{\kappa\over i}}\vert 0\rangle\,\,.
\eqno\eq$$
It only contains operators with the same (Liouville)-exponential
dressing. This is in total agreement with the continuum string theory
situation.

In addition to the tachyon states, the matrix oscillator description
immediately allows a construction of an infinite sequence of discrete
states [\GKN,\AvJe].
They are created by the operators
$$B_{n,\bar n}=\Tr\Bigl((P+M)^n (P-M)^{\bar{n}}\Bigr)\,\,,\eqno\eq$$
with energies and momenta given by
$$i\epsilon=n-\bar{n}\,,\qquad\quad
i p_s =-2+(n+ \bar{n})\,\,.\eqno\eq$$
Comparing this with the discrete spectrum of the string theory given by
$$i \sqrt{2} p = 2m\,,\qquad\quad
i \sqrt{2} p_{\varphi} =-2+2j\,\,,\eqno\eq$$
we find the correspondence
$$m ={n-\bar{n}\over 2}\,,\qquad\quad
j ={n + \bar{n}\over 2} \,\,.\eqno\eq$$
These are indeed half-integers once $n,\bar{n}$ are integers.  The
field theory operators
$$B_{jm}=\int {dx\over 2\pi}\int_{\alpha_-}^{\alpha_+} d\alpha\,
(\alpha + x )^{j+m}\,(\alpha - x )^{j-m}\eqno\eq$$
can be shown (again by using the Poisson brackets or the commutators)
to generate discrete imaginary energy eigenstates of the Hamiltonian
$$\left[ H, B_{jm} \right] = -2i m B_{jm}\,\,.\eqno\eq$$
This commutator shows that the operators $B_{jm}$ are
spectrum-generating operators for the Hamiltonian $H$;
but it also signals the existence of a large symmetry algebra which
operates in this theory [\AvJe,\MoSe,\winf,\WiGR,\KlPo,].

First we had the  sequence of conserved quantities
$$H_l =\Tr\left( P^2 - M^2 \right)^{l+1}\eqno\eq$$
commuting among themselves
$$\left[ H_l , H_{l'} \right] = 0\,\,.\eqno\eq$$
These are  particular cases of the spectrum-generating operators
$B_{jm}$. One is then lead to consider the complete algebra of all the
operators. Introducing the more standard notation
$$O_{JM} = (p+x)^{J+M+1} (p-x)^{J-M+1}\,\,,\eqno\eq$$
with the associated collective field realization
$$O_{JM} = \int {dx\over 2\pi} \int_{\alpha_{-}}^{\alpha_{+}} d\alpha
\,\, (\alpha + x )^{J+M+1} (\alpha - x )^{J-M+1}\,\,,\eqno\eq$$
one checks that they obey the $w_{\infty}$ commutation relations
$$\left[ O_{J_{1}M_{1}} , O_{J_{2} M_{2}} \right] = 4i
\Bigl((J_2+1)M_1-(J_1+1)M_2\Bigr)\,O_{J_{1}+J_{2},M_{1}+M_{2}}\,\,.
\eqn\winfalg$$
We note that this commutator results if no special ordering is taken
for the noncommuting factors.  At the full operator
quantization level, field theory requires special normal ordering.
It is likely that this modifies the simple $w_{\infty}$ algebra to a
$W_{1+\infty}$ algebra.

Recalling the form of tachyon operators $T_n^{\pm}=\Tr(P\pm M )^n$ one sees
a special relationship between the tachyon operators and the
$w_{\infty}$ generators.  A simple computation shows that
$$O_{JM} = {1\over 2i} \, {1\over (J+M+2) (J-M+2)} \, \bigl[
T_{J+M+2}^+ , T_{J-M+2}^- \bigr] \,\,.\eqno\eq$$
This first sheds some light on the nature of higher discrete modes in
the collective formalism:  they are composite states of the tachyon.
More importantly, one then expects a simple realization of the
$w_{\infty}$ algebra on the tachyon sector.

To understand the role played by the scalar collective field with
respect to the $w_\infty$ algebra one can first look at the
following Virasoro subalgebra:
$$O_{l} \equiv O_{{l\over 2},{l\over 2}} =
\int {dx \over 2\pi}\,\int d\alpha\,\, (\alpha+x)^{l+1}\,
(\alpha -x)\,\,,\eqno\eq$$
with
$$[ O_{l},  O_{l'} ]= 2i(l-l')\,O_{l+l'} \,\,.  \eqno\eq$$
We can then determine the transformation property of the tachyon
field under this subalgebra. Actually, the exact tachyon
creation operator $T_n$ can itself be written as an
extension of the whole algebra
$$\tilde T_n^{+} ={1\over n}\,\int {dx \over 2\pi}\,
{(\alpha+x)^{n+1}\over n+1} =
{1\over n}\, O_{{n\over 2}-1, {n\over 2}}\,\,,\eqno\eq$$
and this determines the commutator (with the indices extended  outside
the standard range $|m| \leq j$). Alternatively one  can also directly
use the basic commutation relations to find
$$[O_l, \tilde T_n^{+}] = 2i (n+l)\,\tilde T_{n+l}^{+}\,\,,\eqno\eq$$
which shows that the tachyon transforms as a field of conformal weight 1.
This is understood to be a space-time and not a world sheet feature.
The fact that an infinite space-time  symmetry appears
in the collective field theory explains many similarities that it
has with conformal field theory.
 The $w_\infty$ generators act in a
nonlinear way  on the tachyon field.
This implies that this symmetry can be used to write down Ward
identities for correlation functions and the $S$-matrix.

Let us study in more detail the nonlinearity involved in
the collective representation(here we summarize the results
achieved in [\JRvT]).
One is in general interested in comparison with similar
nonlinearities (and Ward identities) obtainable in the
world sheet conformal field theory analysis.
The latter is only performed in the approximation neglecting
the cosmological constant term $(\mu \to 0)$ which  represents the
strong coupling regime of the field theory
$(g_{\rm st} = 1/\mu \to \infty)$.
In this limit one simply expands
$$\alpha_{\pm}(x,t) =\pm x+{1\over 2x}\,\hat\alpha_{\pm}\,\,,\eqno\eq$$
which is an approximate form when
$$\pi\phi_0(x) =\sqrt{x^2-\mu} \to x = {e^{\tau}\over 2}\,\,.
\eqno\eq$$
The exact tachyon operators reduce
in the leading (linear) approximation to
$$\tilde T_n^{\pm} ={1\over n}\,\int {d\tau \over 2\pi}\,\,
e^{n\tau}\,\hat\alpha_\pm \,\,,\eqno\eq$$
This is as it should be since they are to describe left- and right-moving
waves,
respectively. Consider now the $w_\infty$ generators
in the same approximation. With the
above background shift one easily finds that they reduce to
$$O_{JM}={1\over J-M+2}\,\int {d\tau\over 2\pi}\,\,
e^{2M\tau}\,\hat\alpha_{+}^{J-M+2}
+{(-1)^{J-M}\over J+M+2}\,\int {d\tau\over 2\pi}\,\,
e^{-2M\tau}\,\hat\alpha_{-}^{J+M+2}\,\,.\eqno\eq$$
Here we see that the operator $O_{JM}$ behaves as the
$(J-M+2)$th power of the right-moving tachyon $\alpha_+$ and
also the $(J+M+2)$th power of the left-moving tachyon $\alpha_-$.
These are the leading polynomial powers in the left- and
right-moving components of the tachyon; even in this
strong coupling limit the theory is nonlinear and one has
further higher order terms. Concerning these one can go to the
in (out) fields (where it is likely that only leading terms
remain). The in (out) fields are simply limits of the component fields
$\alpha_\pm$:
$$\alpha_{\rm out}(t-\tau) = \lim_{t\to +\infty} \alpha_+\,,
\qquad\quad \alpha_{\rm in}(t+\tau) = \lim_{t\to -\infty} \alpha_-\,\,,
\eqn\inoutfields$$
Since the operators $O_{JM}$ are conserved (up to a phase),
looking at the $t\to \pm\infty$ limit of $e^{2Mt}\,O_{JM}$ we
obtain an identity (between the in- and out-representation):
$$\eqalign{O_{JM}&={1\over J-M+2}\,\int {dz\over 2\pi}\,\,
\alpha_{\rm out}^{J-M+2} (z)\cr
\noalign{\vskip 0.2cm}
&={(-1)^{J-M}\over J+M+2}\,\int {dz\over 2\pi}\,\,
\alpha_{\rm in}^{J+M+2} (z) \,\,.}\eqno\eq$$
Introducing creation--annihilation operators
$$\alpha_{\rm in}(z)=\int dz\,e^{-ikz}\,\alpha(k)\,,
\qquad\quad \alpha_{\rm out}(z)= \int dz\,e^{-ikz}\,\beta(k)\,\,,
\eqno\eq$$
with $\alpha(k)=a(k)$ and $\alpha(-k)=ka(k)^\dagger$;$\beta(k)=b(k)$ and
$\beta(-k)=kb(k)^\dagger$ we have the expressions
after a continuation $k \to ik$:
$$\eqalign{O_{J,-M}&={1\over J-M+2}\,\int dk_1\ldots dk_{J-M+2}\,\,
\alpha(k_1)\ldots\alpha(k_{J-M+2})
\delta(\sum k_i +2M)\cr
\noalign{\vskip 0.2cm}
&={(-1)^{J-M}\over J+M+2}\,\int dp_1\ldots dp_{J+M+2}\,\,
\beta(p_1)\ldots\beta(p_{J+M+2})
\delta(\sum p_i +2M)\,\,.}\eqno\eq$$
These representations can be compared with analogue expressions
found in conformal field theory [\KlPo].

The Ward identities essentially follow from the in--out
representations of the generators in terms of the tachyon
field. A typical $S$-matrix element is given by
$$S(\{k_i\};\{p_j\})=\langle 0| \prod_j \,\beta(p_j)\,\prod_i \, \alpha(-k_i),
|0\rangle \,\,.\eqno\eq$$
Consider a general matrix element of the $w_\infty$
generator $O_{JM}$,
$$\langle \,\,0|\beta\, O_{JM}\, \alpha |0\,\,\rangle \,\,.$$
It can be evaluated by commuting to the left or to the right
using alternatively the above in--out representations. The
two different evaluations give the identity
$$\langle 0| [\beta, O_{JM}]\, \alpha|0\rangle =
\langle 0| \beta \,[O_{JM}, \alpha]|0\rangle\,\,,\eqno\eq$$
which summarizes the general Ward identities. These, when
written out explicitly using the representations for
$O_{JM}$, have the form of recursion relations reducing the
$N$-point amplitude to lower point ones. Specifically, the
one creation operator term of the $\alpha$-representation
for $O_{JM}$ gives:
$$O_{M+1,-M}\,a^\dagger(k_1)\,a^\dagger(k_2) |0\rangle =
4\pi (k_1+k_2+2M)\,a^\dagger(k_1+k_2+2M) |0\rangle\,\,,\eqno\eq$$
which turns a two-particle state into one-particle state.
Generally,
$$O_{M+N,-M}\,a^\dagger(k_1)\ldots a^\dagger(k_{N+1}) |0\rangle =
2\pi^N\,(N+1)! (\sum k_i+2M)\,a^\dagger(\sum k_i+2M)
|0\rangle\,\,,\eqno\eq$$
showing a reduction of the $(N+1)$-particle state into a
single-particle state.

As an example let us calculate the $3\to 1$ amplitude
$$S_{3,1}=\langle 0|b(p)\,a^\dagger(k_1)\,
a^\dagger(k_2)\,a^\dagger(k_3) |0\rangle\,\,.\eqno\eq$$
The energy-momentum conservation laws imply
(recall that for $B^\pm_n$, $\epsilon=\pm n,
p_s =-2+n$):
$$\eqalign{&k_1+k_2+k_3-p=0\,\,,\cr
&(-2+k_1)+(-2+k_2)+(-2+k_3)+(-2+p)=-4\,\,.}$$
The latter is a specific case of the general
Liouville conservation (or bulk condition):
$$\sum_{i=1}^N = p_s^i = -4\,\,.\eqno\eq$$
The energy-momentum relations specify the momentum of
the 4th particle
$$p=2 \qquad {\rm or} \qquad k_1+k_2+k_3=2\,\,.$$
Use now the operator
$$\eqalign{O_{{1\over 2},-{1\over 2}} &= {1\over \sqrt{\mu}}\,
\int dk_1 dk_2 dk_3\,\, k_1 \,a^\dagger(k_1)\,
a(k_2)\,a(k_3)\,\delta(k_1-k_2-k_3+1) \cr
\noalign{\vskip 0.2cm}
&= {\sqrt{\mu}}\, \int dp_1 dp_2 \,\, p_1\,
b^\dagger(p_1)\,b(p_2)\,\delta(p_1-p_2+1) }$$
to deduce
$$\eqalign{S_{3,1}&=\langle \,b(2)\,a^\dagger(k_1)\,
a^\dagger(k_2)\,a^\dagger(k_3)\,\rangle\cr
\noalign{\vskip 0.2cm}
&={\pi\over \mu}\,(k_1+k_2-1)\,
\langle \,b(1)\,a^\dagger(k_1+k_2-1)\,
a^\dagger(k_3)\,\rangle\cr
\noalign{\vskip 0.2cm}
&+{\pi\over \mu}\,(k_2+k_3-1)\,
\langle \,b(1)\,a^\dagger(k_2+k_3-1)\,
a^\dagger(k_1)\,\rangle\cr
\noalign{\vskip 0.2cm}
&+{\pi\over \mu}\,(k_1+k_3-1)\,
\langle \,b(1)\,a^\dagger(k_1+k_3-1)\,
a^\dagger(k_2)\,\rangle \,\,.}$$
Taking the normalized three-point function to be
$S_3 = 1/\mu$, the result
$$S_{3,1} ={\pi\over \mu^2}\,\bigl(2(k_1+k_2+k_3)-3\bigr)=
{\pi\over \mu^2} $$
follows. One can iteratively repeat the same reduction for higher
point amplitudes and find
$$S_{N,1}= {\pi^{N-1}\over (N-2)!}\,\,\mu^{-N+1}\,\,,\eqno\eq$$
which is the $(N+1)$-point  ``bulk'' scattering amplitude.

In describing the infinite symmetry  we have followed the
matrix model approach were the appearance of the symmetry structure
is most natural. The features described arise also in the continuum
conformal field theory language where the Ward identities take a
particularly elegant form.

Of crucial importance in establishing continuum
quantities that are analogous with those of the matrix model is Witten's
identification of the ground ring [\WiGR]. This consist of ghost
number zero, conformal spin zero operators ${\cal O}_{JM}$
which are closed under operator products
${\cal O'}\cdot {\cal O''} \sim{\cal O''}$ (up to BRST
commutators). The basic generators are
$${\cal O}_{0,0}=1\,,\qquad
{\cal O}_{{1\over 2},\pm{1\over 2}} =
\Bigl[ c\,b \pm {i\over\sqrt{2}}\,\partial X -
{1\over\sqrt{2}}\,\partial\varphi\Bigr]\,
e^{(\pm iX+\varphi)/\sqrt{2}}\,\,.\eqno\eq$$
The suggestion (of Witten) was that
${\cal O}_{{1\over 2},\pm{1\over 2}}$ are the
variables which correspond to the phase space coordinates of the
matrix model
$$\eqalign{
{\cal O}_{{1\over 2},+{1\over 2}} &=a_+ \equiv p+x\,\,,\cr
\noalign{\vskip 0.2cm}
{\cal O}_{{1\over 2},-{1\over 2}} &=a_- \equiv p-x\,\,,}\eqno\eq$$
with ${\cal O}_{0,0}=1$ being the cosmological constant
operator. Once the (fermionic) matrix eigenvalue coordinates have
been identified one could study the action of discrete states
vertex operators upon them. They turn out to act as vector
fields on the scalar ring
$$\Psi_{JM}={\partial h\over \partial a_+}\,\,
{\partial\over\partial a_-} -
{\partial h\over \partial a_-}\,\,
{\partial\over\partial a_+} \,\,,\eqno\eq$$
with the familiar matrix model form
$h_{JM}=a_+^{J+M}\,a_-^{J-M}$. In the continuum
approach the $w_\infty$ generators are integrals of
conserved currents which are for closed string theory
constructed as
$$\eqalign{
&Q_{JM}=\oint {dz\over 2\pi i}\,\,W_{JM}\,(z,\bar z)\,\,,\cr
\noalign{\vskip 0.2cm}
&W_{JM}\,(z,\bar z)=\Psi^+_{J+1,M}(z)\,{\cal O}_{JM}(\bar z)\,\,.}
\eqno\eq$$
One can study the action of these operators on the
the tachyon vertex operators.
A formula derived by Klebanov [\Kl] reads
$$Q_{M+N-1}\,T^+_{k_1}(0)\int T^+_{k_2}\ldots\int T^+_{k_N}=
F_{N,M}(k_1,\ldots,k_N)\, T^+_{-\sum k_i +M}\,\,,\eqno\eq$$
where
$$F_{N,M}(k_1,\ldots,k_N)=2\pi^{N-1}\,N!\,k\,
{\Gamma(2k)\over \Gamma(1-2k)}\,
\prod_{i=1}^N\,{\Gamma(1-2k_i)\over \Gamma(2k_i)}\,\,,\eqno\eq$$
with a similar formula for the action of
$Q_{-M+N-1}$ on $N$ oppositely moving $T^-$ tachyons.
These representations of the $w_\infty$ generators on
tachyon vertex operators are clearly comparable to the
direct representation obtained in the matrix model
(or more precisely collective field formalism). The
comparison and agreement of these representations
 is the closest one comes in being
able to identify the two approaches.

The conformal (vertex operator) formalism gives a very
elegant summary of Ward identities in the form of general
(master) equation. We end this section with a short description
of this equation [\ZW,\Ve,\KlPa]. It  follows from BRST
invariance of the discrete state vertex operators
$$\{ Q_{\rm BRST}, c(z) W_{JM}(z)\}=0\,\,,\eqno\eq$$
which implies that for general tachyon correlation function
$$\big\langle \{ Q_{\rm BRST}, c\, W_{JM}\}\,
V^\pm_{k_1} \ldots V^\pm_{k_n}\,\big\rangle =0\,\,.\eqno\eq$$
Changing to operator formalism
$$\sum_{{\rm perm}} \big\langle V^\pm_{k_1}\,
\{ Q_{\rm BRST}, c\, W_{JM}\}\Delta V^\pm_{1} \ldots
\Delta V^\pm_{1}\,V^\pm_{k_n} \big\rangle =0\,\,,\eqno\eq$$
allows one to eliminate $ Q_{\rm BRST}$. The vertex operators are all
BRST invariant while the propagator is essentially the
inverse of $Q$:
$$[Q,\Delta]=\Pi_{L_0 -\bar L_0}\,b^-_0\,\,,\eqno\eq$$
where the $\Pi$ projects on the subspace
$(L_0 -\bar L_0)|\Phi_i \rangle =0$. The final form of the
Ward identity then follows
$$\sum_{{\rm partitions}\atop {m+m'=n}}\,
\langle V_{i_1}\ldots  V_{i_m}\,\Phi \rangle
\langle \Phi  V_{j_1}\ldots  V_{j_{m'}}\,c W_{JM} \rangle =0\,\,.
\eqno\eq$$

Most of the considerations of this section and most of the studies
of the $w_\infty$ symmetry are performed in the extreme limit where the
cosmological term is ignored. This is particularly the case for the
continuum, conformal field theory approach. Some attempts to
extensions and inclusion of the nontrivial cosmological constant
effect were made however. In the matrix model the cosmological
term is introduced in a simple and elegant way corresponding
 to nonzero Fermi energy
$$h_0={1\over 2}\,(p^2 -x^2)\quad \to \quad
h_\mu={1\over 2}\,(p^2 -x^2)+\mu\,\,.\eqno\eq$$
Since the ground ring generators $a_\pm$ were identified
to be analogues to $p\pm x$ it is then expected that an
equivalent deformation from $a_+ a_- =0$ can be established in
the continuum conformal field theory approach. This is seen
by considering the action of a ground ring on tachyons. For
$\mu =0$ it reads
$$\eqalign{&a_+\,c\,\bar c \,\tilde T^+_k = c\,\bar c \,
\tilde T^+_{k+1}\,\,,\cr
\noalign{\vskip 0.2cm}
&a_-\,c\,\bar c \,\tilde T^+_k = 0\,\,.}\eqno\eq$$
The effect of cosmological perturbation
$\mu\,T^+_{k=0}$ is found by evaluating the first order
perturbation theory contribution
$$a_-\,c\,\bar c \,\tilde T^+_k =
-a_-\,c\,\bar c \,\tilde T^+_k(0)\,
\Bigl( \mu\int d^2 z\,,\tilde T^+_{k=0}(z) \Bigr)\,\,.\eqno\eq$$
On the right-hand side $a_-$ essentially fuses the two
tachyon operators into one giving (to first order)
$$a_- \,\tilde T^+_k = -\mu \,\tilde T^+_{k-1}\,\,.\eqno\eq$$
This replaces the second relation above and now the nonzero
Fermi level condition $a_+ a_- =-\mu$ results. To first
order [\WiGR,\Ba] one then has an agreement with the matrix model.
This is encouraging and one would clearly like to establish the complete
agreement at an exact level.

\bigskip
\chapter {$S$-matrix}
\medskip

Let us now describe the complete tree-level $S$-matrix of the
$c=1$ theory. In the previous sections we have seen the ``bulk''
scattering amplitudes which follow from the Ward identities or
are computed in conformal field theory. The complete
$N$-point scattering amplitude
$S_N=\langle T_{p_1}  T_{p_2}\ldots  T_{p_N} \rangle$
takes the (factorized) form
$$S_N=\prod_{i=1}^N (-)\mu^{i p_i}\,
{\Gamma(-ip_i)\over \Gamma(+ip_i)}\,\,A_{\rm coll}(p_1,\ldots,p_N)\,\,.
\eqno\eq$$
The external leg factors are associated with a field
redefinition [\GrKl] of vertex operators
$$T_k^\pm = {\Gamma(\mp k)\over \Gamma(\pm k)}\,\,
\tilde T_k^\pm\,\,.\eqno\eq$$
It is the redefined tachyon vertex operator $\tilde T$
that found its natural role in collective field theory.

The external leg factors of the full $S$-matrix have a
very relevant physical meaning which we now discuss. In
Minkowski space-time, $(k=ip)$, one has
$$\Delta =\mu^{\mp ip}\,\,
{\Gamma(\pm ip)\over \Gamma(\mp ip)}\,\,.\eqno\eq$$
So, the factors $\Delta =e^{i\theta_p}$ are pure phases.
As such they give no contributions to the actual
transition amplitudes and could be ignored.
The fact is however that they carry physical information
on the nature of tachyon background. The factors exhibit
poles at discrete imaginary energy
$$p\sqrt{2}=in\,,\quad n=1,2,3,\ldots \eqno\eq$$
If we consider a process with an incoming tachyon and
$N$ outgoing ones, the discrete imaginary value of the
incoming momenta signifies the resonant on-shell
process in which a certain number  $r$ of Liouville
exponentials participate
$$\langle T_-\,(\mu\,e^{-\sqrt{2}\varphi})^r\, T_+ \ldots
t_+ \rangle\,\,.\eqno\eq$$
The on-shell condition in this case indeed gives
$$i\sqrt{2}\,p =-(r+N-1) \,\,,\eqno\eq$$
in agreement with the discrete imaginary energy poles noted above.

In collective field theory the
external leg factors are associated with
a field redefinition given by an integral transformation. The transformation
comes from the change of coordinates between the Liouville  and the
time-of-flight variables. It is the later that appears naturally in the
collective field formalism  and as we have seen provides a simple description
of the theory.
Let us recall the basic  (Wilson) loop operator of the matrix model
with its Laplace transform
$$\hat W(\ell,t)\equiv \Tr\,(e^{-\ell M}) =
W_0 +\int dx\,e^{-\ell x}\,\partial_x \eta \,\,.\eqno\eq$$
After the change to the time of flight coordinate
$x=\sqrt{2\mu}\,\cosh\tau$ and the explicit identification of the Liouville
$\ell =2e^{-\varphi/\sqrt{2}}$ the integral
transformation results
$$\hat W(\ell,t)=\int_0^\infty d\tau\,\,
{\rm exp}\,\Bigl[-2\sqrt{2\mu}\,e^{-\varphi/\sqrt{2}}\,
\cosh \tau\Bigr]\,\partial_\tau\,\eta(\tau,t)\,\,,\eqno\eq$$
We have seen in our earlier study of the linearized theory that
this integral transformation takes the
Liouville operator into a Klein-Gordon operator
$$(\partial_t^2 -\partial_\tau^2)\,\eta
\quad \Longleftrightarrow \quad
\bigl(\partial_t^2 -{1\over 2}\,\partial_\varphi^2
+4\mu e^{-\sqrt{2}\,\varphi}\,\bigr) \hat W\,\,.\eqno\eq$$

The integral transformation therefore expresses the tachyon field in
terms of a simple Klein-Gordon field $\eta(\tau,t)$:
$$T(\varphi,X)\equiv e^{-\sqrt{2}\,\varphi}\,
\hat W(\ell,t) = \int_0^\infty d\tau\,\,
{\rm exp}\,\,\Bigl[-2\sqrt{2\mu}\,e^{-\varphi/\sqrt{2}}\,
\cosh \tau\Bigr]\,\partial_\tau\,\eta \,\,,\eqno\eq$$
with the expected relation between the matrix model and
string theory times $X=\sqrt{2}\,t$.
The correlation functions of the
tachyon field $T$  are then expressible in terms of correlation
functions of the collective field
$\eta$. The transformation described takes plane wave solutions of the
Klein-Gordon equation
$$\eta(\tau,t)=\int_{-\infty}^\infty\,{dp\over p}\,\,
\tilde\eta (p)\,e^{-ipt}\,\sin (p\tau) \eqno\eq$$
into Liouville solutions
$$T(\varphi,t)=\int dp\,e^{-ipt}\,\gamma(p)\,
K_{ip}(2\sqrt{\mu}\,e^{-\varphi/\sqrt{2}}\,)\,
\tilde\eta(p)\,\,.\eqno\eq$$
The above redefinition of the
in--out fields does have an effect in supplying external leg
factors. The asymptotic behavior of $T(\varphi,t)$ reads
$$T\sim \int dp\,e^{-ipt}\,\Bigl(\Gamma(ip)\mu^{-ip/2}\,
e^{ip\varphi/\sqrt{2}} +
\Gamma(-ip)\mu^{ip/2}\, e^{-ip\varphi/\sqrt{2}} \Bigr)\,\,,\eqno\eq$$
giving the reflection coefficient
$$R(p)=-\mu^{ip}\,\,{\Gamma(-ip)\over \Gamma(ip)} \eqno\eq$$
for each external leg of the $S$-matrix.

After this redefinition the problem is reduced to
calculating amplitudes in collective field theory:
$ A_{\rm coll}\,(p_1,\ldots,p_N)$. There, as we have already seen,
one can derive an exact relationship between the in- and
out-field which contains the complete information about the $S$-matrix.
The solution to the scattering problem
can also be directly deduced from the exact oscillator
states. It is this procedure that turns out to be the most
straightforward and we now describe it in detail.

Consider the exact (tachyon) creation--annihilation operators
in collective field theory
$$B_{\pm ip}^\pm = \int {dx\over 2\pi}\,
\left\{ {(\alpha_+ \pm x)^{1\pm ip}\over 1\pm ip} -
{ (\alpha_- \pm x)^{1\pm ip}\over 1\pm ip} \right\}\,\,.
\eqno\eq$$
Previously we have seen that at fixed time these operators serve
as exact creation--annihilation operators of the
nonlinear collective Hamiltonian. Let us now follow the
time-dependent formalism (we describe here the derivation given
in [\JRvT]). The exact creation--annihilation operators have a simple
time evolution
$$B_{\pm ip}^\pm (t)= e^{-ipt}\, B_{\pm ip}^\pm (0)\,\,.
\eqno\eq$$
Consequently the quantity
$$\hat B_{\pm ip}^\pm \equiv e^{ipt}\, B_{\pm ip}^\pm (t)
\eqno\eq$$
is time-independent. We can simply look at the operator
$\hat B_{\pm ip}^\pm$ at asymptotic times $t= \pm\infty$
and obtain a relationship between the in and out fields~\inoutfields.
The  operator
$$B_{\pm ip}^\pm = \int {dx\over 2\pi}\,
\left\{ {(\alpha_+ \pm x)^{1\pm ip}\over 1\pm ip} -
{ (\alpha_- \pm x)^{1\pm ip}\over 1\pm ip} \right\}
\eqno\eq$$
contains contributions from both $\alpha_+$ and
$ \alpha_-$. At $t=\pm \infty$ only one of the terms
survives and we have an identity
(recall that $\hat B$ is time-independent),
$\hat B(+\infty) =\hat B(-\infty)$, which reads
$$\int {dx\over 2\pi}\,(\alpha_\pm \pm x)^{1\pm ip} =
\int {dx\over 2\pi}\,(\alpha_\mp \pm x)^{1\pm ip} \,\,.
\eqn\scatteq$$
It relates in and out fields ($\alpha_\pm$ now represent the
asymptotic fields). This is the scattering equation. It contains the
full specification of the $S$-matrix.

One can evaluate
and expand the left- and right-hand side of the Eq.~\scatteq.
Shifting by the static background
$$\alpha_\pm(t,x) \approx \pm \bigl(x-{1\over 2x}\bigr)
+{1\over 2x}\,\hat\alpha_\pm(t\mp \tau)\,\,,\eqno\eq$$
the left-hand side becomes
$$L=\int dx\,(-2x)^{1\pm ip}\,\Bigl\{
1-{(1\pm ip)\over 4x^2}\,(1\pm \hat\alpha_\pm) +
{\cal O}({1\over x^4})\Bigr\}\,\,.\eqno\eq$$
After a change of integration variable
$x=\cosh \tau \approx e^\tau/2$ the
${\cal O}(1/x^4)$ terms are seen to decay away
exponentially and what remains is only the term linear in
$\hat\alpha_\pm$. For the right-hand side we
simply find
$$R=\int dx\,\Bigl(-{1\over 2x}\Bigr)^{1\pm ip}\,
(1\pm \hat\alpha_\mp)^{1\pm ip}\,\,.\eqno\eq$$
The scattering equation then becomes
$$\int dz\,e^{-ipz}\,{1\over \mu}\,\alpha_\pm (z)=
{1\over 1\pm ip}\,\int dz\, e^{-ipz}\,
\left\{ \Bigl(1\pm {1\over\mu}\,\alpha_\mp \Bigr)^{1\mp ip}
-1 \right\}\,\,,\eqno\eq$$
giving the solution for the in-field as a function of the out-field
and vice versa. We have also explicitly restated the string
coupling constant $g_{\rm st}= 1/\mu$. This solution was
originally obtained [\MoPl] by explicitly solving the functional
relationships between the left and right collective
field components $\alpha_+$ and $\alpha_-$ given
earlier. We see here that it directly follows from the
exact oscillator states.

Before proceeding with the consideration of the $S$-matrix
let us note that the solution found has a reasonable
strong coupling limit. Indeed, for $\mu \to 0$ one is
lead to choose $ip$ to be an integer, $ip=N$, and the strong coupling
relation
$$\int dz\,e^{-Nz}\,\alpha_+ (z)=
{1\over 1+N}\,\int dz\, e^{-Nz}\,\Bigl({1\over\mu}\,
\alpha_- \Bigr)^{N+1}\eqno\eq$$
results. It is recognized as a statement specifying the bulk
amplitude where the $S_{1,N}$ and $S_{N,1}$ amplitudes
were nonzero with the momenta of the first (last) particle
being equal to $-1+N$. We have already used
relations of the above type (and their $w_\infty$ generalizations)
in our discussion of the Ward identities at (strong coupling)
$\mu=0$.

One can explicitly perform the series expansion in
$g_{\rm st}=1/\mu$, it reads
$$\hat\alpha_\pm (z) =\sum_{l=1}^{\infty}\,
{(-g_{\rm st})^{l-1}\over l!}\,\,
{\Gamma(\mp\partial +1)\over \Gamma(\mp\partial +2-l)}\,\,
\hat\alpha_\mp^l (z)\,\,.\eqno\eq$$
The $S$-matrix is defined in terms of momentum space
creation--annihilation operators
$$\pm\alpha_\pm (z) =\int {dp\over 2\pi}\,e^{-ipz}\,
\tilde\alpha_\pm (p)\,,\qquad\quad
[\tilde\alpha_\pm (p),\tilde\alpha_\pm (p')]=p\,\delta(p+p')\,\,,
\eqno\eq$$
with $\tilde\alpha_- (-p)$ and  $\tilde\alpha_+ (-p)$
being the in--out creation operators, respectively
($\alpha(-p)$ and $\beta(-p)$ in our earlier notation).
In momentum space
$$\tilde\alpha_\mp (p)= \sum_{l=1}^{\infty}\,
{(-g_{\rm st})^{l-1}\over l!}\,\,
{\Gamma(1 \pm ip)\over \Gamma(2\pm ip -l)}\,\,
\int dp_i\,\,\delta(p-\sum p_i)\,
\tilde\alpha_\pm(p_1)\ldots \tilde\alpha_\pm(p_l)\,\,,\eqn\momsp$$
and the $n\to m$ $S$-matrix element is defined by
$$A_{\rm coll} (\{p_i\}\to\{p'_j\}) =
\langle 0|\,\prod_{j=1}^m\,\tilde\alpha_+ (p'_j)\,\,
\prod_{i=1}^n\,\tilde\alpha_- (p_i)\,|0\rangle\,\,.\eqno\eq$$
Consider for example $n=1, m=3$ which is the four-point amplitude
$$A_{1,3}=\langle 0| \alpha_+(p'_1)\, \alpha_+(p'_2)\,
\alpha_+(p'_3)\,\alpha_-(-p_1)\,|0\rangle\,\,.\eqno\eq$$
It is given by the cubic term ${\cal O}(g_{\rm st}^2)$
in the expansion of \momsp\ which equals
$$\alpha_- (-p_1)= {1\over 3!}\,\,g_{\rm st}^2\,\,
{\Gamma(1-ip_1)\over \Gamma(-1-ip_1)}\, \int dp'_i\,
\delta(p_1-\sum p'_i)\,
\alpha_+(p'_1)\, \alpha_+(p'_2)\, \alpha_+(p'_3)\eqno\eq$$
and
$$A_{1,3}(p_1;p'_1,p'_2,p'_3)=
i\, g_{\rm st}^2 \,p_1 p'_1 p'_2 p'_3\,(1+ip_1)\,\,.
\eqno\eq$$
In general, an arbitrary amplitude is given in [\MoPl] to read
$$A_{n,m}=i\,(- g_{\rm st})^{n+m-2}\,
\Bigl( \prod_{i=1}^n p_i \Bigr) \Bigl( \prod_{j=1}^m p'_j \Bigr)\,
{\Gamma(-ip_n)\over \Gamma(1-m-ip_n)}\,\,
{\Gamma(1-m-i\Omega)\over \Gamma(-3-n-m-i\Omega)}\eqno\eq$$
where $\Omega=\sum_{i=1}^n\, p_i$ and the result is valid
in the kinematic region $p_n> p'_k > \sum_{j=1}^{n-1}\,p_j$.
This completes the derivation of the collective tree-level
amplitudes.

\bigskip
\chapter {Black Hole}
\medskip

Two-dimensional string theory possesses another interesting curved
space solution taking the form of a black hole. It is
described exactly by the $SL(2,\IR)/U(1)$
nonlinear $\sigma$-model
$$S_{\rm WZW} = {k\over 8\pi}\int d^2 z\,\Tr\left( g^{-1}
\partial g g^{- 1} \bar{\partial} g\right) -
ik \Gamma_{\rm WZW} + {\rm Gauge}\,\,,\eqno\eq$$
with $k = {9\over 4}$. This then gives the required central charge
$c = {3k\over k-2} -1 = 26$.
As such the model should be thought of as a different classical solution of the
same theory. We have in the earlier lectures seen that the flat
space-time string theory is very nicely and very completely described by a
matrix model. The black hole solution is however markedly different from the
$c=1$ theory. It is characterized by the absence of tachyon
condensation and a nontrivial metric and dilaton field:
$$\eqalign{&T(X) = 0\,\,,\cr
\noalign{\vskip 0.1cm}
&(ds)^2= - {k\over 2} \, {dudv\over M-uv}\,\,\cr
\noalign{\vskip 0.1cm}
&D = \log (M-uv)\,\,. }\eqno\eq$$
Here a particular $SL(2,\IR)$ parametrization is chosen:
$g=\left( {\alpha\atop -v} \,\,{u\atop \beta}\right) $,
$\alpha\beta +uv=1$ and $M$
is the black hole mass.  There is actually a parametrization (related
to the $c=1$ theory) in which the $\sigma$-model Lagrangian reads
$$\eqalign{S_{\rm eff}={1\over 8\pi} \int d^2 z\,\, \Bigl\{
&(\partial X')^2 + (\partial\varphi')^2 - 2\sqrt{2} \phi' R^{(2)} \cr
&+ M \vert{1\over 2\sqrt{2}} \partial\varphi' + i \sqrt{{k\over2}}
\partial X'\vert^2\, e^{-2\sqrt{2} \varphi'} \Bigr\}\,\,.}\eqno\eq$$
This parametrization corresponds to a linear dilaton but
in contrast to the $c=1$ theory, one has a black hole mass term
perturbation represented by a gravitational vertex operator instead of
the cosmological constant term given by a
tachyon operator $e^{-\sqrt{2} \varphi}$.
One of the surprising facts is however that there exists a classical
duality transformation that can be used to relate the two sigma models to
each other [\MaSh]. From this there arises a hope that one could
possibly be able to describe the black hole by a matrix model also.
More generally from a string field theory viewpoint one would hope to
be able to describe different classical solutions in the same setting.
In what follows we will present some joint work done with T. Yoneya
on this subject [\JeYo]. For other different attempts see
[\DDMW, \MuVa].

First insight into the black hole problem is gained by
considering the linearized tachyon [\DVV] field in the external background.
In the conformal field theory  this is given by the zero mode Virasoro
condition. The Virasoro operator $L_0(u,v)$ consists of
two parts,
$$L_0= -\Delta_0 + {1\over 4}\,(u\partial_u-v\partial_v)^2\,\,,
\eqno\eq$$
where $\Delta_0$ is the Casimir operator of $SL(2,\IR)$.
The Virasoro condition for the linear tachyon field (vertex operator)
reads:
$$L_0(u,v)T\equiv {1 \over k-2}\Bigl[(1-uv)\partial_u \partial_v -
{1 \over 2}(u\partial_u + v\partial_v)-{1 \over 2k}
(u\partial_u-v\partial_v)^2 \Bigr]\,T=T\,\,.\eqn\Vircond$$

The on-shell  tachyon corresponds to the continuous
representation of $SL(2,\IR)$ which has eigenvalues
$\Delta_0= -\lambda^2 - {1 \over 4} \, \quad (\lambda = {\rm  real})$
and  $-i\partial_t= 2i\omega$ with the on-shell condition
$\lambda^2 = 9\omega^2$ at $k=9/4$.
The above equation can be interpreted as corresponding to a covariant
Laplacian $L_0=-{1 \over 2e^{D}\sqrt{G}}\,\partial^{\mu}\,e^{D}\,
\sqrt{G}\,G_{\mu\nu}\,\partial_{\nu}$ in the  background space-time metric
$G_{\mu\nu}$ and dilaton $D$,
which can be read off  from Eq.~\Vircond
$$\eqalign{
&ds^2={k-2 \over 2}\,[\,dr^2 - \beta^2(r)\,d\bar t^2\,]\,\,,\cr
\noalign{\vskip 0.15cm}
&D = \log \,\bigl(\sinh {r\over\beta(r)}\,\bigr) + a\,\,,\cr
\noalign{\vskip 0.15cm}
&\beta(r)= 2\,(\coth^2{r \over 2} -{2 \over k})^{-1/2}\,\,.}\eqno\eq$$
These are  then candidates for the
``exact'' background.
Here the new coordinate $r$ and time $\bar t$ are defined by
$$u= \sinh{r \over 2} \,e^{\bar t}\,,\qquad\quad
v= -\sinh{r \over 2} \,e^{-\bar t}\,\,.\eqno\eq$$
These variables describe the static exterior region outside the
event horizon located  at $r=0$. The constant $a$
determines the  mass of the black hole
$$M_{{\rm bh}}= \sqrt{{2 \over k-2}}\,\,e^a\,\,.\eqno\eq$$

The  exact metric can be shown to be free of curvature singularity.
However, one still has a ``dilaton singularity'' at $uv=1$
where the string coupling $g_{\rm st}\sim e^{-D/2}$
diverges. In terms of the variables $u$ and $v$, the dilaton reads
$$D=\log\,\Bigl[4\Bigl(-uv(1-uv)(-{(1-uv) \over uv}-{2 \over k})
\Bigr)^{1/2}\Bigr]+a\,\,,\eqno\eq$$
and the region $uv >1$ corresponds to a disjoint
region with a naked singularity.

The free parameter $a$ can be eliminated by a scale transformation
$$u \rightarrow M^{-1/2}\,u\,,\qquad v \rightarrow M^{-1/2}\,v\,,
\qquad M\equiv e^a\,\,.\eqno\eq$$
This introduces the black hole mass parameter in more explicit way,
where one replaces $(1-uv)$ by $(M-uv)$ in the expressions for the
dilaton and the metric. An important relation is the connection of the
string coupling constant with the parameter
$a$, or rather the black hole mass. In general, the dilaton field
determines the string coupling constant and in the present case one
obtains
$$g_{\rm st}\,(r=0)  \propto e^{-a/2}= M^{-1/2}\,\,.\eqno\eq$$
This is to be compared  with the dependence of
$g_{\rm st} \propto \mu^{-1}$
on the cosmological constant in flat space-time. One notes the
different power which comes from the different scaling dimensions
of the two parameters.
The two backgrounds become identical in the asymptotic region.
Consider the asymptotic  behavior of the Virasoro operator
and the dilaton  when $r\rightarrow \infty$. Using
$u\sim e^{{r\over 2}+\bar t}\,, \,\,
v \sim e^{{r\over 2}-\bar t}$ one finds
$$\eqalign{&L_0\sim {1 \over 4(k-2)}\,(\partial_r^2+\partial_r)+
{1 \over 4k} \,\partial_{\bar t}^2\,\,,\cr
\noalign{\vskip 0.1 cm}
&D\sim r + a - \log 4\,\,.}\eqno\eq$$
This is the form of Virasoro operator in the linear dilaton case,
the parameters $r, \bar t$ are identified asymptotically with the
$\varphi$ and $t$ for the linear dilaton background as
$$\eqalign{\bar t &\leftrightarrow\sqrt{{1 \over 2k}}\,\,t=
{\sqrt{2} \over 3}\,\,t\,\,,\cr
\noalign{\vskip 0.1 cm}
r &\leftrightarrow\sqrt{{2 \over k-2}}\,\,\varphi=2\sqrt{2}\,\varphi\,\,.}
\eqno\eq$$
For the conjugate  momentum and energy, the correspondence is then
$$\eqalign{&ip_{\varphi}= -\sqrt{2}+i2\sqrt{2}\,\lambda = -\sqrt{2} +
{i \over \sqrt{2}}\,\,p_{\tau}\,\,,\cr
\noalign{\vskip 0.1 cm}
&ip = i{2\sqrt{2} \over 3}\,\omega={i \over\sqrt{2}}\,p_{t}\,\,.}
\eqn\enmom$$
This implies a  one-to-one correspondence of tachyon states in the
black hole and linear dilaton backgrounds. There is also a correspondence
between the discrete states spectra in the two theories.

In the Minkowski metric, the spectrum of the
discrete states  for the black hole is isomorphic to that in the linear
dilaton background. In particular,
the first nontrivial discrete state with zero energy
($j=1, m=0$ or $ip_{\varphi}= -2\sqrt{2}, \, p =0$) is
identified with the operator associated with the mass
of black hole, as can be seen from  the first correction to the
asymptotic behavior of the exact space-time metric
$$ds^2 \sim {k-2 \over 2}\,\Bigl[dr^2 - {4k \over k-2}\,\Bigl(1-
{4k \over k-2}\,e^{-r} + {\cal O}(e^{-2r}) \Bigr)dt^2\Bigr]\,\,.
\eqno\eq$$
It is important to note that the $\varphi$ momentum
is twice that of the operator corresponding to
tachyon condensation.

The solutions of the tachyon Virasoro conditions describe the scattering of
a single tachyon on the black hole. It represents
one of the few quantities that has been rigorously computed
in black hole string theory [\DVV]. The amplitude provides some nontrivial
physical insight and is obtained as follows. One writes an  integral
representation for the solution with definite energy $\omega$ and momentum
$\lambda$ as
$$\int_C {dx\over x}\, x^{-2i\omega}\,\bigl(\sqrt{M-uv}
+{u \over x}\bigr)^{-\nu_-}\, \bigl(\sqrt{M-uv} -vx\bigr)^{-\nu_+}\,\,,
\eqn\intrep$$
with $\nu_{\pm}={1 \over 2}-i(\lambda \pm \omega)$. In general, one has
four different contours of integration with two linearly independent
solutions corresponding, for example, to the coutours
$C_2\equiv[-u\sqrt{M-uv}, 0\,],\,
C_4\equiv(-\infty,\nu^{-1}\sqrt{M-uv}\,]$ as
($y\equiv uv =-\sinh^2 {r \over 2}$):
$$\eqalign{&T_{C_2}= U_{\omega}^{\lambda}=e^{-2i\omega \bar t}\,
F_{\omega}^{\lambda} (y)\,\,,\cr
\noalign{\vskip 0.2cm}
&T_{C_4}= V_{\omega}^{\lambda}=e^{-2i\omega \bar t}\,
F_{-\omega}^{\lambda} (y)\,\,,}\eqno\eq$$
where
$$F_{\omega}^{\lambda}(y)= (-y)^{-i\omega}B(\nu_+, \nu_-)
F(\nu_+, \nu_-, 1-2i\omega, y)\,\,.\eqno\eq$$
The asymptotic behaviors of the solutions are,
for $r \rightarrow 0 \, ({\rm horizon})$:
$$\eqalign{&U_{\omega}^{\lambda} \sim \beta(\lambda, \omega)
({u \over \sqrt{M}})^{-2i \omega}\,\,,\cr
\noalign{\vskip 0.1cm}
&V_{\omega}^{\lambda} \sim \beta(\lambda, -\omega)
(-{v \over \sqrt{M}})^{-2i \omega}\,\,}\eqno\eq$$
while for null-infinity $r \rightarrow \infty$:
$$F_{\omega}^{\lambda}\sim \alpha (\lambda, \omega)(-y)
^{-{1 \over 2}+ i\lambda} + \alpha (-\lambda,\omega)(-y)
^{-{1 \over 2}- i\lambda}\,\,,\eqn\asymp$$
where
$$\eqalign{\alpha(\lambda, \omega) &={\Gamma(\nu_+)\,
\Gamma(\bar\nu_- - \nu_+)\over \Gamma(\bar\nu_-)}\,\,,\cr
\noalign{\vskip 0.1cm}
\beta(\lambda, \omega) &= B(\nu_+, \bar \nu_-)\,\,.}\eqno\eq$$
We  see  that $U_{\omega}^{\lambda}$ describes
a wave coming from past null-infinity scattering on
the black hole, while $V_{\omega}^{\lambda}$ describes a wave emitted
by the white hole crossing the past event horizon.
The solution $U_{\omega}^{\lambda}$ gives the $S$-matrix elements of
tachyons, incoming from the asymptotic flat region at past null-infinity and
scattered out to future null-infinity. On-shell
$\omega = 3\lambda \,(>0)$, and this solution gives the reflection
and transmission coefficients as ratios of the coefficients appearing
in the above asymptotic forms:
$$\eqalign{R_B(\lambda)&= {\alpha(\lambda, \omega)\over
\alpha (-\lambda, \omega)}\,\,,\cr
\noalign{\vskip 0.1cm}
T_B(\lambda)&={\beta(\lambda, \omega)\over\alpha (-\lambda,\omega)}\,\,.}
\eqn\rtcoeff$$
The reflection and abosorbtion coefficients satisfy the unitarity relation
$$|R_B|^2 + {\omega \over \lambda} |T_B|^2 = 1\,\,.\eqno\eq$$
This describes the two-point correlation function and it
is of major interest to formulate a full quantum field theory in
the presence of a black hole which would be capable of giving
general $N$-point scattering amplitudes and correlation functions.
One is also very interested in being able to evaluate loop effects
and even to discuss formation and evaporation of black holes in
the general field theoretic framework.
In the absence of a general theory one can try to follow the analogy
with the $c=1$ theory and attempt to guess the structure required for the
black hole. This is what was done in [\JeYo]. Persuing the above
analogy we can postulate again that string theory in the black hole
background is described by
a factorized $S$-matrix. It is then reasonable to expect that the
external leg factors of the full $S$-matrix
are again determined through a non-local field redefinition whose role
is to connect
the Virasoro equations in the black hole background with the free
massless Klein-Gordon equation. The main part of the $S$-matrix is then
to be determined. The suggestion based on the analogy with the $c=1$
theory is that one again has a description  in terms of a
matrix model and the associated  collective field theory.
To simulate the black hole background the matrix model is expected to
include a deformation from the standard inverted oscillator potential.
 There as yet exist no general principles for constructing the theory but
one can make certain concrete suggestions on the eventual form of the
matrix model.
Let us describe first  the expected form for the external leg factors.
These are supplied by a field redefinition whose
purpose is to reduce the black hole background Virasoro condition
to the scalar free field equation. We have summarized the  black hole
Virasoro equation and its solutions in detail, so let us
consider the integral representation \intrep\
with the contour $C_2$ which is appropriate for
the scattering problem in the exterior region $(u>0, v<0)$:
$$U_{\omega}^{\lambda}(u,v)=
\int_{C_2}{dx \over x} x^{-2i\omega}(\sqrt{M-uv}+ {u \over x})^{-\nu_-}
(\sqrt{M-uv}-vx)^{-\nu_+}\,\,.\eqno\eq$$
Since the spectrum of the on-shell solution has a one-to-one
correspondence through \enmom\ with that of the free Klein-Gordon
equation, it is natural to make the following change of integration
variable:
$$\eqalign{&(\sqrt{M-uv} + {u \over x})^{-1}\,(\sqrt{M-uv} - vx)
= e^{-4t /3} \,\,,\cr
\noalign{\vskip 0.1cm}
&(\sqrt{M-uv} + {u \over x})\,(\sqrt{M-uv} - vx)
=e^{-4\tau} \,\,.}\eqno\eq$$
The integral formula for the solution takes  now the form
$$U_{\omega}^{\lambda}
=\int_{-\infty}^{\infty}dt \int_0^{\infty} d\tau\,
\delta({u\,e^{-2t/3}+v\,e^{2t/3} \over 2}-\sqrt{M}\cosh 2\tau)\,
e^{-4i\omega t/3}\cos 4\lambda \tau\,\,.\eqno\eq$$
This is seen to be an integral transform of a  Klein-Gordon plane wave with
momentum and energy
$$p_{\tau}=4\lambda\,,\qquad\quad p_{t}={4\over 3}\,\omega\,\,.
\eqno\eq$$
Since the plane waves are recognized as natural eigenstates of the
linearied collective field  terms of time-of-flight variable
we have the candidate for the non-local field redefinition
$$T(u,v) =
\int_{-\infty}^{\infty}dt \int_0^{\infty} d\tau\,
\delta({u\,e^{-2t/3}+v\,e^{2t/3} \over 2} - \sqrt{M} \cosh 2\tau)
\gamma (i\partial_{t})\partial_{\tau}\eta (t, \tau)\,\,,
\eqn\bhfieldred$$
where $\gamma(i\partial_{t})^* = \gamma (-i\partial_{t})$
is an arbitrary weight function to be fixed by normalization
condition.

In terms of the Fourier decomposition
$$\eta(t,\tau)=\int_{-\infty}^\infty {dp\over p}\,\,
\tilde\eta (p)\,e^{-ipt}\,\sin\,p\tau\,\,,\eqno\eq$$
it reads
$$T(u,v) = \int_{-\infty}^{\infty}dp\,\,
\tilde\eta(p)\,\gamma (p)\,U_{\omega (p)}^{\lambda(p)}(u,v)\,\,,
\eqno\eq$$
with $\omega (p) = 3p/2, \lambda (p) = p/2$. In particular, the
asymptotic behavior for $y\rightarrow \infty$ is
$$\eqalign{T(u,v) = \int_{-\infty}^{\infty}dp\,\,
\tilde\eta(p)\,\gamma(p)
\Bigl[&(-y)^{-{1 \over 2}+i\lambda(p)}\,\alpha\bigl(\lambda (p),\omega(p)
\bigr) +\cr
&(-y)^{-{1 \over 2} - i \lambda (p)}\,\alpha\bigl(-\lambda (p),\omega(p)
\bigr)\Bigr]\,\, e^{-2i\omega(p) t}\,\,.}\eqno\eq$$
This shows that an asymptotic wave packet of $\eta$ field is
transformed into a deformed wave packet of the tachyon field.
The integral transformation that we have given suplies the leg factors of the
conjectured black hole $S$-matrix.
Even if the factorization becomes only an
approximate feature of the full theory one could expect that the
factorization holds near the poles of the $S$-matrix.

Let us now study the possible resonance poles produced by the external leg
factors. It turns out that studying the location of these poles
gives useful and nontrivial constraints on the full $S$-matrix.
 From the asymptotic behavior of \asymp\ and the associated
reflection coefficient \rtcoeff, we see that the positions
of the resonance poles are
$$i4\lambda = i{4 \over 3}\omega=i\sqrt{2}\,p_t =-2,-4,-6,\ldots\,\,.
\eqno\eq$$
This contrasts with the case of the usual $c=1$ model where we
have poles at all negative integers of the corresponding energy.
On the other hand, if we consider an amplitude for
an incoming tachyon with producing $N-1$ outgoing
tachyons, the energy and momentum conservation laws
are satisfied when the energy of incoming tachyon obeys
$$i\sqrt{2}p_t = -(2r + N-2)\,\,,\eqno\eq$$
where $r$ now counts the number of insertions of the
black hole mass operator. The factor $2$ multiplying  $r$
comes about because the momentum carried by the
black hole mass is twice that of the tachyon condensation.
Comparing next the two expressions for the location of the poles
we seee that these are consistent only if $N$ is even. More precisely,
only the even $N=2k$ point amplitudes are to be nonzero while the odd
$N=2k+1$ point amplitudes should vanish. This represents a strong requirement
on the form of the complete theory.

We are than lead to the main problem of specifying the full dynamics
in the form of a generalized matrix model.
In the limit of vanishing black hole mass, the black hole
background reduces to the linear dilaton vacuum.
This is a singular limit in the sense that
the string coupling diverges, corresponding to the $c=1$ matrix
model with vanishing 2d cosmological constant $\mu = 0$,
or zero Fermi energy. Since, according to our
hypothesis, the deformation corresponding to non-vanishing black hole
mass cannot be described by the usual matrix model,
we have to seek for other possible
deformations than the one given by  the
Fermi energy.  We  assume that the Fermi energy is
kept exactly at zero, while the Hamiltonian itself is modified.

{}From the earlier analysis we have several hints or constraints which the
modified Hamiltonian should obey. The first is that there is a double scaling
limit and that the resulting string coupling constant squared should be given
by the black hole mass $M$. The second constraint is the required vanishing
of all odd $N$-point amplitudes. Finally, in agreement with the world sheet
description of the black hole string theory one should have a natural
$SL(2,\IR)$ symmetry.

Consider a general modification of the inverted oscillator Hamiltonian
$$h(p,x) \quad\rightarrow \quad h_M (p, x)={1\over 2}\,(p^2 -  x^2)
+ M \delta h (p, x)\,\,.\eqno\eq$$
We have assumed that the deformation is
described by a term linear in $M$.
The first requirement for $\delta h$ is a scaling property
to ensure that the string coupling is proportional to $M^{-1/2}$.
In collective field theory after a shift by the classical ground state one has
that the string coupling  generally
proportional to $({dx \over d\tau})^{-2}$. Thus the above
requirement is satisfied if the deformation operator
$\delta h$ scales as
$\delta h(p, x) \rightarrow \rho^{-2}\delta h(p,x)$
under scale transformations $(p, x) \rightarrow
(\rho p, \rho x)$. This leads to
$$\delta h(p,x)={1\over 2 x^2}\,\,f({p \over x})\,\,.\eqno\eq$$

To further specify the  general function $f(p/x)$,
one invokes the requirement of $SL(2,\IR)$ symmetry. We have seen
in sect.~4, that the usual $c=1$
Hamiltonian $h= (p^2 - x^2)/2$ allows a set of
eigenoperators $O_{j,m}$ satisfying  the
the $w_{\infty}$ algebra~\winfalg.
The origin of this  algebraic structure, which is supposed to encode
the extended nature of strings, can be traced to  existence
of an $SL(2,\IR)$ algebra consisting of
$$\eqalign{&L_1={1 \over 4}\,(p^2-x^2)= h(p,x)\,\,,\cr
&L_2= -{1 \over 4}\,(px + xp)\,\,,\cr
&L_3= {1 \over 4}\,(p^2+x^2)\,\,.}\eqno\eq$$
The  eigenoperators satisfying the $w_{\infty}$
algebra are constructed in terms of the  $SL(2,\IR)$ operators according to
$$O_{j,m}=L_+^{{j+m \over 2}}\,L_-^{{j-m \over 2}}\,,
\qquad L_{\pm}=L_3 \pm L_2\,\,,\eqno\eq$$
which close under the Poisson bracket since the Casimir invariant has
a fixed value
$$L_1^2 + L_2^2 - L_3^2= {3\hbar \over 16}\,\,.\eqno\eq$$
(the Planck constant indicates the effect of operator ordering).

Since the spectrum of  discrete states in the
black hole background is expected to be the same
as that of the usual $c=1$ model in the Minkowski metric,
it is natural to require that the deformed model
should also share a similar algebraic structure.

There is actually a very simple model with the above structure. It is given
for $f=1$ in which case one has the extra term represented by a well known
singular potential. One has  the $SL(2,\IR)$ generators of the form:
$$\eqalign{&L_1(M)= {1\over 2}\,h_M (p,x)
={1 \over 4}\,(p^2-x^2+{M \over x^2})\,\,,\cr
&L_2(M)= -{1 \over 4}\,(px + xp)\,\,,\cr
&L_3(M)= {1 \over 4}\,(p^2 + x^2 + {M \over x^2})\,\,,}\eqno\eq$$
which satisfy
$$L_1(M)+L^2_2(M)-L^2_3(M)= -{M\over 2}+{3\hbar\over 16}\,\,.\eqno\eq$$
We  note that because of  different constraint
for the Casimir invariant  the  algebra of eigenoperators is now
modified in an $M$-dependent way. The algebraic properties of the
model with the singular potential have been investigated in detail
in [\AJcmp].
The model is exactly solvable and possesses some features characteristic of
black hole background.

Let us proceed to describe the properties of the  deformed model:
$$h_M(p,x) = {1 \over 2}(p^2 - x^2) + {M \over 2 x^2}\,\,.
\eqn\defham$$
We assume here that $M>0$. Then the genus zero free energy in the limit of
vanishing scaling parameter, $\bar M \rightarrow 0$, behaves like
$F\sim {N^2\over 8\pi\sqrt{2}}\,\bar M \log {\bar M \over \sqrt{2}}$.
The double scaling  limit is thus the limit $\bar M \rightarrow 0,
\,N\rightarrow \infty$
with $M \equiv N^2 \bar M$ being kept fixed. After  the
usual rescaling, $\,x\equiv \sqrt{N}\times$  matrix eigenvalue,
the system is reduced to the free fermion system with the one-body
potential $-{1 \over 2}x^2 + {M \over 2x^2}$.
Note that in the limit $M \rightarrow 0$ the potential
approaches the usual inverted harmonic oscillator potential with
a repulsive $\delta$-function-like singularity.

The solution of the classical equations
with energy $\epsilon$ reads
$$x^2(t) =-\epsilon +  \sqrt{M + \epsilon^2} \cosh 2t\,\,.\eqno\eq$$
The ground state corresponding to zero Fermi energy is obtained
by setting $\epsilon =0$ and replacing the time variable $t$
by the time-of-flight coordinate $\tau$, $x^2=\sqrt{M}\cosh 2\tau$.
This is recognized as  precisely the quantity appearing in
the integral transformation \bhfieldred. The
$\delta$-function present in the transformation gives a relation
between the black hole and the matrix model variables.
It serves to identify the matrix eigenvalue as
$$x^2=(u e^{-2t/3}+v e^{2t/3})/2\,\,.$$
The string coupling is now space dependent
$$g(\tau) \equiv {\sqrt{\pi} \over 12}\,\Bigl({dx \over
d\tau}\Bigr)^{-2} = {1 \over 48}\sqrt{{\pi \over M}}\,
\Bigl({1 \over \sinh^2\tau } +{1 \over \cosh^2 \tau}\Bigr)\,\,,
\eqno\eq$$
with the required relation with the black hole mass and
the asymptotic behavior at large $\tau$.

The tree  level scattering amplitudes are generally obtained
from the exact solution of the classical equations. The exact
solution to the collective  equations  has the following
parametrized  form
$$\eqalign{&x(t,\sigma)=\Bigl[-a(\sigma)+\sqrt{M+a^2(\sigma)}\,
\cosh 2(\sigma-t)\Bigr]^{1/2}\,\,,\cr
&\alpha(t,\sigma)={1 \over x(t,\sigma)}\,\sqrt{M+a^2(\sigma)}\,
\sinh 2(\sigma-t)\,\,.}\eqn\bhparasol$$
It contains an arbitrary function
$a(\sigma)$ describing the deviation of the
Fermi surface from its ground state form.
The asymptotic behavior for large $x$, of the profile
function reads
$$\alpha_{\pm}(t,\tau)=\pm x(\tau)\Bigl(1- {\psi_{\pm}
(t\pm \tau) \over x^2(\tau)}\Bigr) + {\cal O}({1 \over x^2})
\,\,.\eqno\eq$$
The functions $\psi_{\pm}(t\pm \tau)$
represent incoming and outgoing waves, respectively.
In terms of the $\eta$ field, we have
$$(\partial_{t}\pm \partial_{\tau})\eta =
\pm {1 \over \sqrt{\pi}}\,\,\psi_{\pm}(t\pm \tau)\eqno\eq$$
for $t \rightarrow \mp \infty$.

A nonlinear relation between incoming and outgoing fields
$\psi_+$ and $\psi_-$ can be established
by studying the time delay. Take the times at which a parametrized
point $\sigma$ is passed by the incoming and outgoing waves at a fixed
value of large $\tau$ be $t_1 (\rightarrow -\infty)$ and
$t_2 (\rightarrow \infty)$, respectively.
{}From \bhparasol\ we have then
$$\eqalign{\bigl(&M+a^2(\sigma)\bigr)^{1/4}\,e^{\sigma-t_1}=
M^{1/4}\, e^{\tau}\,\,,\cr
\big(&M+a^2(\sigma)\bigr)^{1/4}\,e^{t_2-\sigma}=M^{1/4}\,e^{\tau}\,\,.}
\eqno\eq$$
This implies
$$t_1+\tau=t_2 -\tau +{1\over 2}\log\Bigl(1+{a^2(\sigma)\over M}
\Bigr)\,\,,\eqno\eq$$
and hence
$$a(\sigma) = \psi_+(t_1 +\tau) = \psi_-(t_2 -\tau)\,\,.\eqno\eq$$
This then gives  functional scattering equations connecting
the incoming and outgoing waves
$$\psi_{\pm}(z)= \psi_{\mp}\Bigl(z \mp {1 \over 2}\log\,
(1+ {1 \over M}\,\psi^2_{\pm}(z))\Bigr)\,\,. \eqn\scatt$$
The result is  similar  in form to that of the usual $c=1$ model.
However, one notes a crucial difference that
Eq.~\scatt\ is even, i.e. it is invariant under
the change of sign of $\psi_{\pm}\rightarrow -\psi_{\pm}$.
This  ensures that the number of particles participating in
the scattering is even. All the odd point amplitudes do vanish in
the deformed model.

The explicit power series solution of \scatt\ is
$$\psi_{\pm}(z) = \sum_{p=0}^{\infty}{M^{-p} \over p!\,(2p+1)}\,\,
{\Gamma(1\pm {1 \over 2}\partial_z) \over \Gamma(1-p \pm
{1 \over 2}\partial_z)}\,\,\psi^{2p+1}_{\mp}(z)\,\,,\eqno\eq$$
which shows that the amplitudes are essentially polynomial with respect
to the momenta without any singularity.

The scattering equation \scatt\ can also be derived using directly the
exact states [\DeRo,\Da], as was done in sect.~5 for the $c=1$ model.
First, one recalls the symmetry structure of the collective
theory with Hamiltonian~\defham\ given in [\AJcmp]:
$$\eqalign{
\bigl[ O_{j_1,m_1}^{a_1},  O_{j_2,m_2}^{a_2} \bigr] =
&-4i(j_1 m_2 - m_1 j_2)\,O_{j_1+j_2-2, m_1+m_2}^{a_1+a_2+1}\cr
\noalign{\vskip 0.2cm}
&-4i(a_1 m_2 - m_1 a_2)\,O_{j_1+j_2, m_1+m_2}^{a_1+a_2-1}\,\,,}\eqno\eq$$
where
$$\eqalign{
O_{j,m}^{a} \equiv \int {dx\over 2\pi} \int_{\alpha_-}^{\alpha_+}
d\alpha &\Bigl( \alpha^2 - x^2 +{M\over x^2} \Bigr)^a
\Bigl( (\alpha + x)^2 +{M\over x^2} \Bigr)^{{j+m\over 2}}\cr
\noalign{\vskip 0.1cm}
&\Bigl( (\alpha - x)^2 +{M\over x^2} \Bigr)^{{j-m\over 2}}\,\,.}\eqn\op$$
The operators $T^{(-)}_{-ip}\, (T^{(+)}_{ip})$
which create exact tachyon in (out) states
are obtained by analytic continuation
$j\to \pm ip/2$ of some special operators \op:
$$\eqalign{
&O_{j,j}^{a=0} = \int  {dx\over 2\pi} \int_{\alpha_-}^{\alpha_+}
d\alpha
\Bigl[ (\alpha +x)^2 +{M\over x^2}\Bigr]^j \equiv T_{2j}^{(+)}\,\,, \cr
\noalign{\vskip 0.2truecm}
&O_{j,-j}^{a=0} = \int  {dx\over 2\pi} \int_{\alpha_-}^{\alpha_+}
d\alpha \Bigl[ (\alpha - x)^2 + {M\over x^2}\Bigr]^j
\equiv T_{2j}^{(-)}\,\,.}\eqno\eq$$
Eq.~\scatt\ then easily follows from an asymptotic expansion of
$$T^{(+)}_{ip,+} = -T^{(+)}_{ip,-}\,\,,$$
for large $\tau$, where $T^{(+)}_{ip,+}$ and $T^{(+)}_{ip,-}$
are defined by
$$T^{(+)}_{ip}=T^{(+)}_{ip,+}-T^{(+)}_{ip,-}\,\,.\eqno\eq$$

The scattering equation
can also be rewritten in terms of  energy-momentum tensor
$$T_{\pm\pm}(z) = {1 \over 2\pi}\, \psi^2_{\pm}(z)\,\,,\eqno\eq$$
as
$$\int dz\,\,e^{i\omega z} \,T_{\pm\pm}(z)
={M \over 2\pi}\int dz \,\,e^{i\omega z}
{1 \over 1 \pm {i\omega \over 2}}\,
\Bigl[\Bigl(1+{2\pi\over M}\,T_{\mp\mp}(z)\Bigr)^{1\pm{i\omega \over 2}}
-1\Bigr]\,\,.\eqno\eq$$
One can easily check that this defines a canonical
transformation by confirming that the Virasoro algebra
(at the level of Poisson bracket) is preserved by this transformation.
This relation for  the energy momentum tensor is very
similar to the one obtained recently by Verlinde and Verlinde
[\VV] for the $S$-matrix of the $N=24$ dilaton gravity.A slight difference is
that in the case of dilaton gravity one has the derivative of the energy
momentum tensor participating in the equation.

In conclusion, the framework presented above gives some initial
picture of a black hole in the matrix model. It contains some basic
requirements for a consistent formalism. In particular, the scaling
properties of the black hole mass deformation are in agreement with
the corresponding vertex operators (see also [\Eguchi]). The
particular singular matrix model studied has an interesting double
scaling limit with an $SL(2,\IR)$ algebraic structure. This clearly
is not enough to completely describe black hole and further
generalizations and studies are likely to lead to further interesting
results.

\noindent
{\bf Acknowledgement}

These notes were written while the author was visiting LPTHE, Paris 6,
Paris, France.
He is grateful to the members of the high energy group for
their hospitality.

\refout

\end